\documentclass{bmcart}

\usepackage{natbib} 
\usepackage{float}
\usepackage{subfig}
\usepackage{graphicx}
\usepackage{amsmath}
\usepackage{enumerate}
\usepackage[utf8]{inputenc} 

\startlocaldefs
\endlocaldefs

\begin{document}

\begin{frontmatter}

\begin{fmbox}
\dochead{Research}


\title{Evidence of fast neutron sensitivity for ${^3}$He detectors and comparison with Boron-10 based neutron detectors}


\author[
   addressref={aff1},                   
   corref={aff1},                       
   email={giacomo.mauri@esss.se}   
]{\inits{GM}\fnm{Giacomo} \snm{Mauri}}
\author[
   addressref={aff1,aff2},
]{\inits{FM}\fnm{Francesco} \snm{Messi}}
\author[
   addressref={aff1},
]{\inits{KK}\fnm{Kalliopi} \snm{Kanaki}}
\author[
   addressref={aff1,aff3},
]{\inits{RHW}\fnm{Richard} \snm{Hall-Wilton}}
\author[
   addressref={aff1},
]{\inits{FP}\fnm{Francesco} \snm{Piscitelli}}


\address[id=aff1]{
  \orgname{European Spallation Source ERIC (ESS)}, 
  \street{P.O. Box 176, SE-22100},                     %
  \city{Lund},                              
  \cny{Sweden}                                    
}
\address[id=aff2]{%
  \orgname{Division of Nuclear Physics, Lund University},
  \street{P.O. Box 118, SE-22100 },
  \city{Lund},
  \cny{Sweden}
}
\address[id=aff3]{
  \orgname{Dipartimento di Fisica “G. Occhialini”, University of Milano-Bicocca},
  \city{Milano},
  \cny{Italy}
}


\end{fmbox}


\begin{abstractbox}

\begin{abstract} 
The $\mathrm{^3He}$-based neutron detectors are no longer the default solution for neutron scattering applications. Both the inability of fulfilling the requirements in performance, needed for the new instruments, and the shortage of $\mathrm{^3He}$, drove a series of research programs aiming to find new technologies for neutron detection. The characteristics of the new detector technologies have been extensively tested to prove their effectiveness with respect to the state-of-the-art technology.
\\Among these, the background rejection capability is crucial to determine. The signal-to-background ratio is strongly related to the performance figure-of-merit for most instruments. These are designed to exploit the high flux expected from the new high intensity neutron sources. Therefore, an inadequate background rejection could significantly affect the measurements, leading to detector saturation and misleading events. This is of particular importance for the kind of techniques in which the signals are rather weak.   
\\For the first time, the sensitivity of $\mathrm{^3He}$ detectors to fast neutrons, up to $E_n =10$ MeV, has been estimated. Two independent measurements are presented: a direct calculation based on a \textit{subtraction method} used to disentangle the thermal and the fast neutron contribution, while a further evidence is calculated indirectly through a comparison with the recently published data from a $\mathrm{^{10}B}$-based detector. Both investigations give a characterization on the order of magnitude for the sensitivity. A set of simulations is presented as well in order to support and to validate the results of the measurements. A sensitivity of $4\cdot 10^{-3}$ is observed from the data. This is two orders of magnitude higher than that previously observed in $^{10}$B-based detectors.
\end{abstract}



\end{abstractbox}
%

\end{frontmatter}



\section*{Introduction}
The increasing complexity in science investigations, driven by technological advances, is reflected in the studies of neutron scattering science which enforces a diversification and an improvement of experimental tools, from the instrument design to the detector performance. Complexity means study of realistic and ever more complex heterogeneous samples within extreme or natural environments, and investigation of multiple physical properties of materials~\cite{ESS_TDR}. Therefore, the instruments must be flexible, permitting exchanges between brightness and constrained phase-space: in turn implying better resolution, optimization of signal-to-background ratio, and the use of polarized neutrons when necessary. The increasing performance demands drive for development of neutron detector technologies; not only the enhanced detector response is of a crucial importance, but the background rejection plays a significant role as well.
\\ The new generation, high-intensity neutron sources like the Spallation Neutron Source~\cite{SNS} (Oak Ridge Laboratory, TN, USA), the Japan Spallation Source~\cite{JPARC} (Tokai-mura, Japan), and the European Spallation Source~\cite{ESS_TDR,ESS,ESS-design} (ESS, Lund, Sweden), presently under construction, will ensure an intensity increase of at least one order of magnitude~\cite{Mezei2007, VETTIER_ESS,ESS2011}. The trends in neutron scattering applications point out the importance of exploiting long-wavelength neutrons with energies in the range of 0.1 meV to 80 meV (30\AA- 1\AA). The high brightness, foreseen at ESS, in this energy range will allow faster measurements, increased use of polarized neutrons, investigation of smaller samples and detection of weaker signals. The actual limitations are not only set by the available neutron flux, but also by the detector response.
\\The $\mathrm{^3He}$-based technology has been predominant for thermal neutrons detection. Both the availability~\cite{HE3S_kouzes} and the requirements of higher performance, e.g., of counting rate capability, spatial resolution and background sensitivity, are the reasons why a number of research programs are now aiming to find technologies that would replace $\mathrm{^3He}$~\cite{COLL_icnd,HE3S_karl,HE3S_hurd}. A promising technique is based on solid converter layers ($\mathrm{^{10}B}$, Gd) and either gas proportional counter or solid state material as the sensing medium. Some examples that employ $\mathrm{^{10}B}$ are: the Multi-Grid~\cite{MG_2017,MG_IN6tests,MG_patent,MG_joni}, the Multi-Blade~\cite{MIO_MB2014,MIO_MB16CRISP_jinst,MIO_ScientificMBcrisp}, the Jalousie detector~\cite{DET_jalousie}, BandGEM~\cite{MPGD_GEMcroci, Bgem}, the Boron-coated straw-tubes~\cite{STRAW_lacy2011} and CASCADE~\cite{DET_kohli}. Other technologies are based on Gd, e.g., the Gd-GEM~\cite{DET_doro1,gdgem}, and a whole branch dedicated to the development of solid state neutron detectors coupled with Gadolinium as a converter layer~\cite{Mireshghi_sigd,SCHULTE_sigd,PETRILLO-solidstate,Mauri_psa}. This research phase involves the investigation of all the characteristics of a prototype, not only to prove the effectiveness of the new technology, but to demonstrate their relative performance to $\mathrm{^3He}$ detectors.   
\\The signal-to-background ratio is one of the features that strongly affect the figure-of-merit for most neutron scattering instruments, especially when the detected signals are rather weak and the background discrimination is crucial to disentangle the two effects. The signal is defined by the source brightness and phase space acceptance, and a better background rejection can improve this figure-of-merit, leading to significant impact on instrument's operation, especially at the new high intensity sources.
\\ A previous investigation of the fast neutron sensitivity has been performed on a $^{10}$B-based thermal neutron detector~\cite{MIO_fastn}, the Multi-Blade detector~\cite{MIO_MB2014,MIO_MB16CRISP_jinst}. A sensitivity on the order of $10^{-5}$ has been measured, about $10^3$ times higher than the gamma sensitivity observed with the same detector~\cite{MIO_fastn,MIO_MB2017}. The fast neutrons make a significant contribution to the background, hence, it is important to study the detector response to it. 
\\ This paper investigates, for the first time, the fast neutrons sensitivity of $\mathrm{^{3}He}$-based neutron detector. The discussion is presented in comparison with the Muti-Blade detector, whose response to both $\gamma$-ray~\cite{MIO_MB2017} and fast neutrons~\cite{MIO_fastn} has been already studied. 
The fast neutron sensitivity has been calculated as a function of energy threshold. Moreover, a set of simulations has been performed to verify the \textit{subtraction method} applied to the measurements. From a campaign of measurements perfomed at the CRISP~\cite{CRISP1} reflectometer at ISIS (Science \& Technology Facilities Council in UK~\cite{ISIS}), a direct comparison between an $\mathrm{^{3}He}$ and the Multi-Blade detector~\cite{MIO_MB16CRISP_jinst} is presented as a further evidence of the measured fast neutron sensitivity. Both results are aimed at quantitative analysis on the order of magnitude; higher accuracy cannot be achieved with these kind of measurements.  

\section{Fast neutron sensitivity of $^3$He-tube}

\subsection{Set-up description and methodology}\label{setup}
\begin{figure}[htbp]
\centering
\centering
\subfloat[]{\includegraphics[width=.265\textwidth,keepaspectratio]{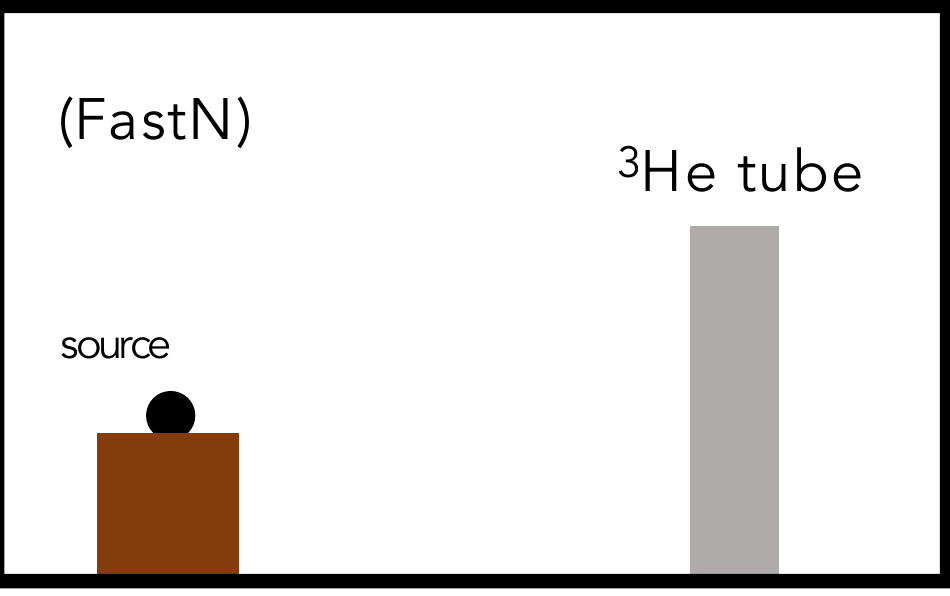}}
\subfloat[]{\includegraphics[width=.30\textwidth,keepaspectratio]{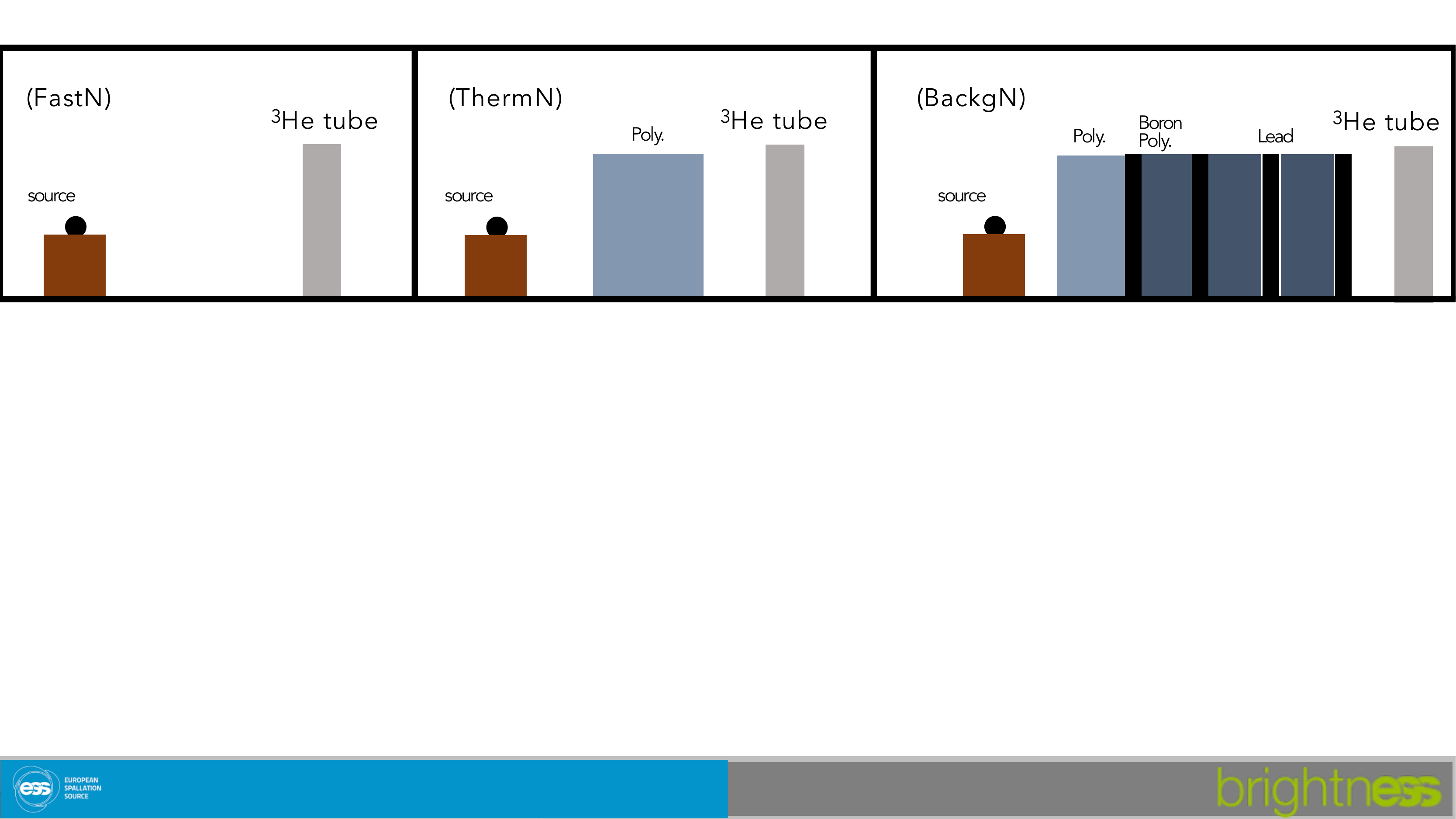}}
\subfloat[]{\includegraphics[width=.375\textwidth,keepaspectratio]{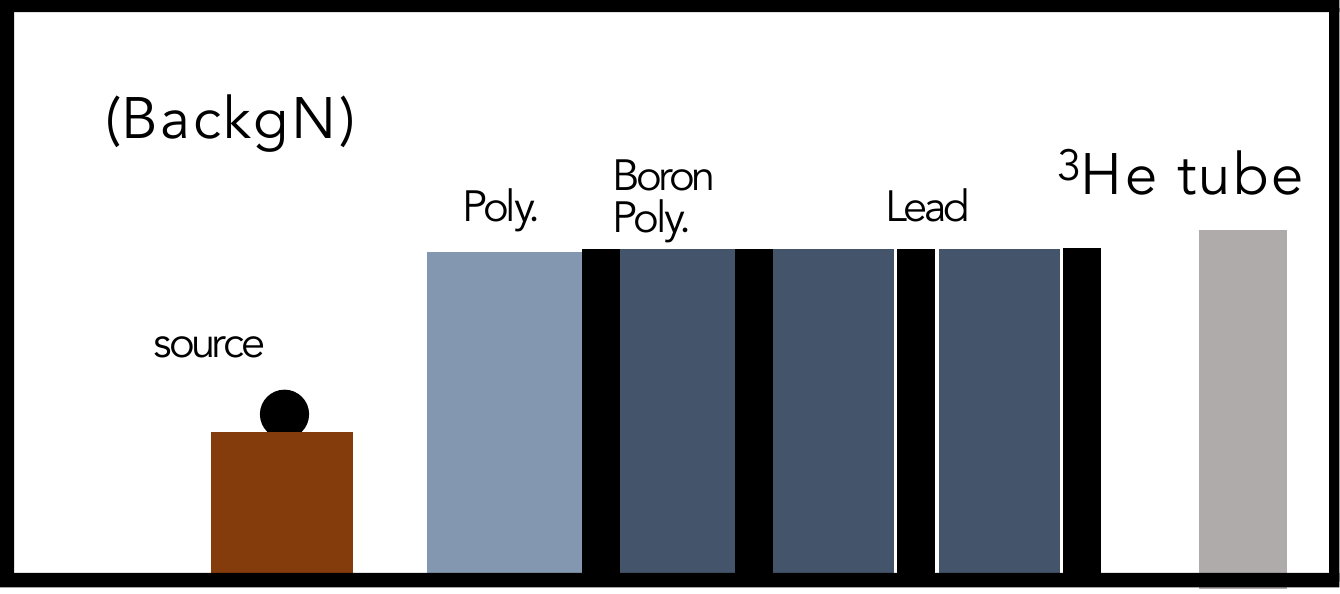}}\\
\caption{\label{he-config} \footnotesize Configurations used in the measurements with the $^3$He-tube: (a) the total flux reaches the detector, (b) flux thermalized through a polyethylene brick, (c) flux highly reduced through polyethylene, borated polyethylene and lead bricks, in order to perform a background measurement with the source.}
\end{figure}

A set of measurements has been carried out at the Source Testing Facility of Lund University~\cite{SF2} with a $^3$He-tube, to investigate the fast neutron sensitivity of this class of devices. The tube used is a RS P4 0810 227, manufactured by General Electric (GE) Reuter Stokes~\cite{ge-reuter}. It is made of stainless steel and it has an active length of 10 inch (250 mm), a radius of 1 inch (25 mm), a gas pressure of 10 bar and an efficiency of 96.4\% (at 2.5\AA). 
\\Two fast neutron radioactive sources have been employed: an $^{241}$Am/Be~\cite{SF1} and a $^{238}$Pu/Be source~\cite{PuBe_AmBe}. The fast neutron energy range is definined by the emission of these sources, it extends up to approximately $E_n = 10$ MeV. A measurement was also performed with a $^{60}$Co gamma source to have a comparison with a previous work on gamma sensitivity performed using the same $^3$He detector~\cite{Rossi-gsHe}.
\\ With the previous evaluation of sensitivity to fast neutron with the $^{10}$B-based detector, excellent separation of the fast neutron and thermal neutron contributions was possible by using identical detectors one containing the $^{10}$B$_4$C layer, sensitive to both contributions, and a second detector without the converter layer, only sensitive to fast neutrons. However, it is not trivial to have a good separation between the different neutron contributions, when using the $^3$He-tube, because of the high efficiency to thermal neutrons. 
\\For this reason the measurements were performed in several configurations, a sketch of which is shown in figure~\ref{he-config}. In configuration (FastN), figure~\ref{he-config}(a), the total flux of the source reaches the detector, i.e., fast and thermal contribution. In configuration (ThermN), figure~\ref{he-config}(b), the flux is partially thermalized through a polyethylene brick, while in (BackgN), figure~\ref{he-config}(c), a shielding of polyethylene, borated-polyethylene and lead is placed between the source and the detector, in order to reduce as much as possible the incoming neutrons both fast and thermal. The latter is a measurement of the background with the source in place.

The detector-source distance is d = 250 mm, the depth of the polyethylene used in configuration (ThermN) is 100 mm, while in the case of (BackN) the depths of polyethylene, borated polyethylene and lead are 40 mm, 25 mm and 2 mm, respectively. These values are used in the simulation to reproduce the actual set-up. The solid angle has been calculated on a rectangular surface of the detector (150x10) mm$^2$ for the fixed distance d. Considering a point-like source, it results in a solid angle of $\sim$0.002 sr ( $\sim$0.2\% of a sphere). It has been estimated that the measurement of the distance between the source and the sensitive region can lead to an uncertainty which is less than a factor two variation for a misplacement of a $\pm$ 25 mm from the fixed location of the source~\cite{MIO_fastn}. 
 
A \textit{subtraction method} has been used to disentangle the thermal and fast contribution. Referring to figure~\ref{he-config}, we define the flux reaching the detector in the three measurements as:

\begin{enumerate}[(a)]
\item $\mathrm{FastN} = \phi_{fn} + \varphi_{tn}$
\item $\mathrm{ThermN} = \varphi_{fn} + \phi_{tn}$
\item $\mathrm{BackN} = \varphi_{fn} + \varphi_{tn}$
\end{enumerate}

Where $\phi$ define the main contribution of the incoming neutron flux, $fn$ in the case of fast neutrons and $tn$ for thermal neutrons; $\varphi$ is the component of flux which is considered as background with respect to the main contribution. By subtracting BackN from the other two contributions, we obtain:

\begin{align}
\mathrm{FastN} - \mathrm{BackN} = \phi_{fn} + \varphi_{tn} - (\varphi_{fn} + \varphi_{tn}) = \phi_{fn} -\varphi_{fn} \label{ffn}\\ 
\mathrm{ThermN} - \mathrm{BackN} = \varphi_{fn} + \phi_{tn} - (\varphi_{fn} + \varphi_{tn}) = \phi_{tn} - \varphi_{tn}
\label{fth}
\end{align} 

The subtraction allows to separate the fast neutron and thermal neutron contributions, as shown in equations~\ref{ffn} and~\ref{fth}. A similar discrimination, based on a more complex analytical procedure using the Inverse Matrix Method can be found in~\cite{fastNmeas}. The \textit{subtraction method} has been applied on both simulations and measurements.
\\ The normalization to the flux, number of neutrons per unit time and area, is obtained multiplying the activity of the source by the solid angle~\cite{MIO_fastn}. Note that, for (ThermN) only a fraction of thermalized flux reaches the detector, therefore we calculate the total and thermalized flux from the simulations. The calculated flux, $C$, is the integral of counts in a selected incident neutron energy range. Note that the solid angle is included in this calculation for the simulation.

\begin{equation}
C = \int_{E_1}^{E_2} C(x) \, \mathrm{d}x
\label{eqnorm}
\end{equation}

Where $C(x)$ is the energy distribution of the counts, while $E_1$ and $E_2$ define the energy range of integration. This range varies for each configuration: in the case of (FastN) the energy limit is set by the detectable energy, $ E_2 \approx 1$ MeV. For (ThermN) it is limited by the thermal neutron energy, for this study we consider $ E_2 = 0.050$ eV.   
\\ Once the normalization is calculated, we can derive the expression for the sensitivity as the integral of the normalized counts, $f(x)$, above a certain energy threshold:

\begin{equation}
\epsilon = \int_{E_{th}}^{E_f} f(x) \, \mathrm{d}x
\label{eqsens}
\end{equation}

Where $E_{f}$ is the maximum detected energy and $E_{th}$ is the selected energy threshold. 

\subsection{Simulations}\label{simulation}

The neutron reactions involved in the case of a $^3$He counter are four: 

\begin{enumerate}
\item $^3$He recoil $^3He + n \rightarrow \, ^3He' + n' $ (elastic scattering), 
\item (n,p) $^3He + n \rightarrow p + T + 764 keV$,
\item (n,d) $^3He + n \rightarrow ^2H + ^2H -3.27 MeV$,
\item $\gamma$-rays.
\end{enumerate}

\begin{figure}[htbp]
\centering
\includegraphics[width=.8\textwidth,keepaspectratio]{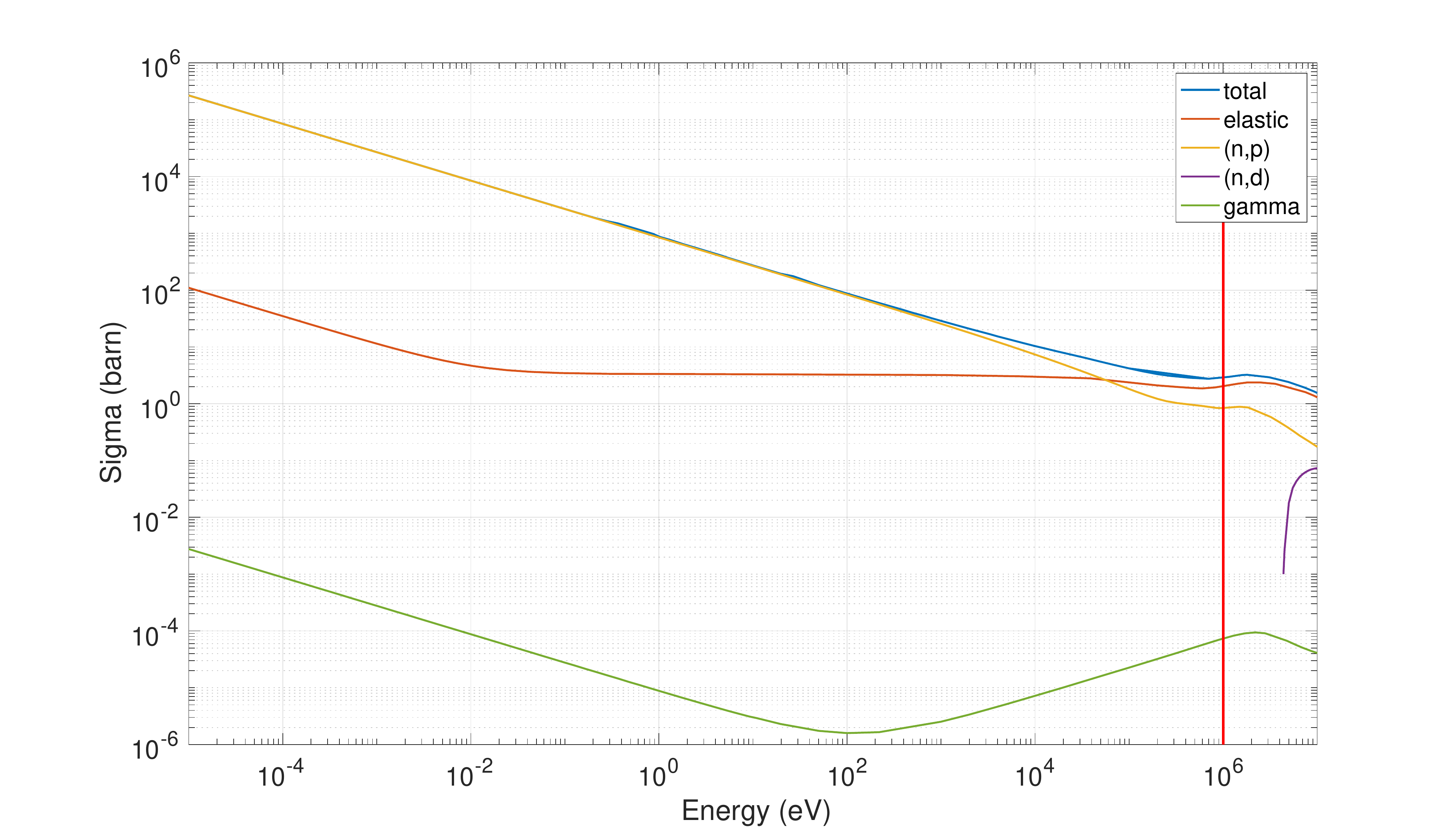}
\caption{\label{sigmaHe} \footnotesize Neutron cross section of $^3$He in the energy range 0-10 MeV for the 4 reactions: $^3$He recoil, (n,p), (n,d) and $\gamma$-rays. The total cross section is shown as well. Data from~\cite{NIST}.}
\end{figure}

The contribution of the (n,d) reaction is negligible because of the high threshold energy ($>$ 4 MeV) and the small cross section for this process. Below 1 MeV the (n,p) reaction has the highest cross section, i.e., contributes mostly to the energy deposition in a $^3$He tube. The cross section for this process is proportional to $1/v$, where $v$ is the velocity of the neutron. It is used in the detection of thermal neutrons, because the cross section increases as the kinetic energy of the neutron decreases. Considering the energy region above 1 MeV, the $^3$He recoil is the predominant process. In figure~\ref{sigmaHe} the cross sections for the different mechanisms between $10^{-5}$ and $10^7$ eV are shown. The vertical line indicates the separation at 1 MeV. The data refers to the NIST web database~\cite{NIST}.

\begin{figure}[htbp]
\centering
\subfloat[]{\includegraphics[width=.25\textwidth,keepaspectratio]{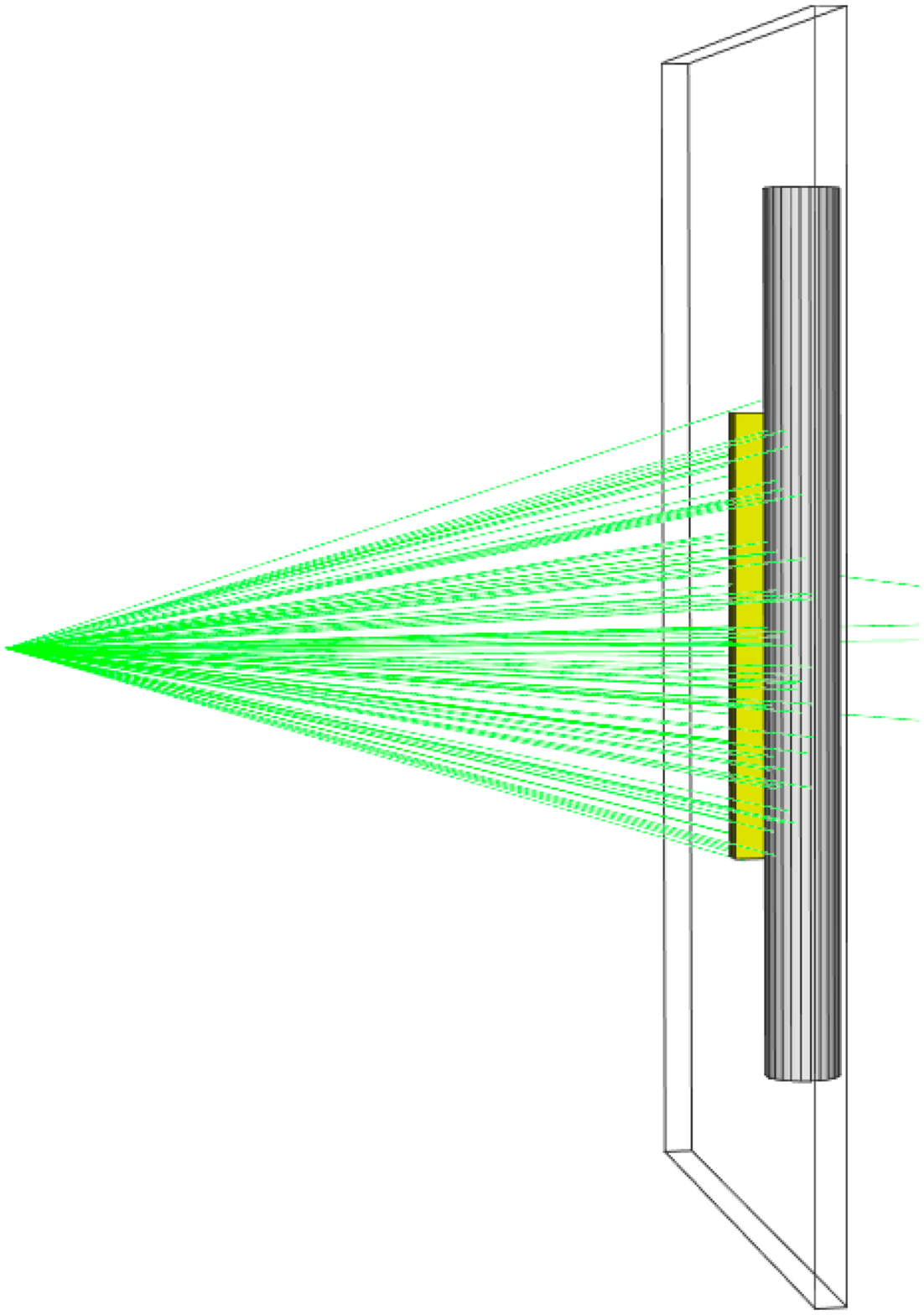}}
\subfloat[]{\includegraphics[width=.23\textwidth,keepaspectratio]{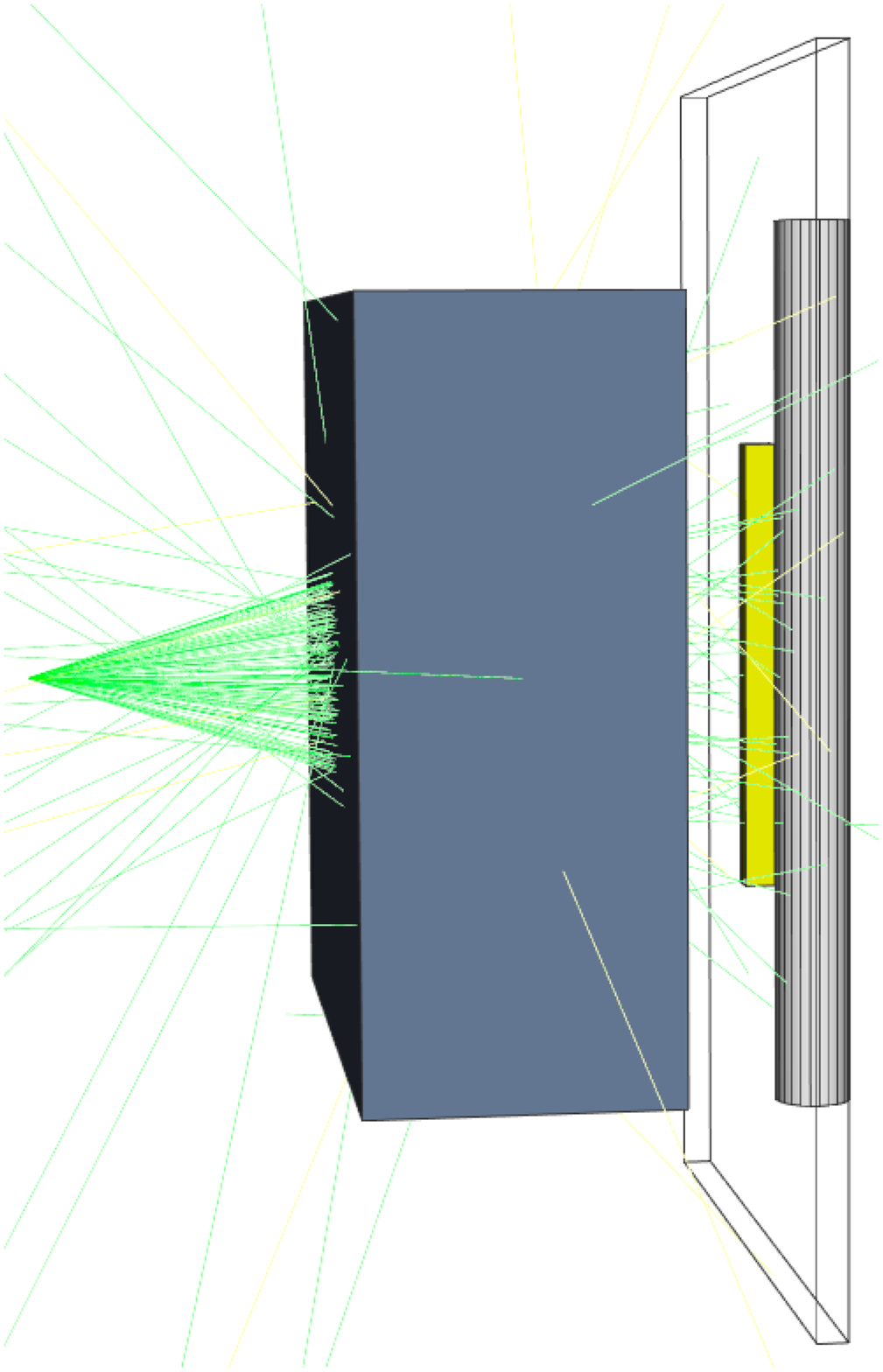}}
\subfloat[]{\includegraphics[width=.26\textwidth,keepaspectratio]{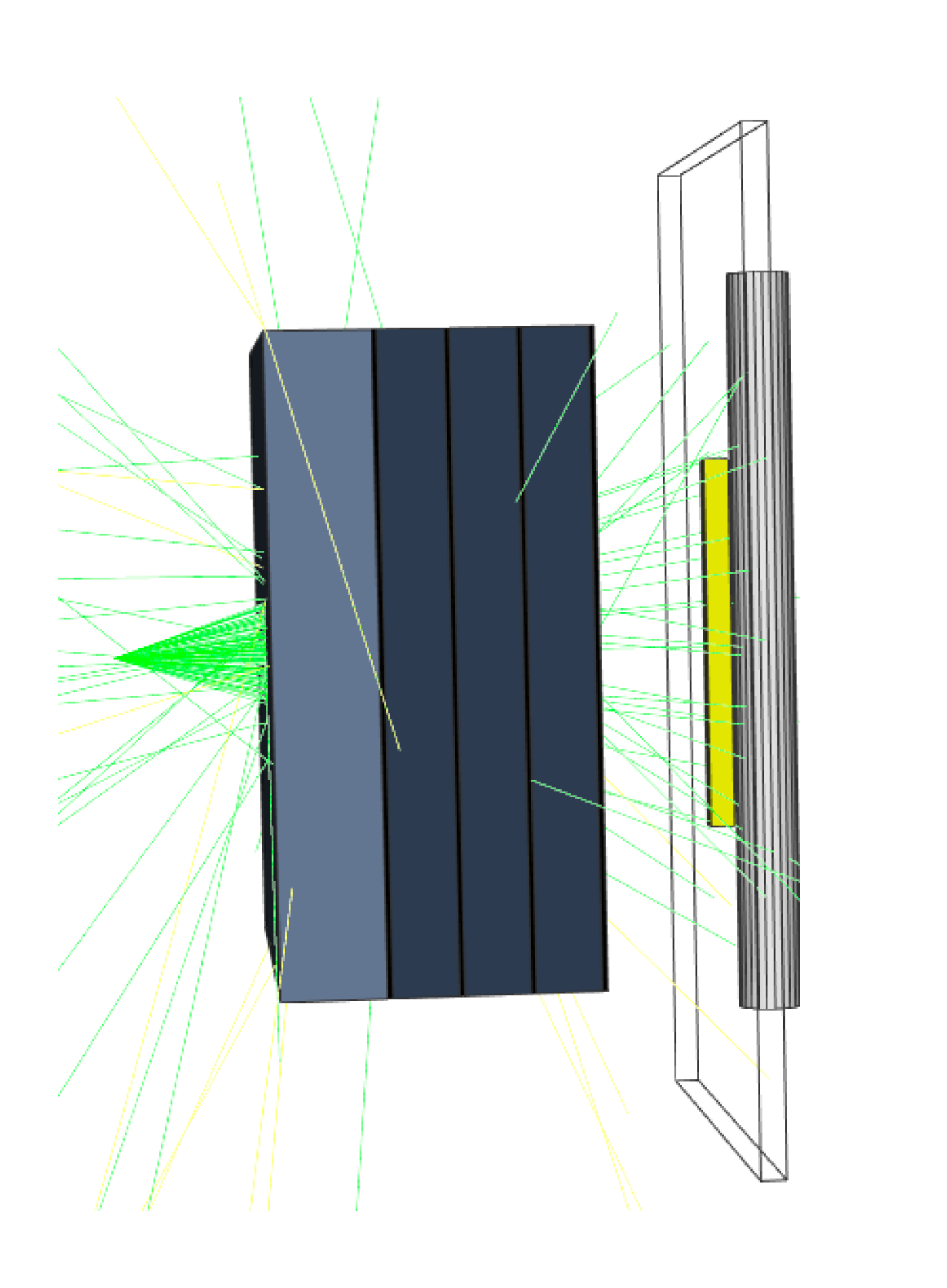}}\\
\caption{\label{sim-sketch} \footnotesize Simulated geometry of configurations (FastN) (a), (ThermN) (b) with the polyethylene brick and (BackgN) (c) with the polyethylene brick and the alternate lead and borated polyethylene slabs in black and dark blue respectively. The detector appears in gray. The wireframe volume acts as the neutron shielding used in the experimental set-up. The yellow window lets the neutrons into the detector active area. Generated neutrons are depicted in green.}
\end{figure}

To support the experimental analysis a Geant4~\cite{geant4a,geant4b,geant4c} simulation model is built within the ESS Detector Group simulation framework~\cite{ess_coding_framework}. The geometry matches the experimental set-up for all three configurations, as seen in figure~\ref{sim-sketch}. The counting gas inside the tube is pure $^3$He, while the material of the steel vessel is represented by $\gamma$-iron with the help of the NXSG4 library~\cite{nxsg4}, which treats the thermal neutron transport. The thickness of the steel is 0.4 cm. The wireframe volume appearing in the geometry represents the shielding used in the set-up. Every neutron incident on it is killed. The yellow volume is the window that shapes the active area of the detector.
\\The neutron generator approximates the Pu/Be source energy distribution but instead of a 4$\pi$ emission, a conical pattern is adopted for better exploitation of statistics. A custom physics list is used to enable correct thermalization in the polyethylene~\cite{backscattering}. 

\begin{figure}[htbp]
\centering
\includegraphics[width=.8\textwidth,keepaspectratio]{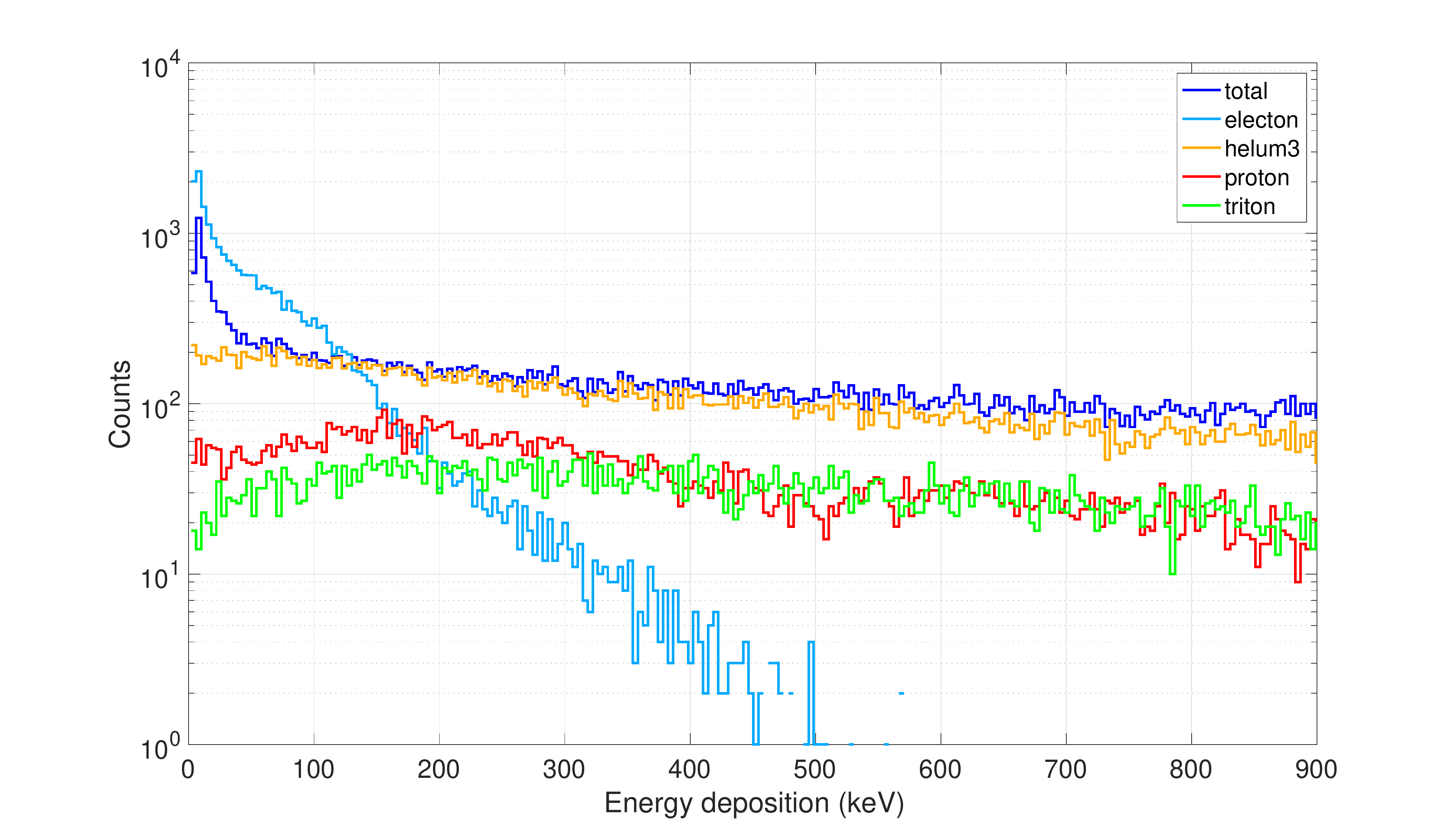}
\caption{\label{PHSimA} \footnotesize Simulated energy deposition for the reaction products in configuration (FastN).}
\end{figure}

\begin{figure}[htbp]
\centering
\includegraphics[width=.8\textwidth,keepaspectratio]{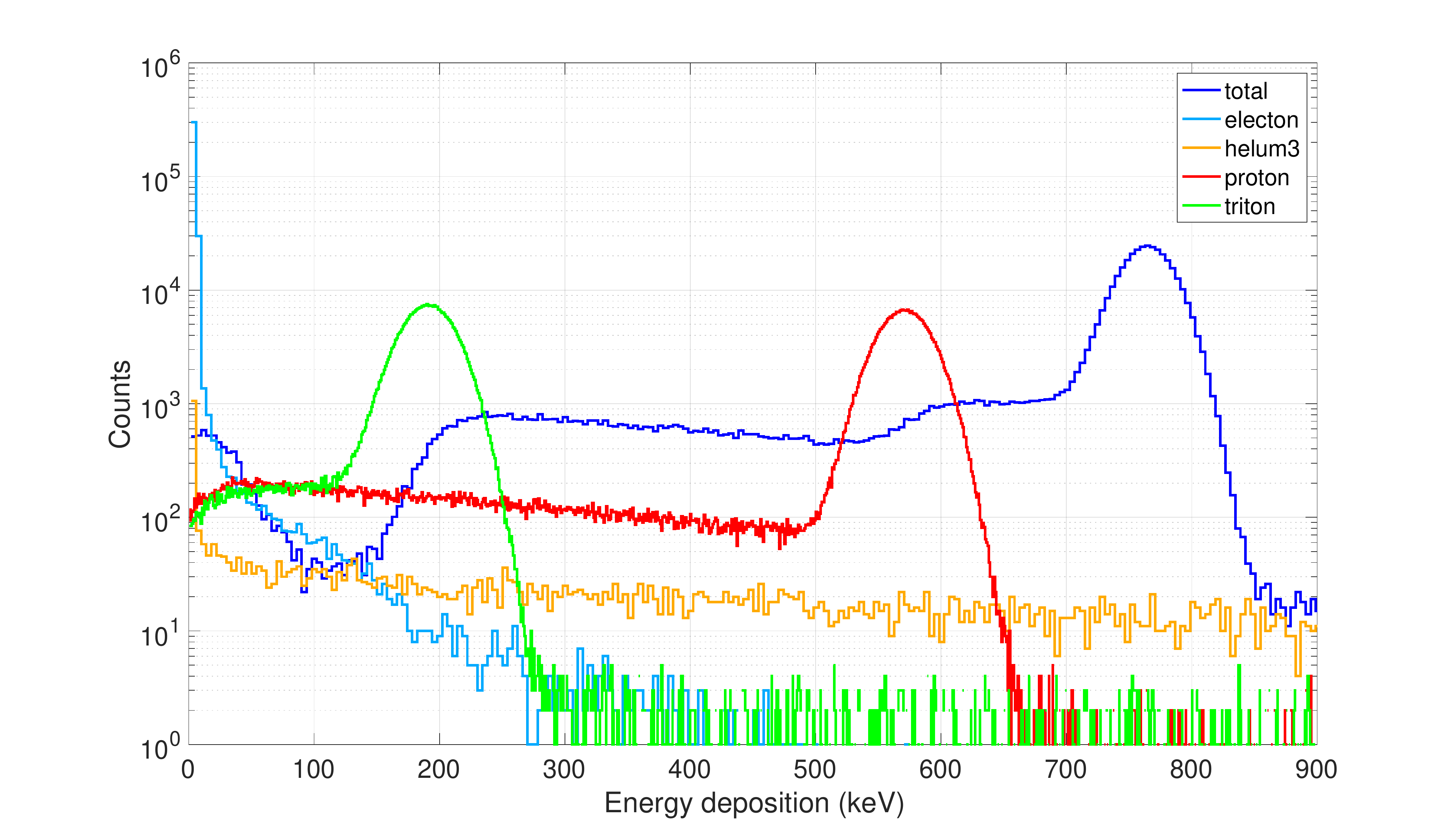}
\caption{\label{PHSimB} \footnotesize Simulated energy deposition for the reaction products in configuration (ThermN).}
\end{figure}

The energy released in a $^3$He detector, with the characteristics described in the previous section, is shown for the configuration (FastN) in figure~\ref{PHSimA} and for configuration (ThermN) in figure~\ref{PHSimB}. 

\begin{figure}[htbp]
\centering
\includegraphics[width=.8\textwidth,keepaspectratio]{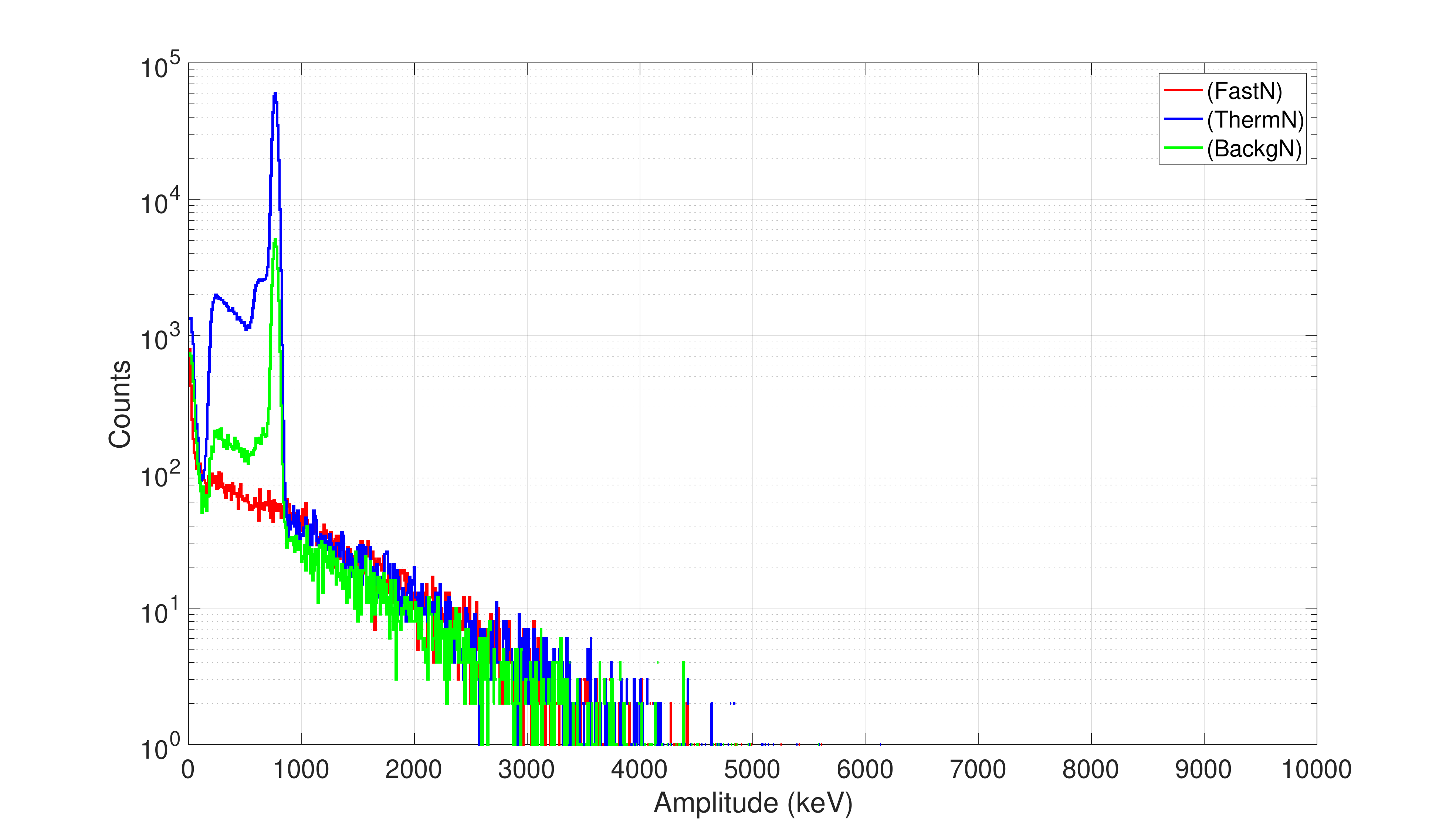}
\caption{\label{edeptot} \footnotesize Simulated total energy deposition for configurations (FastN), (ThermN) and (BackgN) in the full energy emission range of the Pu/Be source.}
\end{figure}

In both cases, the contribution for each of products of the reactions are shown in the energy range below 900 keV, which is the detectable range. As expected, the elastic scattering, $^3$He recoil, contributes mostly to the total Pulse Height Spectrum (PHS) when the fast neutron energies (1-10 MeV in case of the used sources) are taken into account in the simulation, as shown in figure~\ref{edeptot}. On the other hand, for the thermalized flux, the products of the (n,p) process, proton and tritium, give the highest contribution to the energy distribution.  

\begin{figure}[htbp]
\centering
\includegraphics[width=.8\textwidth,keepaspectratio]{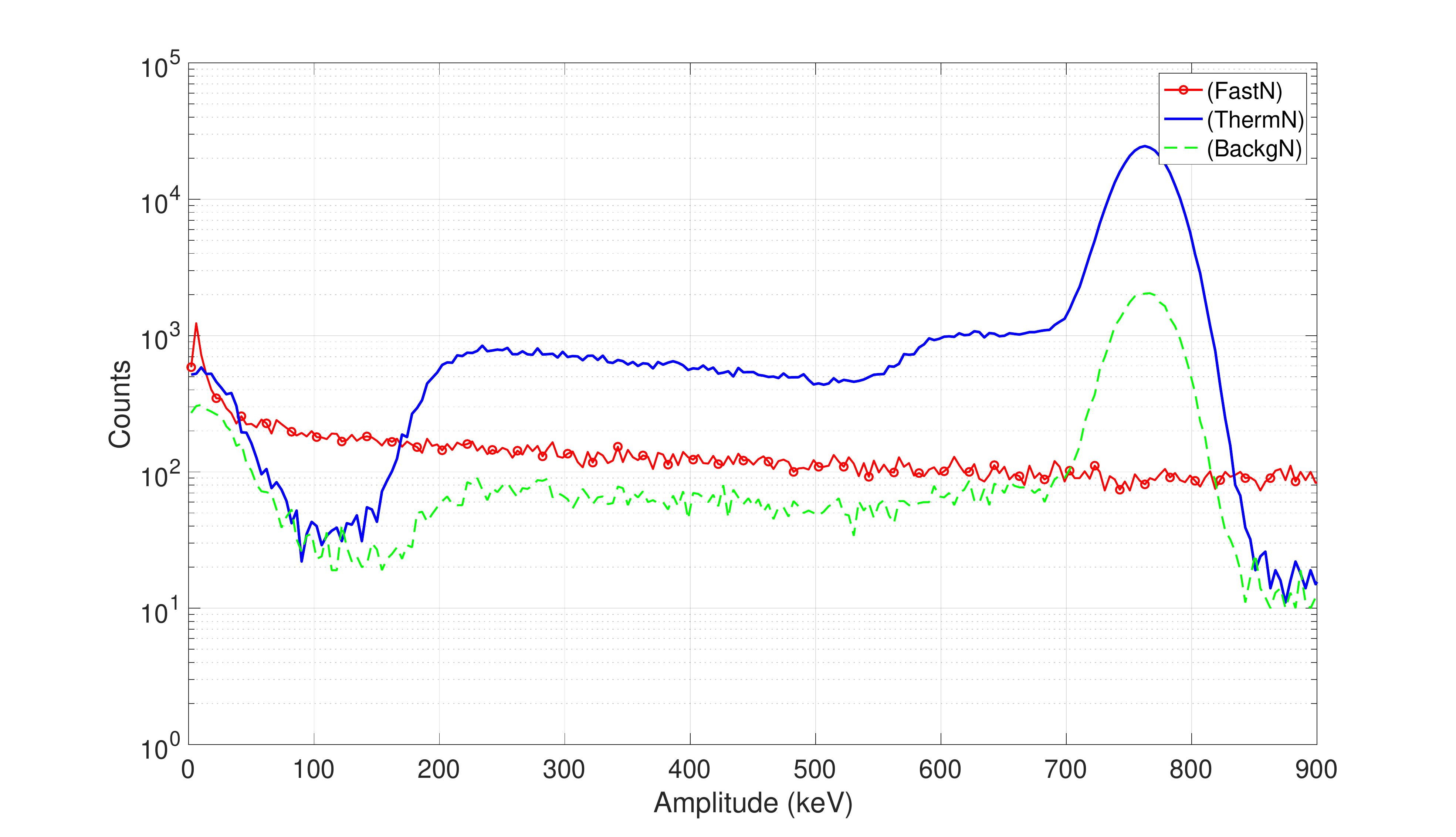}
\caption{\label{PHSimtot} \footnotesize Simulated total energy deposition for configurations (FastN), (ThermN) and (BackgN).}
\end{figure}

The full spectrum of the simulated energy deposition for the three simulation set-ups is shown in figure~\ref{PHSimtot}. With the shielding used in configuration (BackN) the thermal peak is about 20\% of the peak recorded (ThermN) configuration. From the simulations we calculate $\varphi_{tn} = 3\% \, \phi_{th}$, therefore we can assume that with these measurements we identify not only the neutrons that pass through the shielding and thermalize at the detector, but also the fast neutron counts that thermalize from indirect line of view, i.e.\, scattering from around the detector. We estimate the total contribution as about 17\% of the total incoming neutrons calculated in the (FastN) configuration. Note that no normalization is taken into account yet. The estimation of the flux of neutrons reaching the active area of the detector is fundamental to calculate the flux normalization in the different sets of measurements.

\begin{figure}[htbp]
\centering
\subfloat[]{\includegraphics[width=.8\textwidth,keepaspectratio]{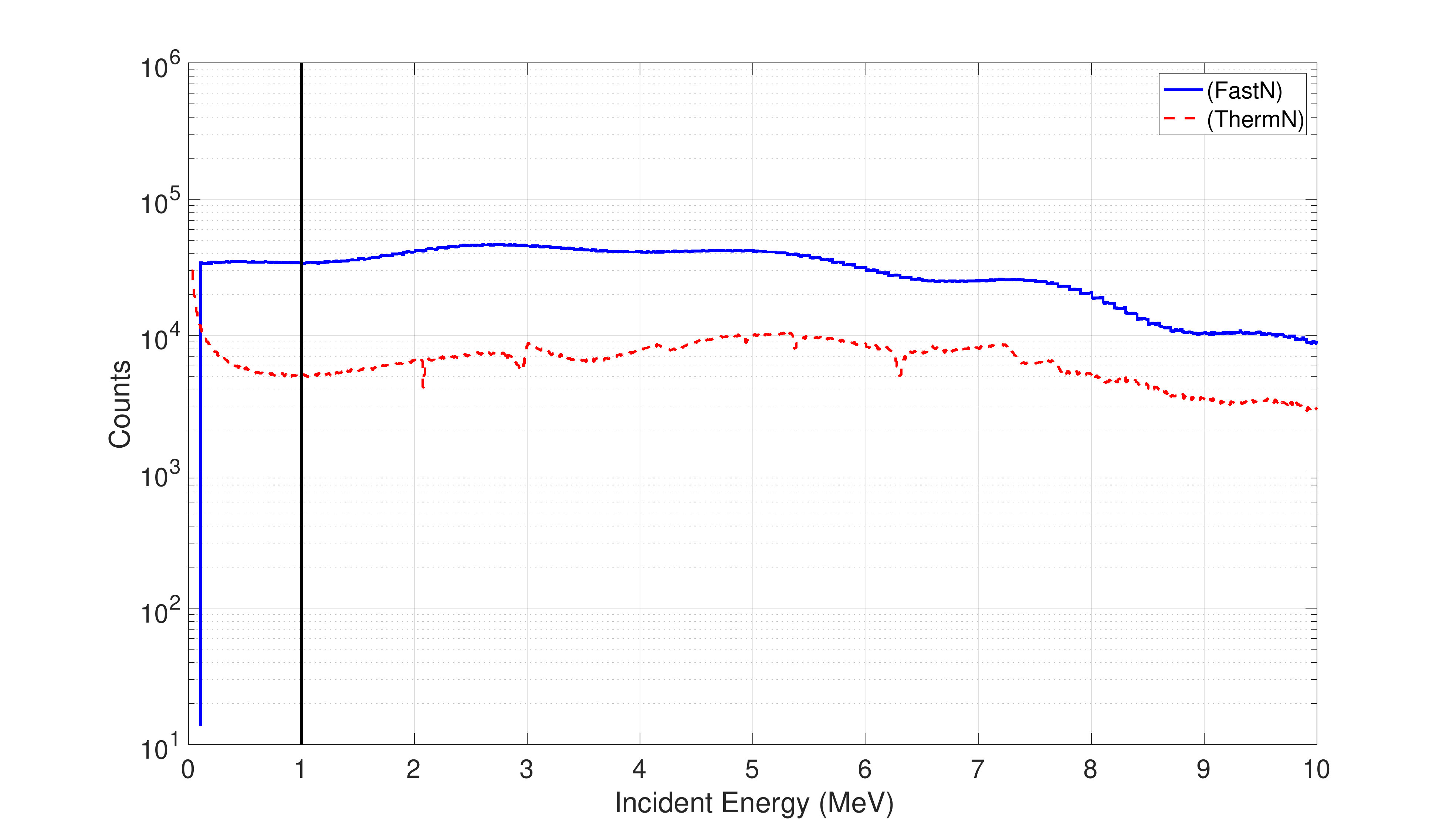}}\\
\subfloat[]{\includegraphics[width=.8\textwidth,keepaspectratio]{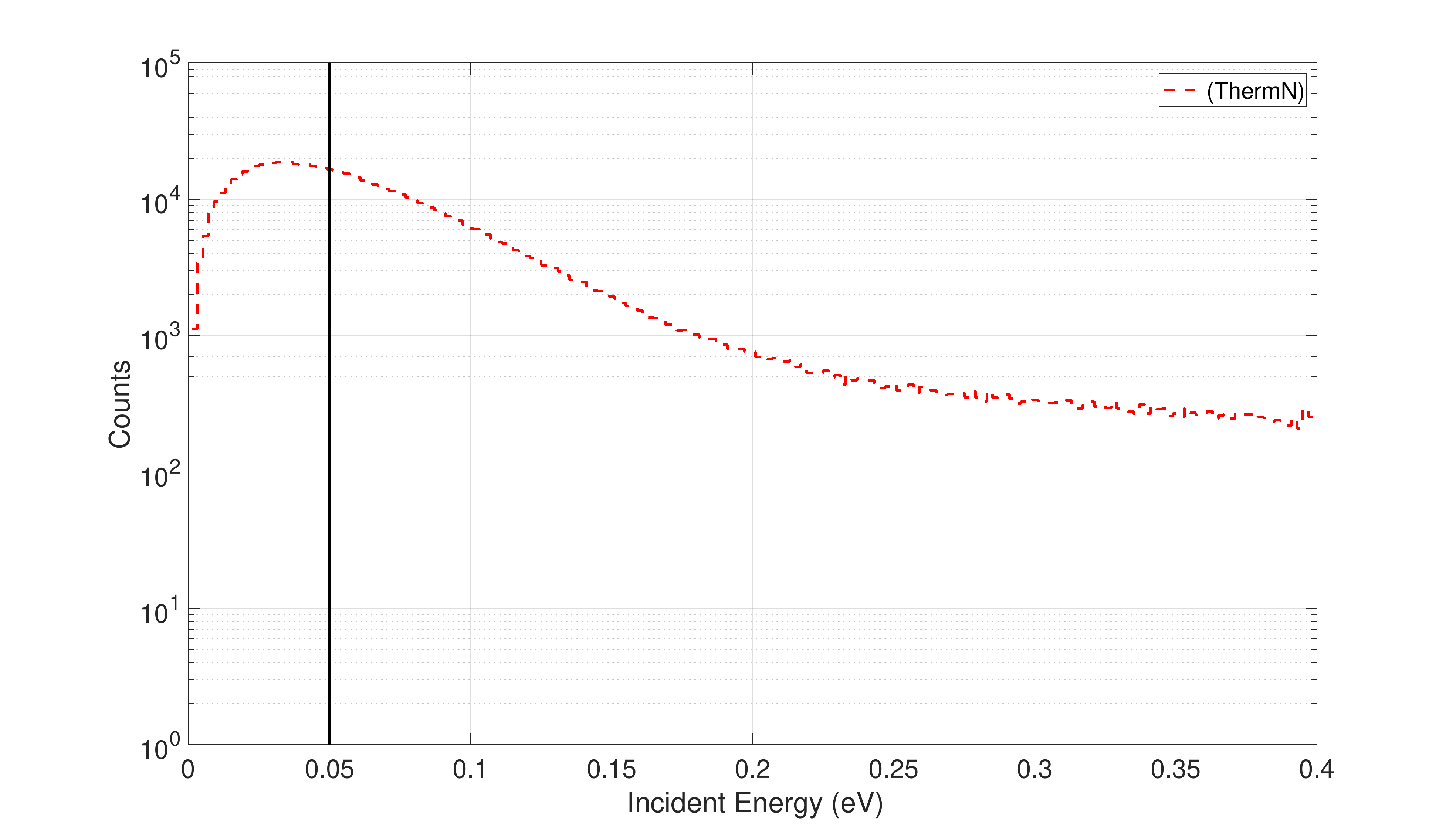}}
\caption{\label{Ekin} \footnotesize (a) Incident neutron energy in the detector active area for configurations (FastN) and (ThermN). The energy range 0-10 MeV is the $^{238}$Pu/Be source emission spectrum. (b) The counts in $0 < E_n < 0.4$ eV are shown. The vertical lines represent the cut at the $E_2 = 1$~MeV and $E_2 = 0.050$~eV used to calculate the fast and thermal flux respectively.}
\end{figure}

The energy distribution of incident neutrons in the detector active area are shown in figure~\ref{Ekin}, both for the (FastN) and (ThermN) configurations. We estimate that the thermalized flux ($E_n < 0.1$~eV) passing through a 100 mm layer of polyethylene is $\approx$8\% of the total neutron counts that reach the detector for the full energy range considered for the Pu/Be source. This value is in agreement with~\cite{GeantThnflux}. 
\\ From the simulation we can derive the thermalized flux obtained in configuration (ThermN) sketched in figure~\ref{sim-sketch}(b) as described in the previous section following equation~\ref{eqnorm}. The integrated counts, $\approx 3.5 \cdot 10^5$, have been calculated in an energy range below $E_2 =0.050$ eV, as shown in panel (b) of figure~\ref{Ekin}. In the case of (FastN), figure~\ref{sim-sketch}(a), the flux has been calculated based on equation~\ref{eqnorm} up to $E_2=1$ MeV, the cut is shown in figure~\ref{Ekin}, panel (a) (vertical line). This is about 1 order of magnitude higher than the thermalized flux. The ratio between the two calculated fluxes will be used to normalize the measured pulse height spectra presented in section~\ref{meashe3}. As described in the previous section for the measurements, the flux is calculated by multiplying the activity and the solid angle for the fast neutron configuration. The ratio calculated from the simulations ($\approx 10\%$) is needed to normalize the (ThermN) measurement. Indeed, this information cannot be extracted from the data.     
\\Before applying the normalizations to the PHS, we apply the \textit{subtraction method} to disentangle the fast and the thermal neutrons. After the subtraction, we proceed with the normalization.

\begin{figure}[htbp]
\centering
\includegraphics[width=.9\textwidth,keepaspectratio]{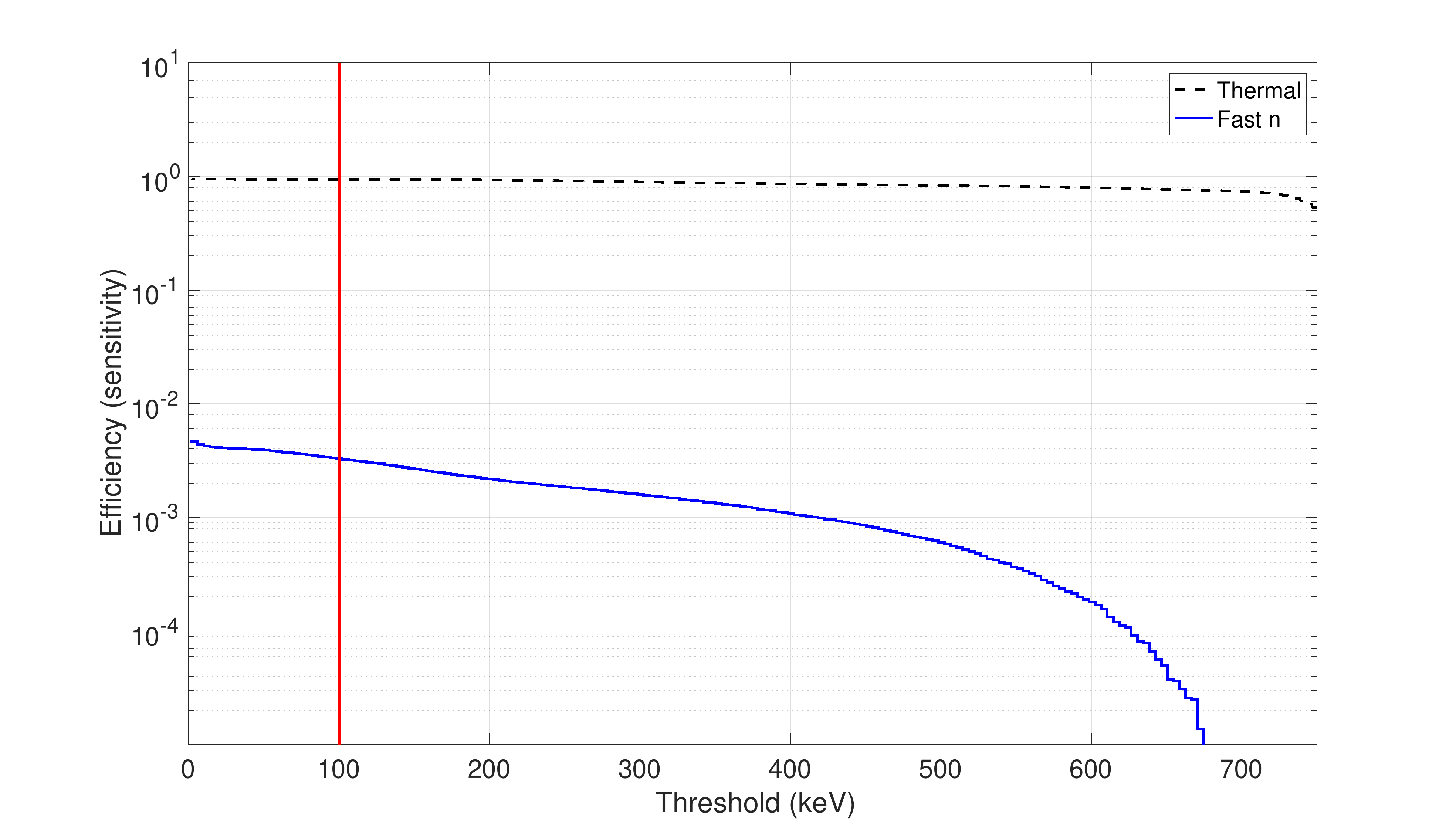}
\caption{\label{sensim} \footnotesize Fast neutron sensitivity of the $^3$He-detector as a function of an energy threshold $ E_{th}=100$~keV. The threshold value has been taken in agreement to the threshold used for background discrimination based on previous works~\cite{MIO_MB2017,MIO_fastn}.}
\end{figure}

The efficiency of the detector can be calculated by integrating the counts of the normalized PHS above a certain energy threshold as described in equation~\ref{eqsens}. In figure~\ref{sensim} the sensitivity both to thermal and fast neutrons is shown. For a threshold of 100~keV, which is usually chosen to discriminate the background~\cite{MIO_MB2017,MIO_fastn}, a fast neutron sensitivity of about $3.3\cdot10^{-3}$ is obtained and an efficiency of 94\% for thermal neutrons, in agreement with the specification of the detector. Moreover, the sensitivity is in the order of $10^{-3}$ up to a threshold of $E_{th} = 400$~keV.

\subsection{Fast neutron sensitivity of $^3$He detector}\label{meashe3}

\begin{figure}[htbp]
\centering
\includegraphics[width=.9\textwidth,keepaspectratio]{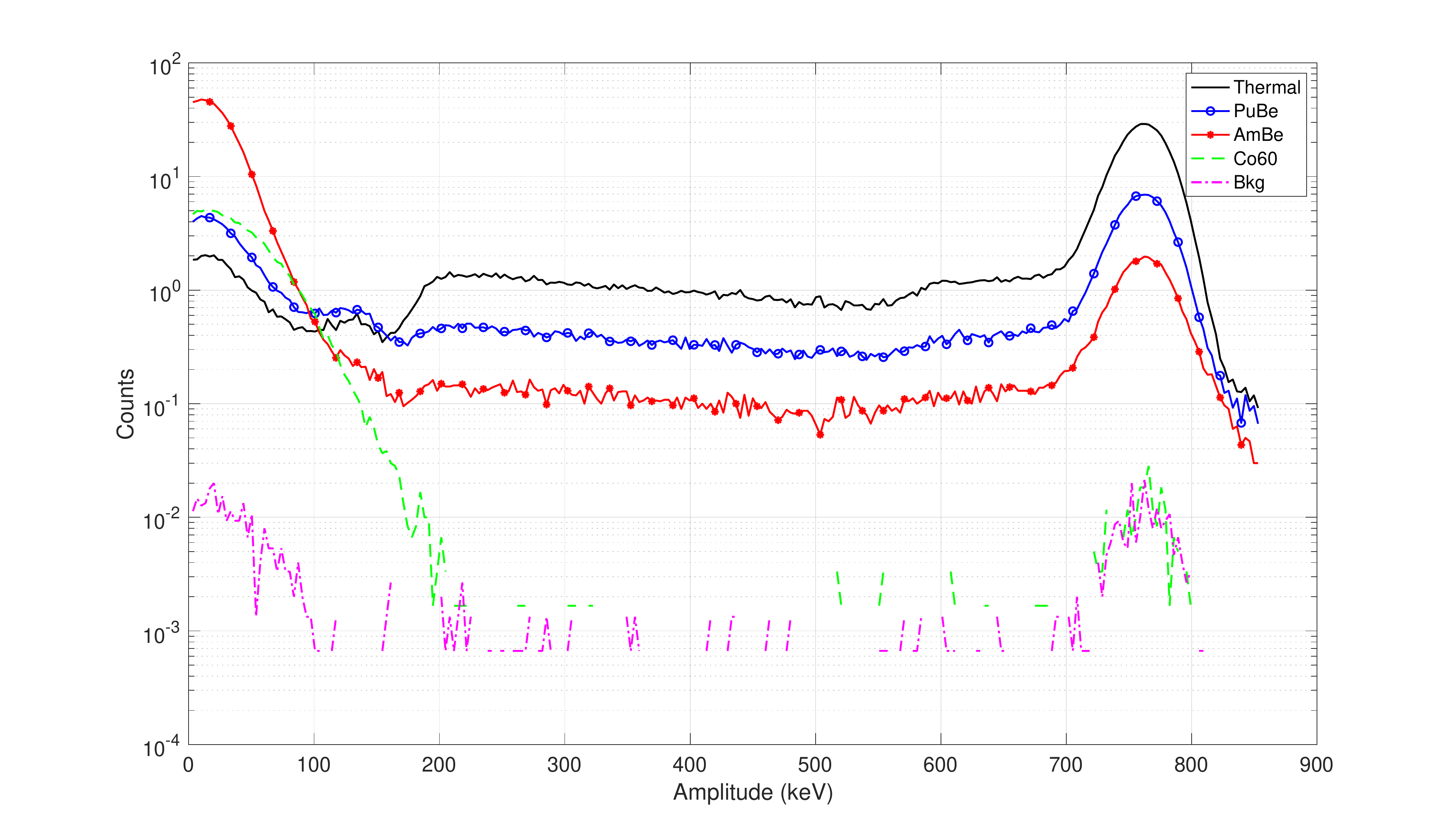}
\caption{\label{he-gen} \footnotesize Pulse Height Spectrum (PHS) normalized over time for different sources: Am/Be and Pu/Be for the fast neutrons and Co-60 for $\gamma$-rays, and the environmental background.}
\end{figure}

As described in section~\ref{setup}, several sources have been used: Am/Be and Pu/Be to measure the fast neutrons, the Pu/Be source has been employed to perform the thermal neutron measurement as well. A Co-60 source was used to measure the $\gamma$-ray sensitivity, as a comparison with previous work~\cite{Rossi-gsHe}. Measurements of the environmental background, without any source, were performed as well. The PHS, normalized over time, are depicted in figure~\ref{he-gen}; the centre of the peak corresponding to an energy of 764 keV is used to convert the X-axis from ADC values to energy. The same method is applied to each PHS reported below. 

\begin{figure}[htbp]
\centering
\includegraphics[width=.9\textwidth,keepaspectratio]{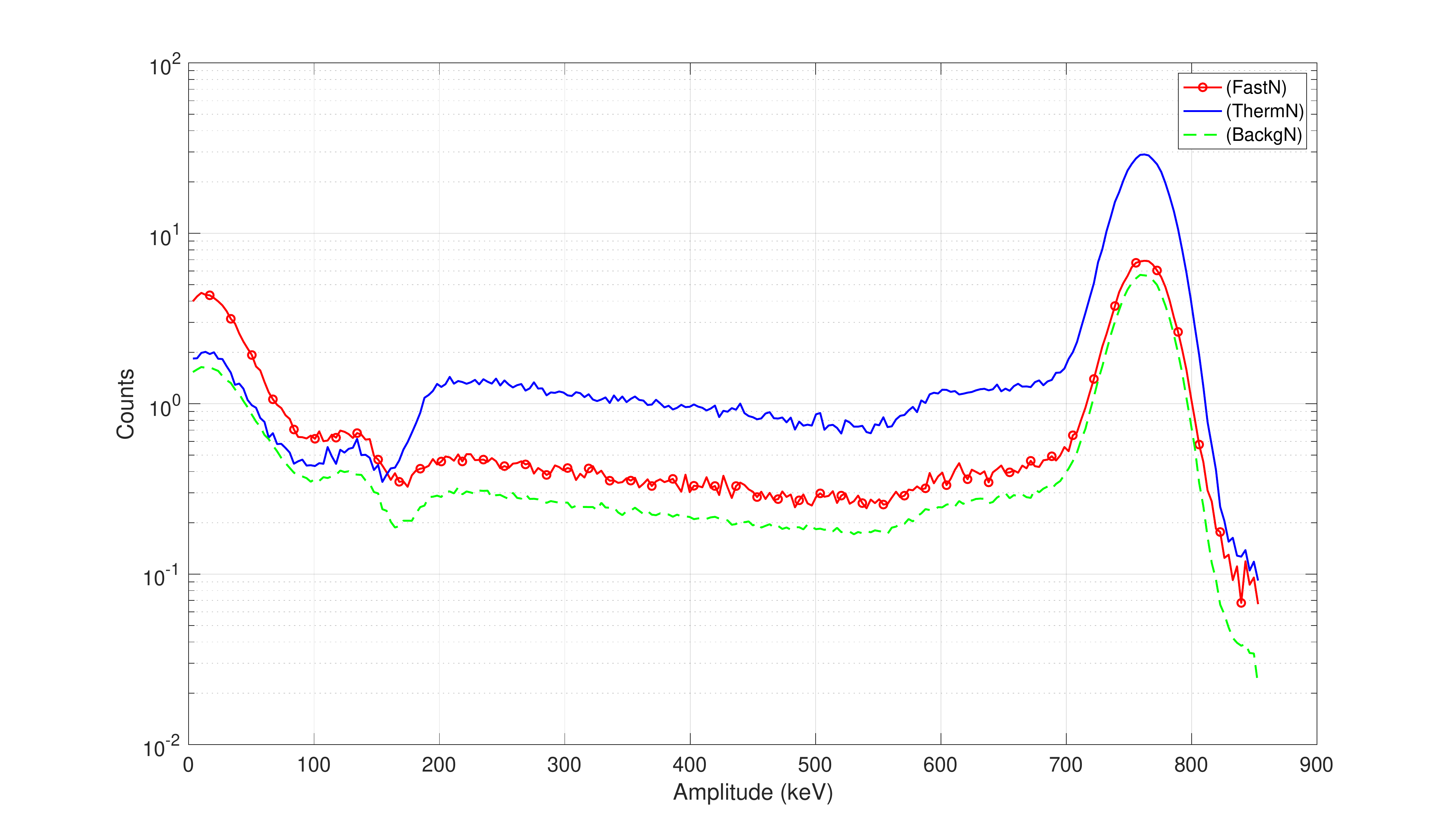}
\caption{\label{he-pube} \footnotesize Measured PHS normalized by time and solid angle, obtained in configuration (FastN), (ThermN) and (BackgN).}
\end{figure}

The Am/Be source has a larger $\gamma$ emission with respect to the Pu/Be source~\cite{PuBe_AmBe, MIO_fastn}. The effect is clearly visible, indeed, the energy released by $\gamma$-rays spreads up to about 200 keV (green curve). Below this region ($<$100 keV) the PHS measured with the Am/Be source is about one order of magnitude more intense than that recorded with the Pu/Be source. The lower $\gamma$-rays emission and the highest activity are the reasons why the presented measurements were carried out with the Pu/Be source.   
\\In figure~\ref{he-pube} the PHS measured in the configurations (FastN), (ThermN) and (BackgN) are shown. The background counts due to the environment, figure~\ref{he-gen}, have been subtracted from this data sets. The spectra are normalized by time and solid angle. In configuration (BackgN) the reaction product peak (764 keV) is only one order of magnitude less intense than in case of the thermalized flux configuration (ThermN). This value is in agreement with the simulation, see figure~\ref{PHSimtot}. The same peak is detected in the (FastN) measurement, which confirms the difficulty to decouple the fast and the thermal neutron contributions. The difference of the integrated peak in (FastN) and (BackgN) configurations is approximately 20\%, which is in agreement with the ratio of total incoming neutrons calculated in the simulations. Therefore, the peaks are normalized by this difference before applying the \textit{subtraction method} needed to disentangle the two outputs. From the simulation we know that $\varphi_{tn} = 3\% \, \phi_{th}$, hence we normalize according to this ratio the integral of the thermal peak obtained for the (BackgN) and the (ThermN) measurement, before applying the subtraction.

\begin{figure}[htbp]
\centering
\includegraphics[width=.9\textwidth,keepaspectratio]{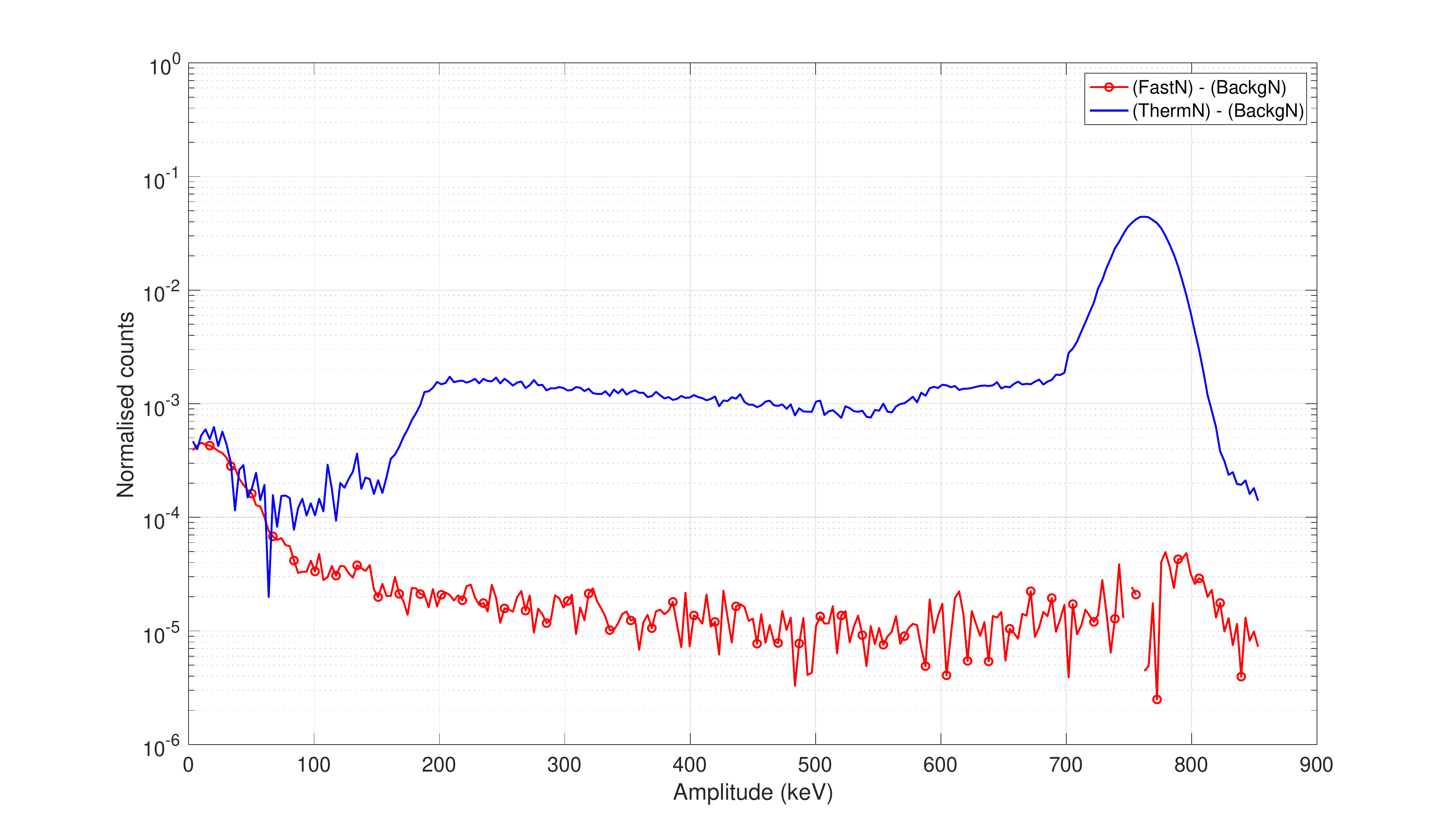}
\caption{\label{he-fth} \footnotesize PHS of thermal neutrons (blue) and fast neutrons (red). The (BackgN) measurement has been subtracted in the case of (FastN), in order to ensure that the spectrum refers to fast neutron events.}
\end{figure}

Based on the previous discussion, the flux in (BackgN) measurement can be considered as a background component, both due to the thermalized component passing after the shielding and fast neutrons that scatter around the detector thermalizing at the detector. Thus, this contribution can be subtracted in both scenarios. The resulting PHS are depicted in figure~\ref{he-fth}. The fast neutron spectrum is, as expected from the theory, mostly flat after applying the \textit{subtraction method}. 

Note that, a slight shape of the thermal peak is still distiguishable, but it is the best discrimination achievable in this measurements. When the energy range of the incoming source is in the fast neutron range, up to $E_n =10$ MeV in the case under study, the pulse height distribution for the $^3$He recoil reaction is continuous from $E=0$ to $(3/4) E_n$ (figure~\ref{edeptot}), while for the (n,p) reaction the pulse height spectrum has a peak at a value proportional to $E_n + Q$ and a continuous wall effect up to $E_n + Q$, in case a monochromatic energy is considered~\cite{fastNmeas}. Since the pulse height spectrum is shown for energies below 900 keV a continuous distribution is expected. This result is in agreement with the simulated energy deposition shown in figure~\ref{PHSimtot}. Therefore, the \textit{subtraction method} can be considered as a useful tool to disentangle thermal and fast neutron events. 

\begin{figure}[htbp]
\centering
\includegraphics[width=.89\textwidth,keepaspectratio]{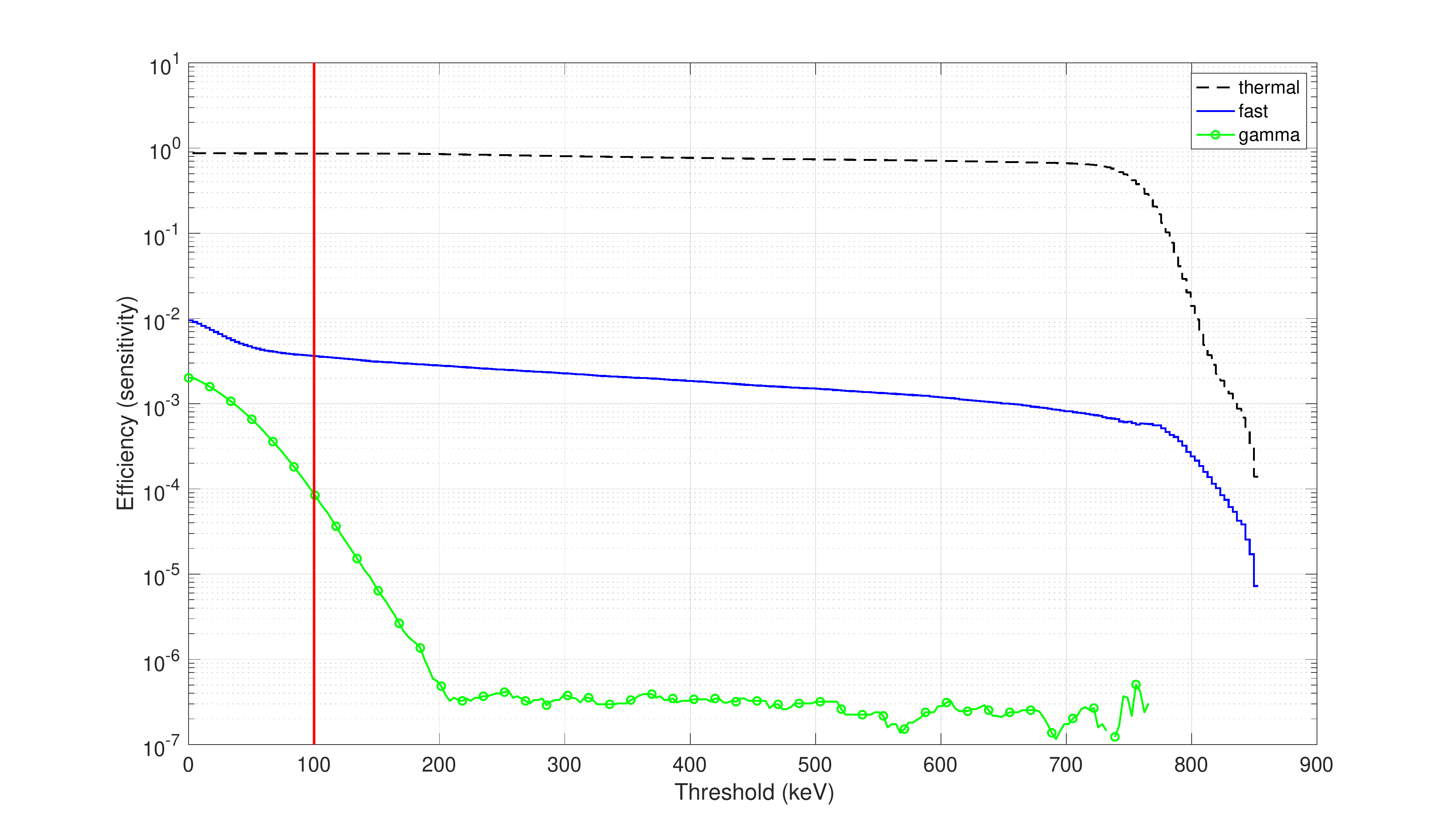}
\caption{\label{he-sens} \footnotesize Fast neutron sensitivity of the $^3$He-detector as a function of the applied threshold. For a software threshold of 100 keV a fast neutron sensitivity on the order of $10^{-3}$ is obtained.}
\end{figure}

The PHS in figure~\ref{he-fth} are normalized by time and flux. The latter has been derived from the simulation. As described in section~\ref{simulation}, we calculated the integrated flux after the thermalization through the polyethylene brick, the total incoming neutron flux, and the same for the fast neutrons, based on equation~\ref{eqnorm}. For the measurement, the normalization is obtained from the activity of the source and the solid angle in the case of fast neutrons. The calculated ratio, $\approx 10\%$, between the thermal and fast flux from the simulation is used to normalize the PHS in (ThermN) configuration. 
\\ In figure~\ref{he-sens} is shown the sensitivity for configuration (FastN) and (ThermN), calculated referring to equation~\ref{eqsens}. The $E_{th}=100$ keV threshold, red vertical line, is taken according to the simulation and the background discrimination based on previous works with some similar characterizations~\cite{MIO_MB2017,MIO_fastn}. A sensitivity of  $\approx 4\cdot 10^{-3}$ is observed. The same order of magnitude is obtained both for measurements and simulations.

\section{Further evidence of fast neutron sensitivity of $^3$He detector in comparison with the Multi-Blade B10-based detector}\label{25hz}

\subsection*{Background study from the test at the CRISP reflectometer at ISIS}

\begin{figure}[htbp]
\centering
\includegraphics[width=0.9\textwidth,keepaspectratio]{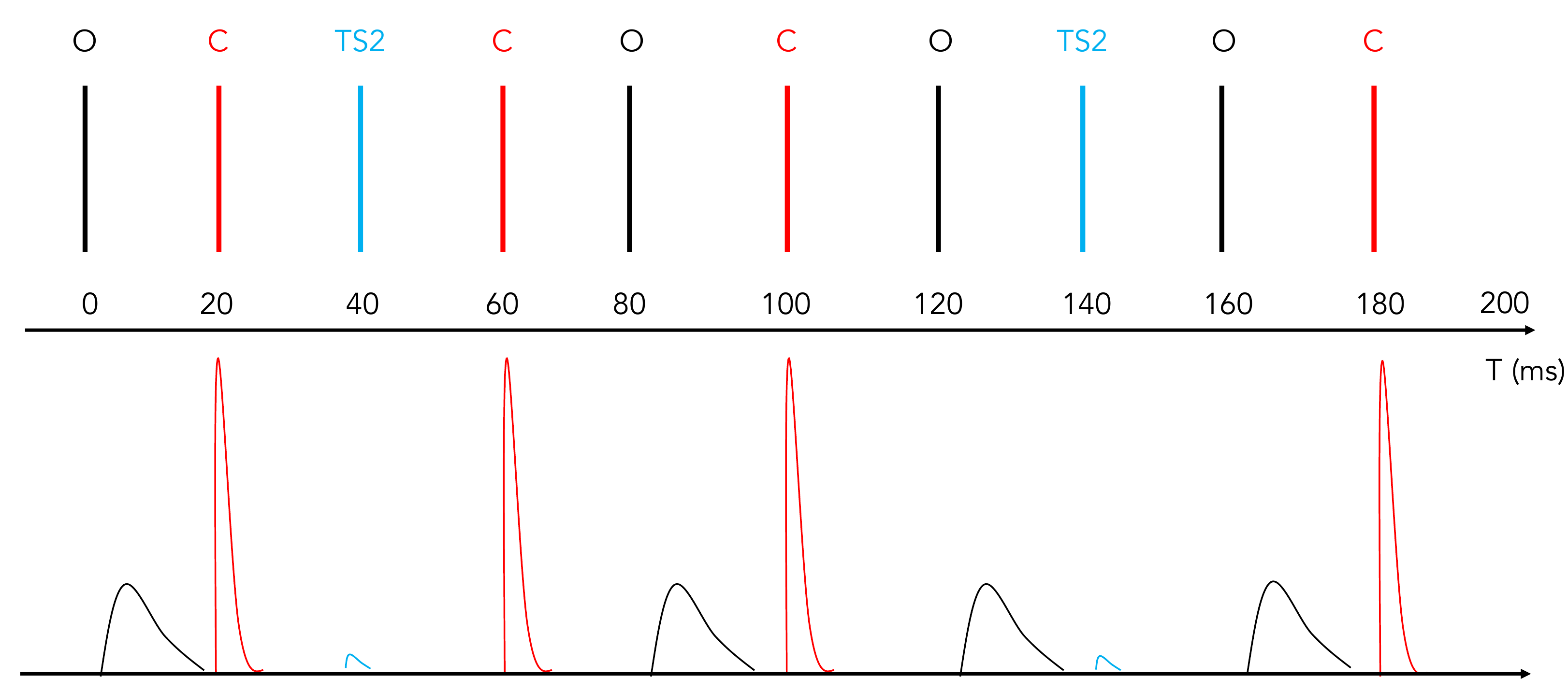}
\caption{\label{fig25hz} \footnotesize Sketch of the arrival signals in a window of 200 ms, when the chopper works at 25 Hz (CRISP). The black identifies the signal passing throw the chopper (O = open), the red colour is used when the neutron beam hits the chopper without passing through (C = close). In this case a strong spike is expected at the detector. One every five pulses is sent to the Target Station 2 (TS2), blue in figure. A low background can be detected.}
\end{figure}

The measurements presented in this section refer to a test performed at the CRISP reflectometer~\cite{CRISP1} at ISIS, with a Boron-10-based detector, the Multi-Blade~\cite{MIO_MB2017}, which is under development at the European Spallation Source, ESS~\cite{ESS,ESS_TDR}. The full description of the experiment can be found in~\cite{MIO_MB16CRISP_jinst, MIO_ScientificMBcrisp}. A background study has been performed on a particular set of measurements, and a direct comparison between the $^3$He-tube installed at the instrument and the Multi-Blade detector employed for the tests, is reported in this paper.  
\\For those measurements, the two detectors were placed in the direct beam with the chopper working at half of the nominal speed, i.e., 25 Hz instead of 50 Hz. Choppers are rotating disk with at least one aperture, which allow only neutrons of a certain velocity (energy) pass through it. When the neutron beam hits the chopper thermal neutrons are absorbed, $\gamma$-rays attenuated, but a fast neutron component can still pass through the chopper and reach the detector.  
 \\The CRISP reflectometer uses a broad band neutron time-of-flight method for determining the wavelength, $\lambda$, in a range of 0.5-6.5 \AA\ when operating at the source frequency of 50 Hz, and can be extened up to a maximum of 13 \AA, if operated at 25 Hz~\cite{CRISP1}. The larger wavelengths, lower neutron energies (cold range), will be exploited at the new high intensity source ESS. In this range, the efficiency of the Multi-Blade is comparable with that of the $^3$He detectors~\cite{MIO_MB16CRISP_jinst}. Moreover, at high q where the reflectivity intensity is very low, down to $10^{-6}$ at the current instruments, the signal-to-background ratio is crucial to perform an adequate data analysis. 
\\ As the instrument is working at half of the nominal speed, every second pulse the neutron beam find the chopper closed to let the longer wavelengths of the previous pulse to reach the sample, together with this, an intense prompt pulse is generated by the scattering between the neutrons and the chopper material. A sketch of the spectrum in a time window of $T=200$ ms is shown in figure~\ref{fig25hz}. There are three possible scenarios: the chopper is in phase with the proton pulse, thus the beam passes through it (O in the picture), the chopper is not in phase with the frequency of the proton pulse (C in the picture), so the neutron beam hits the chopper producing a high background spike. Furthermore, every five pulses, one pulse is sent to the Target Station 2 (TS2 in picture), a negligible background is expected in this case, because there is no beam impinging the chopper disc. 

The detector response can be analysed in different time windows of this spectrum. An indirect estimation of the fast neutron sensitivity is obtained focusing the analysis on the background contributions when the chopper is closed (C) in figure~\ref{fig25hz}. A prompt pulse is expected in the case the chopper is open as well. All these events are mainly due to $\gamma$-rays, fast neutrons, environmental neutron counts (thermal and epithermal) and spurious scattering. 
\\We define three different regions of studying: the full spectrum $t=0-200$ ms (T), the thermal spectrum region (O) in $t=120-160$ ms and the background peak region (C) $t=100-105$ ms. Note that the time window for regions (O) and (C) have been arbitrarily selected between the four possible choices, see figure~\ref{fig25hz}. For each region we calculate the detected flux as the integral of the PHS:

\begin{enumerate}
\item $\Phi$: total flux integrated in $t=0-200$ ms
\item $\Phi_{tn}$ : thermal flux integrated in $t=120-160$ ms
\item $\Phi_{p}$: background flux integrated in $t=100-105$ ms
\end{enumerate}

As both the detectors are located in the same place, the same incident flux is impinging on both. Therefore, the indirect calculation consists in extracting the sensitivity of the $^3$He detector by comparing the measured flux with both detectors and the well-known Multi-Blade thermal neutron efficiency~\cite{MIO_MB16CRISP_jinst} and fast neutron sensitivity~\cite{MIO_fastn}, following the expression~\ref{eq1}.

\begin{equation}
\Phi^{MB}_{i}: \epsilon^{MB}_{i} = \Phi^{He}_{i} : \epsilon^{He}_{i}
\label{eq1}
\end{equation}

It is possible to calculate the $^3$He efficiency either in the thermal spectrum region or for the background, the index $i$ in equation~\ref{eq1} refers to the two possibilities. The calculation of thermal efficiency serves as a proof of the reliability of this method, whereas the $^3$He efficiency is widely known~\cite{sans-frm2-he3}. 
\\ A further step is needed in order to separate the fast neutron and the $\gamma$-ray background contributions. Based on previous studies on the Multi-Blade detector, a threshold discrimination can be applied~\cite{MIO_MB16CRISP_jinst}. It has been proved, indeed, that by setting an appropriate software threshold, it is possible to reject most of the $\gamma$-ray contribution from the data. Usually, the $\gamma$-rays give rise to less energetic signals, and only few are detected above this value, on the order of  $10^{-7}-10^{-8}$~\cite{MG_gamma,MIO_MB2017, MIO_fastn}. Instead, the fast neutrons have a much broader energy spectrum, which leads to events with higher energy. This contribution can be highlighted by selecting the data above the threshold with a good discrimination to the $\gamma$-ray background. We derive the fast neutron flux, $\Phi_{fn}$, by the following relation:

\begin{equation}
\Phi_{fn} = \frac{\Phi-(4*\Phi_{tn})}{4}
\label{eq2}
\end{equation}

\begin{figure}[htbp]
\centering
\includegraphics[width=1\textwidth,keepaspectratio]{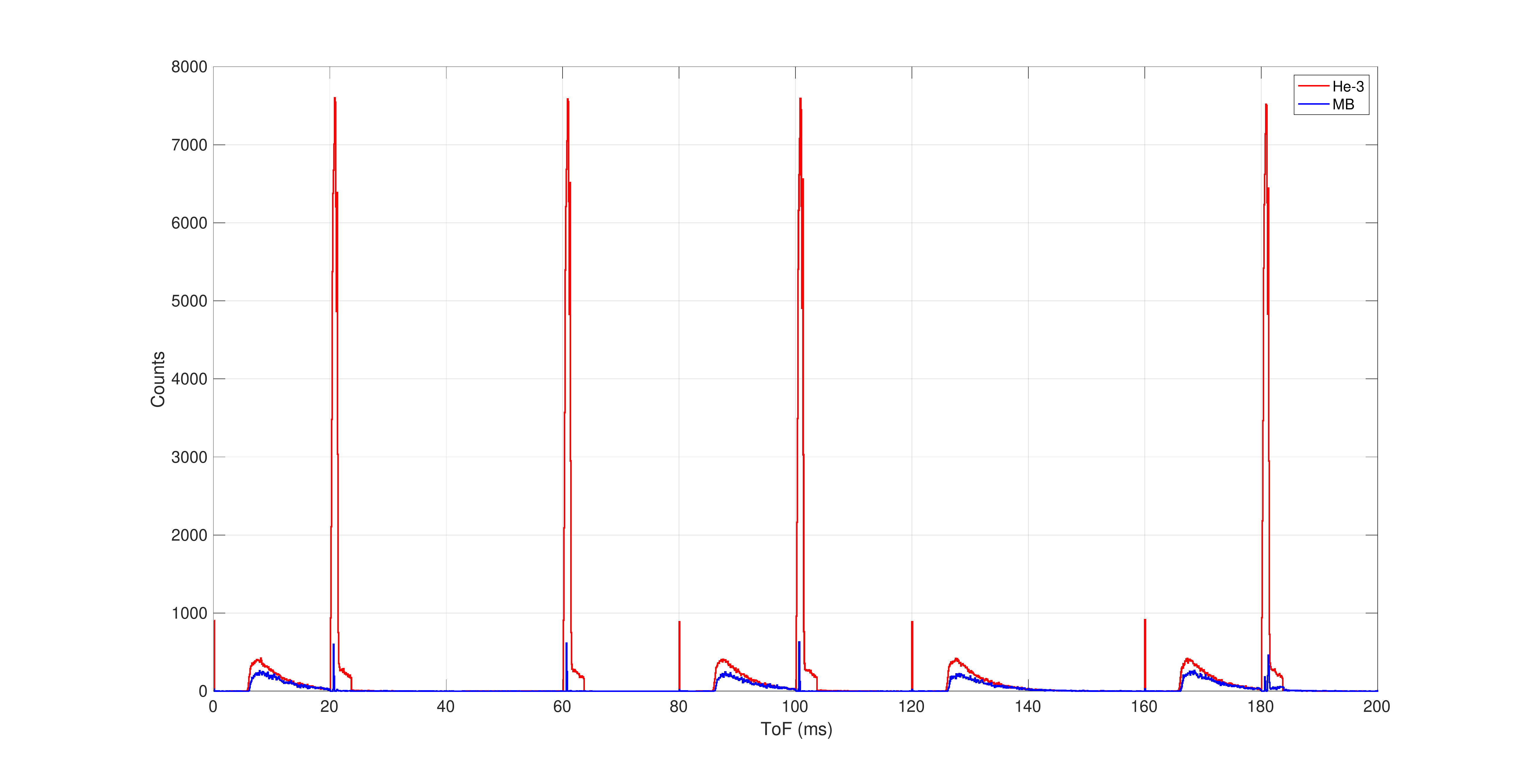}
\caption{\label{spec25hz} \footnotesize ToF spectrum obtained with the Multi-Blade detector (blue line) and the $^3$He-tube (red line) in a time window of 200 ms.}
\end{figure}

Note that the fluxes in equation~\ref{eq2} are calculated after the software threshold discrimination, while the factor 4 is given by the repetition of the regions (O) and (C) in the spectrum (T), as shown in figure~\ref{25hz}. Moreover, $\Phi_{fn} \sim \Phi_{p}$ when $\Phi_{p}$ is calculated above threshold. The difference is that $\Phi_{fn}$ is the average value of background in a time window $t=20$ ms, while $\Phi_{p}$ is integrated around the peak of region (C) in an interval $t=5$ ms.
\\In order to perform the calculation described above, the Time-of-Flight (ToF) spectrum and the PHS in the three regions of interest (T), (O) and (C) will be presented for both detectors.    
\\ In figure~\ref{spec25hz} the ToF spectrum of full time selected $t = 200$ ms is shown, in red for the $^3$He detector and in blue for the Multi-Blade. This distinction holds to all the subsequent figures. It can be noticed that a much higher background is detected in the case of the $^3$He-tube compared with the Multi-Blade. Despite the high counting, no saturation effects were detected by either detector.

\begin{figure}[htbp]
\centering
\includegraphics[width=.9\textwidth,keepaspectratio]{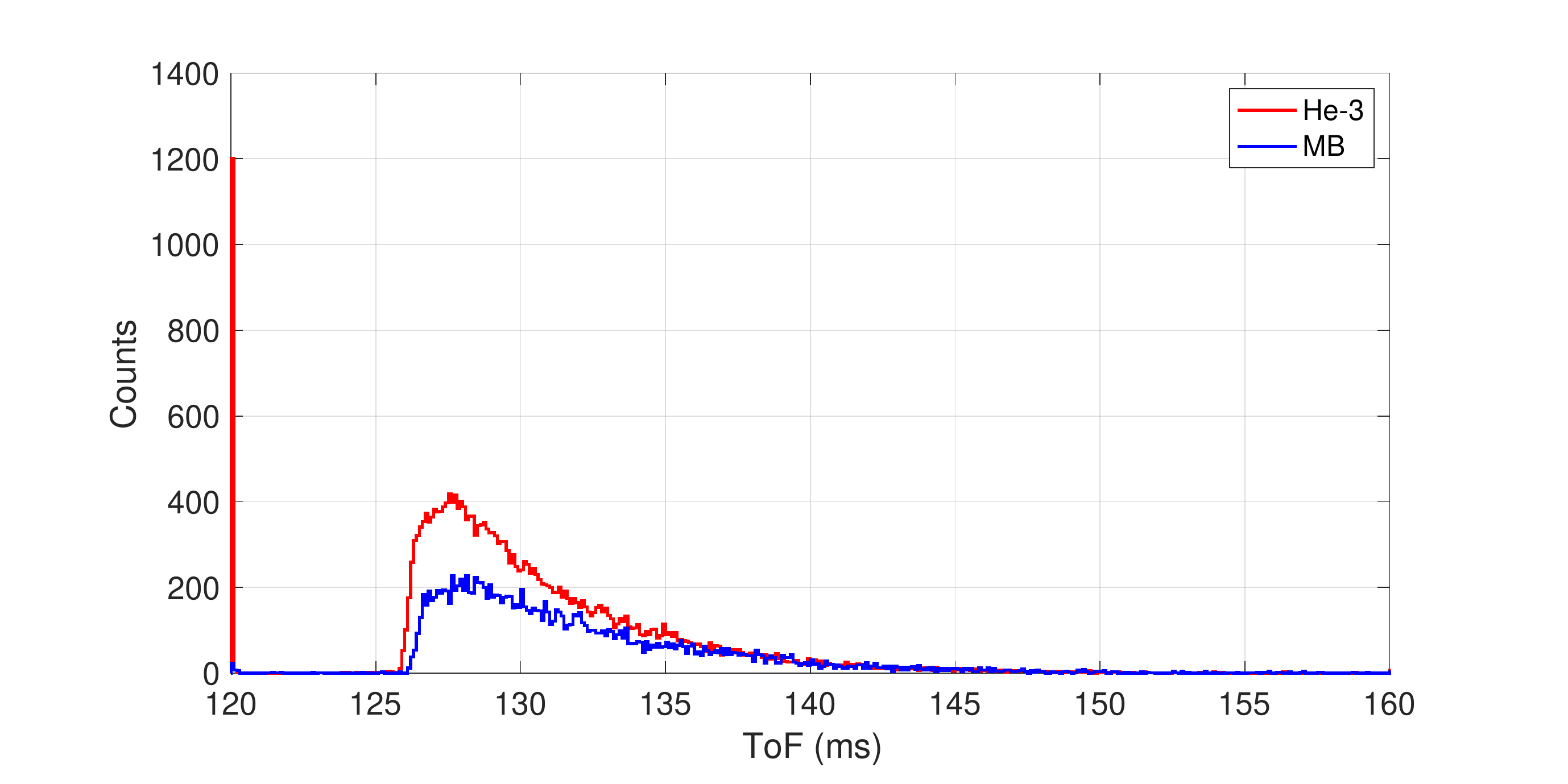}
\caption{\label{specbeam} \footnotesize ToF spectrum obtained with the Multi-Blade detector (blue line) and the $^3$He-tube (red line) for ToF = 120 - 160 ms.}
\end{figure}

For the region (O) ($t = 120 - 160$ ms) the recorded ToF spectrum reproduces the beam profile, as shown in figure~\ref{specbeam}. The spike at the beginning of this spectrum, and for each of the (O) repetitions shown in figure~\ref{fig25hz} ($t = 0, 80, 120$ and 160 ms), is significantly higher in the case of the $^3$He-tube than the one recorded with the Multi-Blade detector by about 2 order of magnitudes. The intensity at the thermal peak ($t=128$ ms in this case) is higher in the case of the $^3$He because CRISP is a thermal neutron reflectometer peaked at $\lambda = 2.5$ \AA, where the Multi-Blade has an efficiency of $\sim 50$ \%, with respect to the $\sim 85$\% efficiency of the $^3$He tube of the instrument~\cite{MIO_MB16CRISP_jinst}. As expected, above 6.5 \AA\ ($t=140$ ms in figure~\ref{specbeam}), the two detectors efficiency are close enough that they count a similar amount of events.

\begin{figure}[htbp]
\centering
\subfloat[]{\includegraphics[width=.9\textwidth,keepaspectratio]{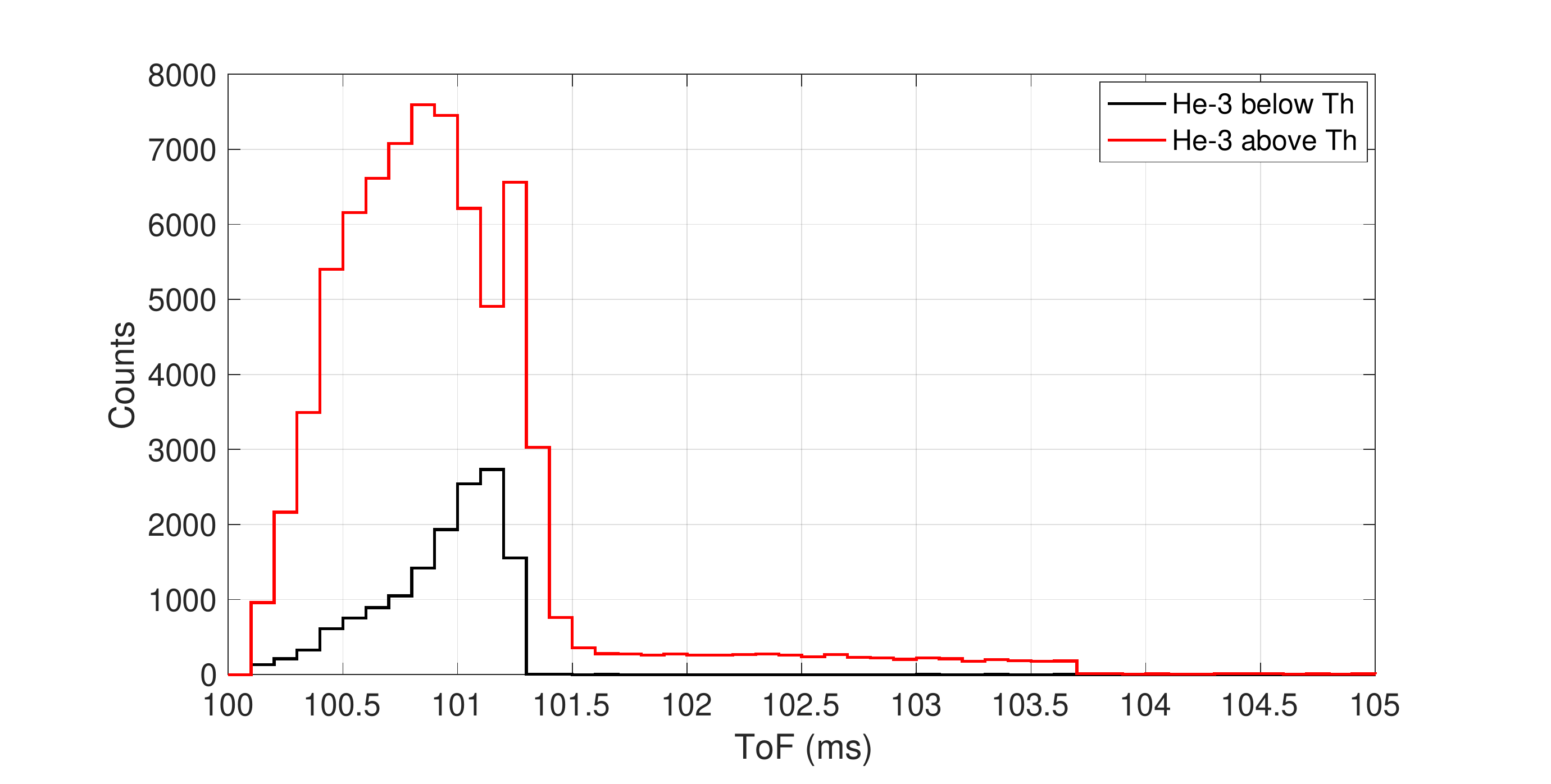}}\\
\subfloat[]{\includegraphics[width=.9\textwidth,keepaspectratio]{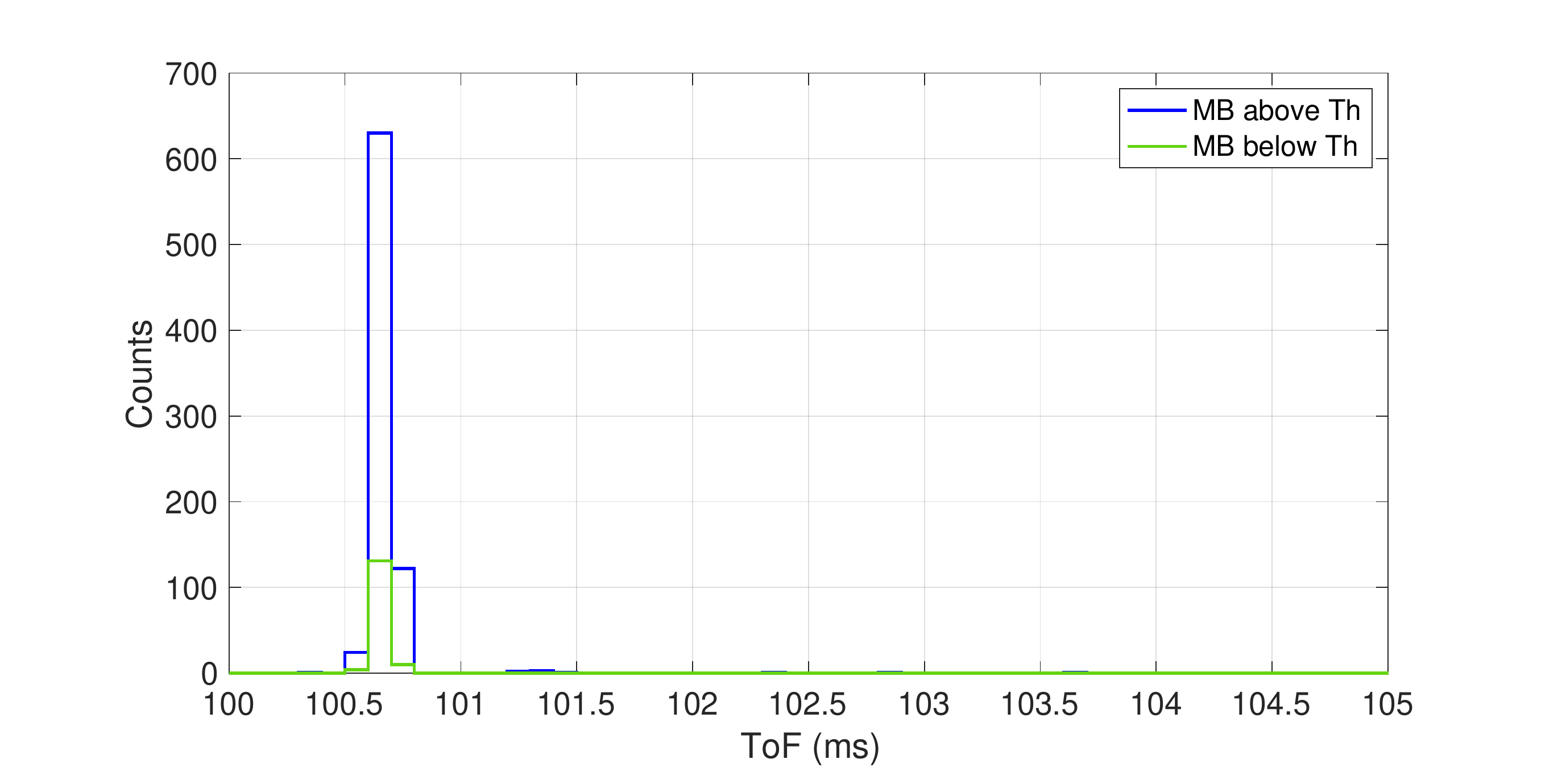}}
\caption{\label{specmbhe} \footnotesize (a) ToF spectrum obtained with the $^3$He-tube below (black line) and above (red line) threshold. (b) ToF spectrum obtained with the the Multi-Blade detector below (green line) and above (blue line) threshold. Both are recorded in ToF range between 100 to 105 ms, i.e., when the neutron beam hits against the chopper.}
\end{figure}

\begin{figure}[htbp]
\centering
\subfloat[]{\includegraphics[width=.49\textwidth,keepaspectratio]{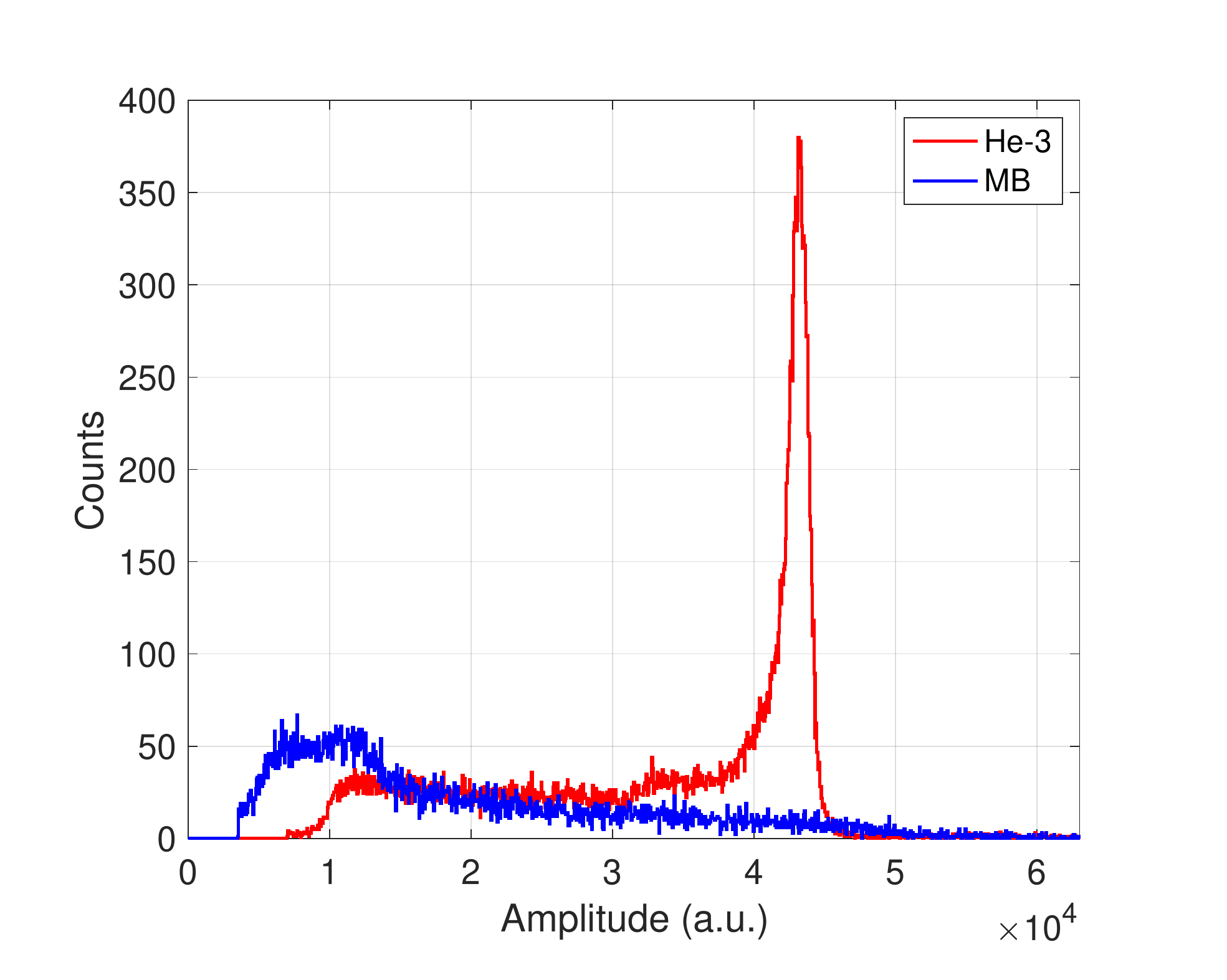}}
\subfloat[]{\includegraphics[width=.49\textwidth,keepaspectratio]{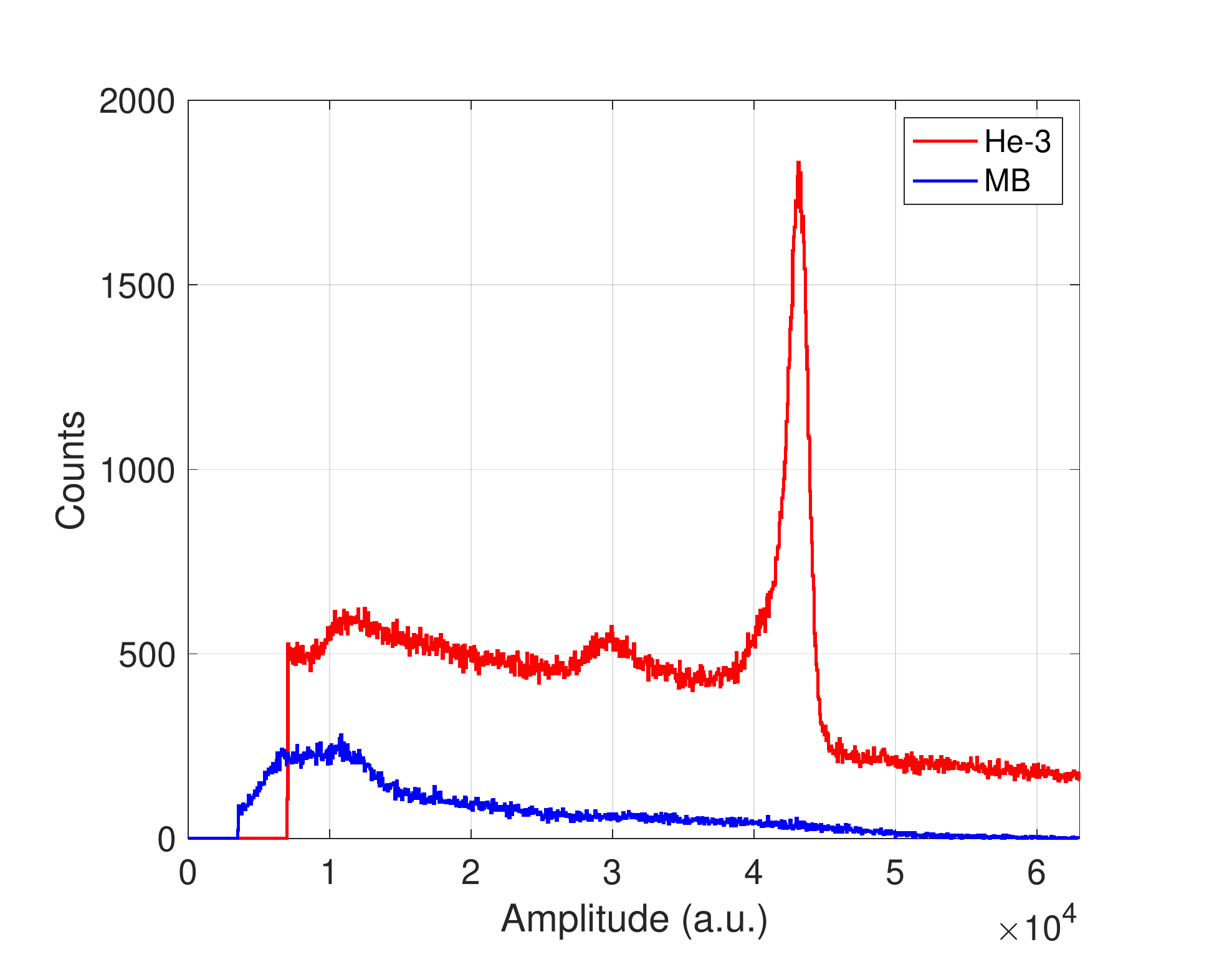}}\\
\subfloat[]{\includegraphics[width=.49\textwidth,keepaspectratio]{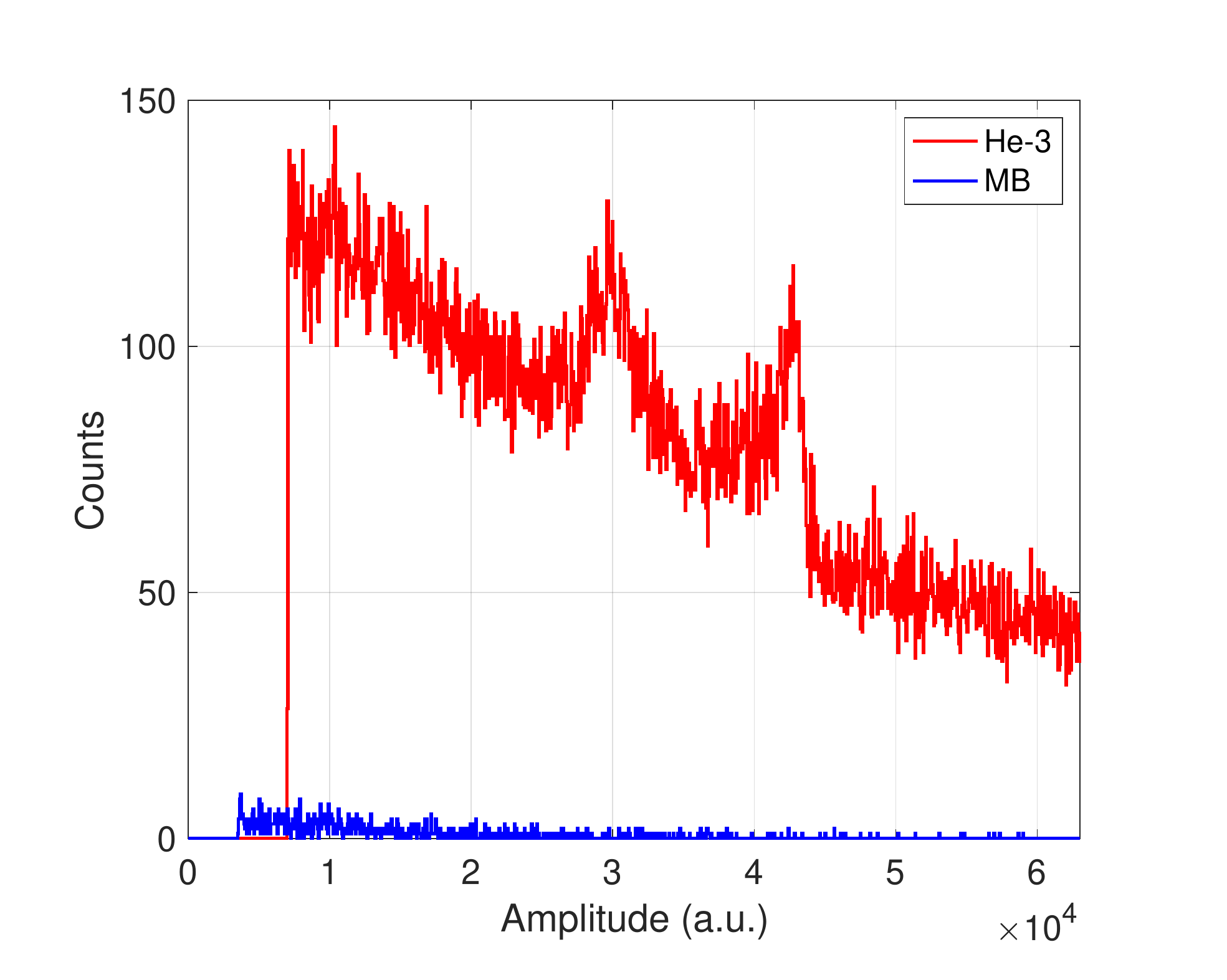}}
\caption{\label{phs25hz} \footnotesize (a) PHS in ToF = $0-200$ ms (chopper O), (b) PHS in the full time window 200 ms  and (c) PHS in ToF = $100-105$ ms.}
\end{figure}

The ToF spectrum in (C) ($t = 100 - 105$ ms) is shown in figure~\ref{specmbhe} separately for the $^3$He detector, panel (a), and the Multi-Blade detector, panel (b), because the threshold discrimination is applied individually. The time structure of the pulses is a further indication of the nature of the various background contributions. The gamma flash induces prompt events, indeed, when events below threshold are taken into account, the black pulse in figure~\ref{specmbhe}(a), it spreads only over 1 ms. The red peak shows, instead, the signals above threshold, this is more intense because, along with the fast neutrons interaction, also a small contribution due to $\gamma$-rays of higher energy and spurious scattering from the environment may occur. Note that this peak is broader and extends up to approximately 4 ms. The latter is a further indication that applying the threshold separation it is possible to discriminate most of the $\gamma$-rays from the fast neutrons. In the case of the Multi-Blade, figure~\ref{specmbhe}(b), the difference between the two analysis is still significant, but the counts are much fewer than in the case of the $^3$He-tube, more than 1 order of magnitude at peak and around 2 orders of magnitude in the integrated region. Not only with the Multi-Blade a lower background is detected, but also for a shorter time, below 1 ms, compared with the $\approx$ 5 ms recorded with the $^3$He detector.
\\In figure~\ref{phs25hz} the Pulse Height Spectrum, normalized over the time, for both detectors is shown in the three regions of study (O) (T) and (C) in panel (a) (b) and (c) respectively. The $^3$He-detector counts many more background events than the Multi-Blade, as discussed for figure~\ref{specmbhe}. The total flux on the detector when the beam passes through the chopper is calculated as the integral of the PHS in figure~\ref{phs25hz}(a). The integral of these PHS is used to perform the calculation of $\Phi_{fn}$ from equation~\ref{eq2}, as described above. 
The integrated counts for both detectors are listed in table~\ref{table1}. 

\begin{table}[htbp]
\centering
\caption{\label{table1} \footnotesize Integrated counts of PHS for figure~\ref{phs25hz} (a) $\Phi$, (b) $\Phi_{tn}$, (c) $\Phi_p$ and the calculated $\Phi_{fn}$ from equation~\ref{eq2}, for both detectors: Multi-Blade (MB) and $^3$He.}
\smallskip
\begin{tabular}{|c|c|c|c|c|}
\hline
 & $\Phi$ & $\Phi_{tn}$ & $\Phi_p$& $\Phi_{fn}$ \\
\hline
MB & $6.6\cdot 10^4 \pm 250$ &$1.56\cdot 10^4 \pm 120$ & $870 \pm 30$ & $900 \pm 140$ \\
\hline
$^3$He & $3.95\cdot10^5 \pm 600$ &$2.45\cdot 10^4 \pm 150$ & $7.38\cdot 10^4 \pm300$ & $7.4\cdot10^4 \pm 200$ \\
\hline
\end{tabular}
\end{table}

Knowing the different $\Phi_{i}$ for both detectors, the indirect calculation of the $^3$He efficiency can be derived following equation~\ref{eq1}. The sensitivity of the Multi-Blade detector to the fast neutrons is $\epsilon^{MB}_{fn} = 1.4\cdot10^{-5}$ with uncertainty not higher than a factor 2, as reported in~\cite{MIO_fastn}. Hence the sensitivity of the $^3$He detector to fast neutrons, $\epsilon^{He}_{fn}$ can be expressed as:

\begin{equation}
\epsilon^{He}_{fn} = \frac{\Phi^{He}_{fn} \cdot \epsilon^{MB}_{fn}}{\Phi^{MB}_{fn}}
\label{eq3}
\end{equation}

By substituting the values in equation~\ref{eq3}, a sensitivity $\epsilon^{He}_{fn} = 1.2\cdot 10^{-3} \pm 6\cdot 10^{-4}$ is obtained. We verify the calculation, considering the Multi-Blade efficiency $\approx 0.6$ at $\lambda = 3$\AA . By equation~\ref{eq1}, an efficiency to thermal neutrons of $\epsilon^{He}_{tn} = 0.94 \pm0.09$ is obtained, in agreement with the measured efficiency of the $^3$He detector, as discussed in~\cite{MIO_MB16CRISP_jinst}. 
\\A factor of $\sim$ 2 orders of magnitude between the Multi-Blade and the $^3$He is obtained in favour of the Multi-Blade. The estimated value is on the order of 10$^{-3}$, in agreement with the fast neutron sensitivity measurements performed at the Source Testing Facility discussed in the previous section~\ref{meashe3}.

\section*{Conclusions}
 
The signal-to-background ratio strongly affects the figure-of-merit of most of the instruments. Therefore, the background rejection is of crucial importance to improve the performance of the detector, the efficiency of the instruments, and hence the quality of the measurements together with the data analysis. This paper is focused on the study of fast neutron sensitivity of a $^3$He-detector. It is the first time this kind of investigation has been performed on a $^3$He-based thermal neutron detector.
\\Together with the measurements using a $^{238}$Pu/Be fast neutron source, a direct comparison with a boron-based neutron detector, the Multi-Blade detector, has been presented. The aim of this study is to provide a further parameter of discrimination between the new classes of neutron detector technology and the current reference solutions based on the $^3$He. 
\\The first measurements have been performed at the Source Testing Facility in Lund University in Sweden, while the comparison with the Multi-Blade was part of an experimental test performed at the CRISP reflectometer at ISIS (UK). Two independent methods have been used to obtain the sensitivity: a \textit{subtraction method} and an indirect calculation. These investigations allow an estimation on the order of magnitude for the sensitivity and are in agreement with each other. Moreover a set of simulations is described in the paper. A good agreement between measurements and simulations is achieved.
\\For the first set of measurements presented in the paper, the efficiency calculation is affected by an uncertainty no more than a factor two, which is dominated by the systematic effects from the calculation of the solid angle. This leads to a detection efficiency for thermal neutrons of 87\%. From the simulations it has been estimated as 94\% with an uncertainty of about 20\% due to statistical fluctuation. Both values are in agreement with the nominal efficiency of the $^3$He detector employed for this measurements, which is 96\% (at 2.5 \AA).
\\The fast neutron sensitivity has been calculated, as in the case of the Multi-Blade detector, for a fixed software threshold of 100 keV, which is representative of the typical real world applications. We obtained a sensitivity of $\approx 4 \cdot10^{-3}$, affected by the systematic uncertainty described before. The value is confirmed by the simulation with a calculated sensitivity of $\approx 3 \cdot 10^{-3}$ with a 10\% uncertainty.  
\\This estimate of the fast neutron sensitivity is also confirmed by the background characterization analysis performed on the campaign of measurements carried out at CRISP. The sensitivity obtained through the indirect calculation applied to the measurements is on the order of 10$^{-3}$, while the efficiency to thermal neutrons obtained using the same calculation is 94\% with an uncertainty of 10\% due to statistical fluctuation. This value is in agreement with the measured efficiency~\cite{MIO_MB16CRISP_jinst}.
\\ Along with the good agreement achieved between the different methods used to investigate the fast neutron sensitivity, the direct comparison with Multi-Blade detector shows that this sensitivity is about 2 orders of magnitudes higher for a $^3$He detector than for a $^{10}$B-based neutron detector. Indeed, the fast neutron sensitivity of the Multi-Blade for the fixed software threshold of 100 keV is approximately $10^{-5}$, as described in a previous study~\cite{MIO_fastn}. 
\\The better background discrimination achievable with a boron-based detector has also an impact from a cost effective point of view. The $^3$He detectors need a lot of shielding material for equivalent performance to prevent from counting fast neutrons. Less shielding can be potentially used with a boron-10-based neutron detector, decreasing the costs, especially for large area applications.


\begin{backmatter}

\section*{Competing interests}
The authors declare that they have no competing interests.

\section*{Author's contributions}
G.M. drafted the manuscript. G.M., F.P. and F.M. conceived the experiment. G.M. analysed the data. K.K. performed the simulation. All the authors contributed to the interpretation of the results. All the authors gave their final approval for publication.

\section*{Availability of data and materials}
Data essential for evaluation is contained within the figures. Authors are open to placing raw data and processed datasets to aid in replication in a convenient location, but it is currently available by direct request only.

\section*{Acknowledgements}
 This work was supported by the BrightnESS project (Horizon 2020, INFRADEV-3-2015, grant number 676548) and carried out as a part of the collaboration between the European Spallation Source (ESS - Sweden), the Lund University (LU - Sweden) and the University of Perugia (Italy).
\\The work was carried out at the Source Testing Facility, Lund University (LU - Sweden).\\
Computing resources provided by DMSC Computing Centre:\\https://europeanspallationsource.se/data-management-software/computing-centre.
\\ The authors would like to thank the ISIS detector group for the support during the tests. 
\\ The authors thank the CRISP instrument scientists R. Dalgliesh and C. Kinane for providing the beam time and the instrument support necessary for this detector test.

\bibliographystyle{bmc-mathphys} 
\bibliography{BIBLIOhefn}      


\begin{thebibliography}{50}
\ifx \bisbn   \undefined \def \bisbn  #1{ISBN #1}\fi
\ifx \binits  \undefined \def \binits#1{#1}\fi
\ifx \bauthor  \undefined \def \bauthor#1{#1}\fi
\ifx \batitle  \undefined \def \batitle#1{#1}\fi
\ifx \bjtitle  \undefined \def \bjtitle#1{#1}\fi
\ifx \bvolume  \undefined \def \bvolume#1{\textbf{#1}}\fi
\ifx \byear  \undefined \def \byear#1{#1}\fi
\ifx \bissue  \undefined \def \bissue#1{#1}\fi
\ifx \bfpage  \undefined \def \bfpage#1{#1}\fi
\ifx \blpage  \undefined \def \blpage #1{#1}\fi
\ifx \burl  \undefined \def \burl#1{\textsf{#1}}\fi
\ifx \doiurl  \undefined \def \doiurl#1{\textsf{#1}}\fi
\ifx \betal  \undefined \def \betal{\textit{et al.}}\fi
\ifx \binstitute  \undefined \def \binstitute#1{#1}\fi
\ifx \binstitutionaled  \undefined \def \binstitutionaled#1{#1}\fi
\ifx \bctitle  \undefined \def \bctitle#1{#1}\fi
\ifx \beditor  \undefined \def \beditor#1{#1}\fi
\ifx \bpublisher  \undefined \def \bpublisher#1{#1}\fi
\ifx \bbtitle  \undefined \def \bbtitle#1{#1}\fi
\ifx \bedition  \undefined \def \bedition#1{#1}\fi
\ifx \bseriesno  \undefined \def \bseriesno#1{#1}\fi
\ifx \blocation  \undefined \def \blocation#1{#1}\fi
\ifx \bsertitle  \undefined \def \bsertitle#1{#1}\fi
\ifx \bsnm \undefined \def \bsnm#1{#1}\fi
\ifx \bsuffix \undefined \def \bsuffix#1{#1}\fi
\ifx \bparticle \undefined \def \bparticle#1{#1}\fi
\ifx \barticle \undefined \def \barticle#1{#1}\fi
\ifx \bconfdate \undefined \def \bconfdate #1{#1}\fi
\ifx \botherref \undefined \def \botherref #1{#1}\fi
\ifx \url \undefined \def \url#1{\textsf{#1}}\fi
\ifx \bchapter \undefined \def \bchapter#1{#1}\fi
\ifx \bbook \undefined \def \bbook#1{#1}\fi
\ifx \bcomment \undefined \def \bcomment#1{#1}\fi
\ifx \oauthor \undefined \def \oauthor#1{#1}\fi
\ifx \citeauthoryear \undefined \def \citeauthoryear#1{#1}\fi
\ifx \endbibitem  \undefined \def \endbibitem {}\fi
\ifx \bconflocation  \undefined \def \bconflocation#1{#1}\fi
\ifx \arxivurl  \undefined \def \arxivurl#1{\textsf{#1}}\fi
\csname PreBibitemsHook\endcsname

\bibitem{ESS_TDR}
\begin{botherref}
\oauthor{\bsnm{Peggs}, \binits{S.}}
ESS Technical Design Report - (ESS-2013-0001).
\url{http://eval.esss.lu.se/cgi-bin/public/DocDB/ShowDocument?docid=274.}
\end{botherref}
\endbibitem

\bibitem{SNS}
\begin{botherref}
{SNS}.
http://neutrons.ornl.gov/facilities/SNS/.
\url{http://neutrons.ornl.gov/facilities/SNS/}
\end{botherref}
\endbibitem

\bibitem{JPARC}
\begin{botherref}
{J-PARC}.
http://j-parc.jp/MatLife/en/index.html.
\url{http://j-parc.jp/MatLife/en/index.html}
\end{botherref}
\endbibitem

\bibitem{ESS}
\begin{botherref}
{European Spallation Source ESS ERIC}.
http://europeanspallationsource.se.
\url{http://europeanspallationsource.se}
\end{botherref}
\endbibitem

\bibitem{ESS-design}
\begin{barticle}
\bauthor{\bsnm{Garoby}, \binits{R.}}:
\batitle{{The European Spallation Source Design}}.
\bjtitle{Physica Scripta}
\bvolume{93}(\bissue{1}),
\bfpage{014001}
(\byear{2018})
\end{barticle}
\endbibitem

\bibitem{Mezei2007}
\begin{barticle}
\bauthor{\bsnm{Mezei}, \binits{F.}}:
\batitle{New perspectives from new generations of neutron sources}.
\bjtitle{Comptes Rendus Physique}
\bvolume{8}(\bissue{7}),
\bfpage{909}--\blpage{920}
(\byear{2007}).
doi:\doiurl{10.1016/j.crhy.2007.10.003}.
\bcomment{Neutron scattering: a comprehensive tool for condensed matter
  research}
\end{barticle}
\endbibitem

\bibitem{VETTIER_ESS}
\begin{barticle}
\bauthor{\bsnm{Vettier}, \binits{C.}},
\bauthor{\bsnm{Carlile}, \binits{C.J.}},
\bauthor{\bsnm{Carlsson}, \binits{P.}}:
\batitle{Progress for the european spallation source in scandinavia}.
\bjtitle{Nuclear Instruments and Methods in Physics Research Section A:
  Accelerators, Spectrometers, Detectors and Associated Equipment}
\bvolume{600}(\bissue{1}),
\bfpage{8}--\blpage{9}
(\byear{2009}).
doi:\doiurl{10.1016/j.nima.2008.11.097}
\end{barticle}
\endbibitem

\bibitem{ESS2011}
\begin{barticle}
\bauthor{\bsnm{Lindroos}, \binits{M.}},
\bauthor{\bsnm{Bousson}, \binits{S.}},
\bauthor{\bsnm{Calaga}, \binits{R.}},
\bauthor{\bsnm{Danared}, \binits{H.}},
\bauthor{\bsnm{Devanz}, \binits{G.}},
\bauthor{\bsnm{Duperrier}, \binits{R.}},
\bauthor{\bsnm{Eguia}, \binits{J.}},
\bauthor{\bsnm{Eshraqi}, \binits{M.}},
\bauthor{\bsnm{Gammino}, \binits{S.}},
\bauthor{\bsnm{Hahn}, \binits{H.}},
\bauthor{\bsnm{Jansson}, \binits{A.}},
\bauthor{\bsnm{Oyon}, \binits{C.}},
\bauthor{\bsnm{Pape-Moller}, \binits{S.}},
\bauthor{\bsnm{Peggs}, \binits{S.}},
\bauthor{\bsnm{Ponton}, \binits{A.}},
\bauthor{\bsnm{Rathsman}, \binits{K.}},
\bauthor{\bsnm{Ruber}, \binits{R.}},
\bauthor{\bsnm{Satogata}, \binits{T.}},
\bauthor{\bsnm{Trahern}, \binits{G.}}:
\batitle{{The European Spallation Source}}.
\bjtitle{Nuclear Instruments and Methods in Physics Research, Section B: Beam
  Interactions with Materials and Atoms}
\bvolume{269}(\bissue{24}),
\bfpage{3258}--\blpage{3260}
(\byear{2011}).
doi:\doiurl{10.1016/j.nimb.2011.04.012}
\end{barticle}
\endbibitem

\bibitem{HE3S_kouzes}
\begin{botherref}
\oauthor{\bsnm{Kouzes}, \binits{R.T.}}:
{The 3He Supply Problem}.
Technical report,
Pacific Northwest National Laboratory, Richland, WA
(2009)
\end{botherref}
\endbibitem

\bibitem{COLL_icnd}
\begin{botherref}
International collaboration on the development of neutron detectors.
http://www.icnd.org.
\url{http://www.icnd.org}
\end{botherref}
\endbibitem

\bibitem{HE3S_karl}
\begin{barticle}
\bauthor{\bsnm{Zeitelhack}, \binits{K.}}:
\batitle{Search for alternative techniques to helium-3 based detectors for
  neutron scattering applications}.
\bjtitle{Neutron News}
\bvolume{23}(\bissue{4}),
\bfpage{10}--\blpage{13}
(\byear{2012}).
doi:\doiurl{10.1080/10448632.2012.725325}.
\arxivurl{http://dx.doi.org/10.1080/10448632.2012.725325}
\end{barticle}
\endbibitem

\bibitem{HE3S_hurd}
\begin{barticle}
\bauthor{\bsnm{Hurd}, \binits{A.J.}},
\bauthor{\bsnm{Kouzes}, \binits{R.T.}}:
\batitle{Why new neutron detector materials must replace helium-3}.
\bjtitle{The European Physical Journal Plus}
\bvolume{129}(\bissue{10}),
\bfpage{1}--\blpage{3}
(\byear{2014}).
doi:\doiurl{10.1140/epjp/i2014-14236-6}
\end{barticle}
\endbibitem

\bibitem{MG_2017}
\begin{barticle}
\bauthor{\bsnm{Anastasopoulos}, \binits{M.}},
\bauthor{\bsnm{Bebb}, \binits{R.}},
\bauthor{\bsnm{Berry}, \binits{K.}},
\bauthor{\bsnm{Birch}, \binits{J.}},
\bauthor{\bsnm{Bry{\'s}}, \binits{T.}},
\bauthor{\bsnm{Buffet}, \binits{J.-C.}},
\bauthor{\bsnm{Clergeau}, \binits{J.-F.}},
\bauthor{\bsnm{Deen}, \binits{P.P.}},
\bauthor{\bsnm{Ehlers}, \binits{G.}},
\bauthor{\bparticle{van} \bsnm{Esch}, \binits{P.}},
\bauthor{\bsnm{Everett}, \binits{S.M.}},
\bauthor{\bsnm{Guerard}, \binits{B.}},
\bauthor{\bsnm{Hall-Wilton}, \binits{R.}},
\bauthor{\bsnm{Herwig}, \binits{K.}},
\bauthor{\bsnm{Hultman}, \binits{L.}},
\bauthor{\bsnm{H{\"o}glund}, \binits{C.}},
\bauthor{\bsnm{Iruretagoiena}, \binits{I.}},
\bauthor{\bsnm{Issa}, \binits{F.}},
\bauthor{\bsnm{Jensen}, \binits{J.}},
\bauthor{\bsnm{Khaplanov}, \binits{A.}},
\bauthor{\bsnm{Kirstein}, \binits{O.}},
\bauthor{\bsnm{Higuera}, \binits{I.L.}},
\bauthor{\bsnm{Piscitelli}, \binits{F.}},
\bauthor{\bsnm{Robinson}, \binits{L.}},
\bauthor{\bsnm{Schmidt}, \binits{S.}},
\bauthor{\bsnm{Stefanescu}, \binits{I.}}:
\batitle{{Multi-Grid detector for neutron spectroscopy: results obtained on
  time-of-flight spectrometer CNCS}}.
\bjtitle{Journal of Instrumentation}
\bvolume{12}(\bissue{04}),
\bfpage{04030}
(\byear{2017})
\end{barticle}
\endbibitem

\bibitem{MG_IN6tests}
\begin{barticle}
\bauthor{\bsnm{Birch}, \binits{J.}},
\bauthor{\bsnm{Buffet}, \binits{J.-C.}},
\bauthor{\bsnm{Clergeau}, \binits{J.-F.}},
\bauthor{\bsnm{Correa}, \binits{J.}},
\bauthor{\bparticle{van} \bsnm{Esch}, \binits{P.}},
\bauthor{\bsnm{Ferraton}, \binits{M.}},
\bauthor{\bsnm{Guerard}, \binits{B.}},
\bauthor{\bsnm{Halbwachs}, \binits{J.}},
\bauthor{\bsnm{Hall-Wilton}, \binits{R.}},
\bauthor{\bsnm{Hultman}, \binits{L.}},
\bauthor{\bsnm{H{\"o}glund}, \binits{C.}},
\bauthor{\bsnm{Khaplanov}, \binits{A.}},
\bauthor{\bsnm{Koza}, \binits{M.}},
\bauthor{\bsnm{Piscitelli}, \binits{F.}},
\bauthor{\bsnm{Zbiri}, \binits{M.}}:
\batitle{{In-beam test of the Boron-10 Multi-Grid neutron detector at the IN6
  time-of-flight spectrometer at the ILL}}.
\bjtitle{Journal of Physics: Conference Series}
\bvolume{528}(\bissue{1}),
\bfpage{012040}
(\byear{2014})
\end{barticle}
\endbibitem

\bibitem{MG_patent}
\begin{botherref}
\oauthor{\bsnm{Guerard}, \binits{B.}},
\oauthor{\bsnm{Buffet}, \binits{J.C.}}:
{Ionizing radiation detector - Patent no. 20110215251.}
Google Patents.
US Patent App. 13/038,915
(2011).
\url{https://www.google.com/patents/US20110215251}
\end{botherref}
\endbibitem

\bibitem{MG_joni}
\begin{barticle}
\bauthor{\bsnm{Birch}, \binits{J.}},
\bauthor{\bsnm{Buffet}, \binits{J.C.}},
\bauthor{\bsnm{Correa}, \binits{J.}},
\bauthor{\bparticle{van} \bsnm{Esch}, \binits{P.}},
\bauthor{\bsnm{Gu{\'e}rard}, \binits{B.}},
\bauthor{\bsnm{Hall-Wilton}, \binits{R.}},
\bauthor{\bsnm{H{\"o}glund}, \binits{C.}},
\bauthor{\bsnm{Hultman}, \binits{L.}},
\bauthor{\bsnm{Khaplanov}, \binits{A.}},
\bauthor{\bsnm{Piscitelli}, \binits{F.}}:
\batitle{{10B4C Multi-Grid as an Alternative to 3He for Large Area Neutron
  Detectors}}.
\bjtitle{IEEE Transactions on Nuclear Science}
\bvolume{60}(\bissue{2}),
\bfpage{871}--\blpage{878}
(\byear{2013}).
doi:\doiurl{10.1109/TNS.2012.2227798}
\end{barticle}
\endbibitem

\bibitem{MIO_MB2014}
\begin{barticle}
\bauthor{\bsnm{Piscitelli}, \binits{F.}},
\bauthor{\bsnm{Buffet}, \binits{J.C.}},
\bauthor{\bsnm{Clergeau}, \binits{J.F.}},
\bauthor{\bsnm{Cuccaro}, \binits{S.}},
\bauthor{\bsnm{Gu{\'e}rard}, \binits{B.}},
\bauthor{\bsnm{Khaplanov}, \binits{A.}},
\bauthor{\bsnm{Manna}, \binits{Q.L.}},
\bauthor{\bsnm{Rigal}, \binits{J.M.}},
\bauthor{\bsnm{Esch}, \binits{P.V.}}:
\batitle{{Study of a high spatial resolution 10 B -based thermal neutron
  detector for application in neutron reflectometry: the Multi-Blade
  prototype}}.
\bjtitle{Journal of Instrumentation}
\bvolume{9}(\bissue{03}),
\bfpage{03007}
(\byear{2014})
\end{barticle}
\endbibitem

\bibitem{MIO_MB16CRISP_jinst}
\begin{barticle}
\bauthor{\bsnm{Piscitelli}, \binits{F.}},
\bauthor{\bsnm{Mauri}, \binits{G.}},
\bauthor{\bsnm{Messi}, \binits{F.}},
\bauthor{\bsnm{Anastasopoulos}, \binits{M.}},
\bauthor{\bsnm{Arnold}, \binits{T.}},
\bauthor{\bsnm{Glavic}, \binits{A.}},
\bauthor{\bsnm{H{\"o}glund}, \binits{C.}},
\bauthor{\bsnm{Ilves}, \binits{T.}},
\bauthor{\bsnm{Higuera}, \binits{I.L.}},
\bauthor{\bsnm{Pazmandi}, \binits{P.}},
\bauthor{\bsnm{Raspino}, \binits{D.}},
\bauthor{\bsnm{Robinson}, \binits{L.}},
\bauthor{\bsnm{Schmidt}, \binits{S.}},
\bauthor{\bsnm{Svensson}, \binits{P.}},
\bauthor{\bsnm{Varga}, \binits{D.}},
\bauthor{\bsnm{Hall-Wilton}, \binits{R.}}:
\batitle{{Characterization of the Multi-Blade 10B-based detector at the CRISP
  reflectometer at ISIS for neutron reflectometry at ESS}}.
\bjtitle{Journal of Instrumentation}
\bvolume{13}(\bissue{05}),
\bfpage{05009}
(\byear{2018})
\end{barticle}
\endbibitem

\bibitem{MIO_ScientificMBcrisp}
\begin{botherref}
\oauthor{\bsnm{Mauri}, \binits{G.}},
\oauthor{\bsnm{Messi}, \binits{F.}},
\oauthor{\bsnm{Anastasopoulos}, \binits{M.}},
\oauthor{\bsnm{Arnold}, \binits{T.}},
\oauthor{\bsnm{Glavic}, \binits{A.}},
\oauthor{\bsnm{H{\"o}glund}, \binits{C.}},
\oauthor{\bsnm{Ilves}, \binits{T.}},
\oauthor{\bsnm{Lopez~Higuera}, \binits{I.}},
\oauthor{\bsnm{Pazmandi}, \binits{P.}},
\oauthor{\bsnm{Raspino}, \binits{D.}},
\oauthor{\bsnm{Robinson}, \binits{L.}},
\oauthor{\bsnm{Schmidt}, \binits{S.}},
\oauthor{\bsnm{Svensson}, \binits{P.}},
\oauthor{\bsnm{Varga}, \binits{D.}},
\oauthor{\bsnm{Hall-Wilton}, \binits{R.}},
\oauthor{\bsnm{Piscitelli}, \binits{F.}}:
{Neutron reflectometry with the Multi-Blade 10B-based detector}.
Proceedings of the Royal Society of London A: Mathematical, Physical and
  Engineering Sciences
\textbf{474}(2216)
(2018).
doi:\doiurl{10.1098/rspa.2018.0266}.
\arxivurl{http://rspa.royalsocietypublishing.org/content/474/2216/20180266.full.pdf}
\end{botherref}
\endbibitem

\bibitem{DET_jalousie}
\begin{barticle}
\bauthor{\bsnm{Henske}, \binits{M.}},
\bauthor{\bsnm{Klein}, \binits{M.}},
\bauthor{\bsnm{K{\"o}hli}, \binits{M.}},
\bauthor{\bsnm{Lennert}, \binits{P.}},
\bauthor{\bsnm{Modzel}, \binits{G.}},
\bauthor{\bsnm{Schmidt}, \binits{C.J.}},
\bauthor{\bsnm{Schmidt}, \binits{U.}}:
\batitle{{The 10B based Jalousie neutron detector -- An alternative for 3He
  filled position sensitive counter tubes}}.
\bjtitle{Nuclear Instruments and Methods in Physics Research Section A:
  Accelerators, Spectrometers, Detectors and Associated Equipment}
\bvolume{686},
\bfpage{151}--\blpage{155}
(\byear{2012}).
doi:\doiurl{10.1016/j.nima.2012.05.075}
\end{barticle}
\endbibitem

\bibitem{MPGD_GEMcroci}
\begin{barticle}
\bauthor{\bsnm{Croci}, \binits{G.}},
\bauthor{\bparticle{et} \bsnm{al.}}:
\batitle{Diffraction measurements with a boron-based {GEM} neutron detector}.
\bjtitle{EPL}
\bvolume{107}(\bissue{1}),
\bfpage{12001}
(\byear{2014}).
doi:\doiurl{10.1209/0295-5075/107/12001}
\end{barticle}
\endbibitem

\bibitem{Bgem}
\begin{barticle}
\bauthor{\bsnm{Albani}, \binits{G.}},
\bauthor{\bsnm{Cippo}, \binits{E.P.}},
\bauthor{\bsnm{Croci}, \binits{G.}},
\bauthor{\bsnm{Muraro}, \binits{A.}},
\bauthor{\bsnm{Schooneveld}, \binits{E.}},
\bauthor{\bsnm{Scherillo}, \binits{A.}},
\bauthor{\bsnm{Hall-Wilton}, \binits{R.}},
\bauthor{\bsnm{Kanaki}, \binits{K.}},
\bauthor{\bsnm{H{\"o}glund}, \binits{C.}},
\bauthor{\bsnm{Hultman}, \binits{L.}},
\bauthor{\bsnm{Birch}, \binits{J.}},
\bauthor{\bsnm{Claps}, \binits{G.}},
\bauthor{\bsnm{Murtas}, \binits{F.}},
\bauthor{\bsnm{Rebai}, \binits{M.}},
\bauthor{\bsnm{Tardocchi}, \binits{M.}},
\bauthor{\bsnm{Gorini}, \binits{G.}}:
\batitle{{Evolution in boron-based GEM detectors for diffraction measurements:
  from planar to 3D converters}}.
\bjtitle{Measurement Science and Technology}
\bvolume{27}(\bissue{11}),
\bfpage{115902}
(\byear{2016})
\end{barticle}
\endbibitem

\bibitem{STRAW_lacy2011}
\begin{barticle}
\bauthor{\bsnm{Lacy}, \binits{J.L.}},
\bauthor{\bsnm{Athanasiades}, \binits{A.}},
\bauthor{\bsnm{Sun}, \binits{L.}},
\bauthor{\bsnm{Martin}, \binits{C.S.}},
\bauthor{\bsnm{Lyons}, \binits{T.D.}},
\bauthor{\bsnm{Foss}, \binits{M.A.}},
\bauthor{\bsnm{Haygood}, \binits{H.B.}}:
\batitle{{Boron-coated straws as a replacement for 3He-based neutron
  detectors}}.
\bjtitle{Nuclear Instruments and Methods in Physics Research Section A:
  Accelerators, Spectrometers, Detectors and Associated Equipment}
\bvolume{652}(\bissue{1}),
\bfpage{359}--\blpage{363}
(\byear{2011}).
doi:\doiurl{10.1016/j.nima.2010.09.011}.
\bcomment{Symposium on Radiation Measurements and Applications (SORMA) \{XII\}
  2010}
\end{barticle}
\endbibitem

\bibitem{DET_kohli}
\begin{barticle}
\bauthor{\bsnm{K{\"o}hli}, \binits{M.}},
\bauthor{\bsnm{Allmendinger}, \binits{F.}},
\bauthor{\bsnm{H{\"a}u{\ss}ler}, \binits{W.}},
\bauthor{\bsnm{Schr{\"o}der}, \binits{T.}},
\bauthor{\bsnm{Klein}, \binits{M.}},
\bauthor{\bsnm{Meven}, \binits{M.}},
\bauthor{\bsnm{Schmidt}, \binits{U.}}:
\batitle{{Efficiency and spatial resolution of the CASCADE thermal neutron
  detector}}.
\bjtitle{Nuclear Instruments and Methods in Physics Research Section A:
  Accelerators, Spectrometers, Detectors and Associated Equipment}
\bvolume{828},
\bfpage{242}--\blpage{249}
(\byear{2016}).
doi:\doiurl{10.1016/j.nima.2016.05.014}
\end{barticle}
\endbibitem

\bibitem{DET_doro1}
\begin{barticle}
\bauthor{\bsnm{Pfeiffer}, \binits{D.}},
\bauthor{\bsnm{Resnati}, \binits{F.}},
\bauthor{\bsnm{Birch}, \binits{J.}},
\bauthor{\bsnm{Etxegarai}, \binits{M.}},
\bauthor{\bsnm{Hall-Wilton}, \binits{R.}},
\bauthor{\bsnm{H{\"o}glund}, \binits{C.}},
\bauthor{\bsnm{Hultman}, \binits{L.}},
\bauthor{\bsnm{Llamas-Jansa}, \binits{I.}},
\bauthor{\bsnm{Oliveri}, \binits{E.}},
\bauthor{\bsnm{Oksanen}, \binits{E.}},
\bauthor{\bsnm{Robinson}, \binits{L.}},
\bauthor{\bsnm{Ropelewski}, \binits{L.}},
\bauthor{\bsnm{Schmidt}, \binits{S.}},
\bauthor{\bsnm{Streli}, \binits{C.}},
\bauthor{\bsnm{Thuiner}, \binits{P.}}:
\batitle{First measurements with new high-resolution gadolinium-{GEM} neutron
  detectors}.
\bjtitle{Journal of Instrumentation}
\bvolume{11}(\bissue{05}),
\bfpage{05011}
(\byear{2016})
\end{barticle}
\endbibitem

\bibitem{gdgem}
\begin{barticle}
\bauthor{\bsnm{Pfeiffer}, \binits{D.}},
\bauthor{\bsnm{Resnati}, \binits{F.}},
\bauthor{\bsnm{Birch}, \binits{J.}},
\bauthor{\bsnm{Hall-Wilton}, \binits{R.}},
\bauthor{\bsnm{H{\"o}glund}, \binits{C.}},
\bauthor{\bsnm{Hultman}, \binits{L.}},
\bauthor{\bsnm{Iakovidis}, \binits{G.}},
\bauthor{\bsnm{Oliveri}, \binits{E.}},
\bauthor{\bsnm{Oksanen}, \binits{E.}},
\bauthor{\bsnm{Ropelewski}, \binits{L.}},
\bauthor{\bsnm{Thuiner}, \binits{P.}}:
\batitle{{The {$\mu$}TPC method: improving the position resolution of neutron
  detectors based on MPGDs}}.
\bjtitle{Journal of Instrumentation}
\bvolume{10}(\bissue{04}),
\bfpage{04004}
(\byear{2015})
\end{barticle}
\endbibitem

\bibitem{Mireshghi_sigd}
\begin{barticle}
\bauthor{\bsnm{Mireshghi}, \binits{A.}},
\bauthor{\bsnm{Cho}, \binits{G.}},
\bauthor{\bsnm{Drewery}, \binits{J.}},
\bauthor{\bsnm{Jing}, \binits{T.}},
\bauthor{\bsnm{Kaplan}, \binits{S.N.}},
\bauthor{\bsnm{Perez-Mendez}, \binits{V.}},
\bauthor{\bsnm{Wildermuth}, \binits{D.}}:
\batitle{Amorphous silicon position sensitive neutron detector}.
\bjtitle{IEEE Transactions on Nuclear Science}
\bvolume{39}(\bissue{4}),
\bfpage{635}--\blpage{640}
(\byear{1992}).
doi:\doiurl{10.1109/23.159678}
\end{barticle}
\endbibitem

\bibitem{SCHULTE_sigd}
\begin{barticle}
\bauthor{\bsnm{Schulte}, \binits{R.L.}},
\bauthor{\bsnm{Swanson}, \binits{F.}},
\bauthor{\bsnm{Kesselman}, \binits{M.}}:
\batitle{The use of large area silicon sensors for thermal neutron detection}.
\bjtitle{Nuclear Instruments and Methods in Physics Research Section A:
  Accelerators, Spectrometers, Detectors and Associated Equipment}
\bvolume{353}(\bissue{1}),
\bfpage{123}--\blpage{127}
(\byear{1994}).
doi:\doiurl{10.1016/0168-9002(94)91617-9}
\end{barticle}
\endbibitem

\bibitem{PETRILLO-solidstate}
\begin{barticle}
\bauthor{\bsnm{Petrillo}, \binits{C.}},
\bauthor{\bsnm{Sacchetti}, \binits{F.}},
\bauthor{\bsnm{Toker}, \binits{O.}},
\bauthor{\bsnm{Rhodes}, \binits{N.J.}}:
\batitle{Solid state neutron detectors}.
\bjtitle{Nuclear Instruments and Methods in Physics Research Section A:
  Accelerators, Spectrometers, Detectors and Associated Equipment}
\bvolume{378}(\bissue{3}),
\bfpage{541}--\blpage{551}
(\byear{1996}).
doi:\doiurl{10.1016/0168-9002(96)00487-1}
\end{barticle}
\endbibitem

\bibitem{Mauri_psa}
\begin{barticle}
\bauthor{\bsnm{Mauri}, \binits{G.}},
\bauthor{\bsnm{Mariotti}, \binits{M.}},
\bauthor{\bsnm{Casinini}, \binits{F.}},
\bauthor{\bsnm{Sacchetti}, \binits{F.}},
\bauthor{\bsnm{Petrillo}, \binits{C.}}:
\batitle{{Development of pulse shape analysis for noise reduction in Si-based
  neutron detectors}}.
\bjtitle{Nuclear Instruments and Methods in Physics Research Section A:
  Accelerators, Spectrometers, Detectors and Associated Equipment}
\bvolume{910},
\bfpage{184}--\blpage{193}
(\byear{2018}).
doi:\doiurl{10.1016/j.nima.2018.09.071}
\end{barticle}
\endbibitem

\bibitem{MIO_fastn}
\begin{barticle}
\bauthor{\bsnm{Mauri}, \binits{G.}},
\bauthor{\bsnm{Messi}, \binits{F.}},
\bauthor{\bsnm{Kanaki}, \binits{K.}},
\bauthor{\bsnm{Hall-Wilton}, \binits{R.}},
\bauthor{\bsnm{Karnickis}, \binits{E.}},
\bauthor{\bsnm{Khaplanov}, \binits{A.}},
\bauthor{\bsnm{Piscitelli}, \binits{F.}}:
\batitle{Fast neutron sensitivity of neutron detectors based on boron-10
  converter layers}.
\bjtitle{Journal of Instrumentation}
\bvolume{13}(\bissue{03}),
\bfpage{03004}--\blpage{171205614}
(\byear{2018})
\end{barticle}
\endbibitem

\bibitem{MIO_MB2017}
\begin{barticle}
\bauthor{\bsnm{Piscitelli}, \binits{F.}},
\bauthor{\bsnm{Messi}, \binits{F.}},
\bauthor{\bsnm{Anastasopoulos}, \binits{M.}},
\bauthor{\bsnm{Bry{\'s}}, \binits{T.}},
\bauthor{\bsnm{Chicken}, \binits{F.}},
\bauthor{\bsnm{Dian}, \binits{E.}},
\bauthor{\bsnm{Fuzi}, \binits{J.}},
\bauthor{\bsnm{H{\"o}glund}, \binits{C.}},
\bauthor{\bsnm{Kiss}, \binits{G.}},
\bauthor{\bsnm{Orban}, \binits{J.}},
\bauthor{\bsnm{Pazmandi}, \binits{P.}},
\bauthor{\bsnm{Robinson}, \binits{L.}},
\bauthor{\bsnm{Rosta}, \binits{L.}},
\bauthor{\bsnm{Schmidt}, \binits{S.}},
\bauthor{\bsnm{Varga}, \binits{D.}},
\bauthor{\bsnm{Zsiros}, \binits{T.}},
\bauthor{\bsnm{Hall-Wilton}, \binits{R.}}:
\batitle{{The Multi-Blade Boron-10-based neutron detector for high intensity
  neutron reflectometry at ESS}}.
\bjtitle{Journal of Instrumentation}
\bvolume{12}(\bissue{03}),
\bfpage{03013}
(\byear{2017})
\end{barticle}
\endbibitem

\bibitem{CRISP1}
\begin{botherref}
{CRISP instrument manual 2010 -
  https://www.isis.stfc.ac.uk/Pages/crisp-instrument-manual-nov-2010.pdf}
(2010).
\url{https://www.isis.stfc.ac.uk/Pages/crisp-instrument-manual-nov-2010.pdf}
\end{botherref}
\endbibitem

\bibitem{ISIS}
\begin{botherref}
{ISIS Neutron and Muon Source - https://www.isis.stfc.ac.uk}.
\url{https://www.isis.stfc.ac.uk}
\end{botherref}
\endbibitem

\bibitem{SF2}
\begin{botherref}
\oauthor{\bsnm{Messi}, \binits{F.}},
\oauthor{\bsnm{Perrey}, \binits{H.}},
\oauthor{\bsnm{Fissum}, \binits{K.}},
\oauthor{\bsnm{Akkawi}, \binits{M.}},
\oauthor{\bsnm{Jebali}, \binits{R.A.}},
\oauthor{\bsnm{Annand}, \binits{J.R.M.}},
\oauthor{\bsnm{Bentley}, \binits{P.}},
\oauthor{\bsnm{Boyd}, \binits{L.}},
\oauthor{\bsnm{Cooper-Jensen}, \binits{C.P.}},
\oauthor{\bsnm{DiJulio}, \binits{D.D.}},
\oauthor{\bsnm{Freita-Ramos}, \binits{J.}},
\oauthor{\bsnm{Hall-Wilton}, \binits{R.}},
\oauthor{\bsnm{Huusko}, \binits{A.}},
\oauthor{\bsnm{Ilves}, \binits{T.}},
\oauthor{\bsnm{Issa}, \binits{F.}},
\oauthor{\bsnm{Jalgen}, \binits{A.}},
\oauthor{\bsnm{Kanaki}, \binits{K.}},
\oauthor{\bsnm{Karnickis}, \binits{E.}},
\oauthor{\bsnm{Khaplanov}, \binits{A.}},
\oauthor{\bsnm{Koufigar}, \binits{S.}},
\oauthor{\bsnm{Maulerova}, \binits{V.}},
\oauthor{\bsnm{Mauri}, \binits{G.}},
\oauthor{\bsnm{Mauritzson}, \binits{N.}},
\oauthor{\bsnm{Pei}, \binits{W.}},
\oauthor{\bsnm{Piscitelli}, \binits{F.}},
\oauthor{\bsnm{Rofors}, \binits{E.}},
\oauthor{\bsnm{Scherzinger}, \binits{J.}},
\oauthor{\bsnm{Soderhielm}, \binits{H.}},
\oauthor{\bsnm{Soderstrom}, \binits{D.}},
\oauthor{\bsnm{Stefanescu}, \binits{I.}}:
{The neutron tagging facility at Lund University}.
arXiv:1711.10286 (submitted to IAEA Technical Report on Modern Neutron
  Detection (2017))
(2017)
\end{botherref}
\endbibitem

\bibitem{ge-reuter}
\begin{botherref}
General Electric (GE) Reuter Stokes.
https://www.industrial.ai.
\url{https://www.industrial.ai}
\end{botherref}
\endbibitem

\bibitem{SF1}
\begin{barticle}
\bauthor{\bsnm{Scherzinger}, \binits{J.}},
\bauthor{\bsnm{Annand}, \binits{J.R.M.}},
\bauthor{\bsnm{Davatz}, \binits{G.}},
\bauthor{\bsnm{Fissum}, \binits{K.G.}},
\bauthor{\bsnm{Gendotti}, \binits{U.}},
\bauthor{\bsnm{Hall-Wilton}, \binits{R.}},
\bauthor{\bsnm{H{\aa}kansson}, \binits{E.}},
\bauthor{\bsnm{Jebali}, \binits{R.}},
\bauthor{\bsnm{Kanaki}, \binits{K.}},
\bauthor{\bsnm{Lundin}, \binits{M.}},
\bauthor{\bsnm{Nilsson}, \binits{B.}},
\bauthor{\bsnm{Rosborge}, \binits{A.}},
\bauthor{\bsnm{Svensson}, \binits{H.}}:
\batitle{{Tagging fast neutrons from an 241Am/9Be source}}.
\bjtitle{Applied Radiation and Isotopes}
\bvolume{98},
\bfpage{74}--\blpage{79}
(\byear{2015}).
doi:\doiurl{10.1016/j.apradiso.2015.01.003}
\end{barticle}
\endbibitem

\bibitem{PuBe_AmBe}
\begin{barticle}
\bauthor{\bsnm{Scherzinger}, \binits{J.}},
\bauthor{\bsnm{Jebali}, \binits{R.A.}},
\bauthor{\bsnm{Annand}, \binits{J.R.M.}},
\bauthor{\bsnm{Fissum}, \binits{K.G.}},
\bauthor{\bsnm{Hall-Wilton}, \binits{R.}},
\bauthor{\bsnm{Koufigar}, \binits{S.}},
\bauthor{\bsnm{Mauritzson}, \binits{N.}},
\bauthor{\bsnm{Messi}, \binits{F.}},
\bauthor{\bsnm{Perrey}, \binits{H.}},
\bauthor{\bsnm{Rofors}, \binits{E.}}:
\batitle{{A comparison of untagged gamma-ray and tagged-neutron yields from
  241AmBe and 238PuBe sources}}.
\bjtitle{Applied Radiation and Isotopes}
\bvolume{127},
\bfpage{98}--\blpage{102}
(\byear{2017}).
doi:\doiurl{10.1016/j.apradiso.2017.05.014}
\end{barticle}
\endbibitem

\bibitem{Rossi-gsHe}
\begin{botherref}
\oauthor{\bsnm{Rossi}, \binits{E.}}:
{Characterisation of the Spatial Resolution and the Gamma-ray Discrimination of
  Helium-3 Proportional Counters}.
Master's thesis
(2017).
\arxivurl{1702.06501}
\end{botherref}
\endbibitem

\bibitem{fastNmeas}
\begin{barticle}
\bauthor{\bsnm{{Tsutomu IIJIMA and Takehiko MUKAIYAMA and Keisho SHIRAKATA}}}:
\batitle{{Measurement of Fast Neutron Spectrum with Helium-3 Proportional
  Counter}}.
\bjtitle{Journal of Nuclear Science and Technology}
\bvolume{8}(\bissue{4}),
\bfpage{192}--\blpage{200}
(\byear{1971}).
doi:\doiurl{10.1080/18811248.1971.9735315}.
\arxivurl{https://www.tandfonline.com/doi/pdf/10.1080/18811248.1971.9735315}
\end{barticle}
\endbibitem

\bibitem{NIST}
\begin{botherref}
NIST National Nuclear Data Center.
https://www.nndc.bnl.gov.
\url{https://www.nndc.bnl.gov}
\end{botherref}
\endbibitem

\bibitem{geant4a}
\begin{barticle}
\bauthor{\bsnm{{S.~Agostinelli, J.~Allison \textit{et al}.}}}:
\batitle{{Geant4 - A Simulation Toolkit}}.
\bjtitle{Nucl. Instrum. Meth. A}
\bvolume{506},
\bfpage{250}--\blpage{303}
(\byear{2003})
\end{barticle}
\endbibitem

\bibitem{geant4b}
\begin{barticle}
\bauthor{\bsnm{{J.~Allison \textit{et al}.}}}:
\batitle{{Geant4 developments and applications}}.
\bjtitle{IEEE Trans. Nucl. Sci.}
\bvolume{53},
\bfpage{07}
(\byear{2006})
\end{barticle}
\endbibitem

\bibitem{geant4c}
\begin{barticle}
\bauthor{\bsnm{{J.~Allison, K.~Amako \textit{et al}.}}}:
\batitle{{Recent Developments in Geant4}}.
\bjtitle{Nucl. Instrum. Meth. A}
\bvolume{835},
\bfpage{186}--\blpage{225}
(\byear{2016})
\end{barticle}
\endbibitem

\bibitem{ess_coding_framework}
\begin{barticle}
\bauthor{\bsnm{{T.~Kittelmann \textit{et al}.}}}:
\batitle{{Geant4 Based Simulations for Novel Neutron Detector Development}}.
\bjtitle{J. Phys. Conf. Ser.}
\bvolume{513},
\bfpage{022017}
(\byear{2014})
\end{barticle}
\endbibitem

\bibitem{nxsg4}
\begin{barticle}
\bauthor{\bsnm{Kittelmann}, \binits{T.}},
\bauthor{\bsnm{Boin}, \binits{M.}}:
\batitle{Polycrystalline neutron scattering for {Geant4}: {NXSG4}}.
\bjtitle{Computer Physics Communications}
\bvolume{189}(\bissue{0}),
\bfpage{114}--\blpage{118}
(\byear{2015}).
doi:\doiurl{10.1016/j.cpc.2014.11.009}
\end{barticle}
\endbibitem

\bibitem{backscattering}
\begin{barticle}
\bauthor{\bsnm{Kittelmann}, \binits{T.}},
\bauthor{\bsnm{Kanaki}, \binits{K.}},
\bauthor{\bsnm{Klinkby}, \binits{E.}},
\bauthor{\bsnm{Cai}, \binits{X.X.}},
\bauthor{\bsnm{Cooper-Jensen}, \binits{C.P.}},
\bauthor{\bsnm{Hall-Wilton}, \binits{R.}}:
\batitle{Using backscattering to enhance efficiency in neutron detectors}.
\bjtitle{IEEE TNS}
\bvolume{64}(\bissue{6}),
\bfpage{1562}--\blpage{1573}
(\byear{2017}).
doi:\doiurl{10.1109/TNS.2017.2695404}
\end{barticle}
\endbibitem

\bibitem{GeantThnflux}
\begin{botherref}
\oauthor{\bsnm{Deiev}, \binits{O.S.}}:
{GEANT 4 simulation of neutron transport and scattering in media}
(3-85/60),
236--241
(2013)
\end{botherref}
\endbibitem

\bibitem{sans-frm2-he3}
\begin{barticle}
\bauthor{\bsnm{M{\"u}hlbauer}, \binits{S.}},
\bauthor{\bsnm{Heinemann}, \binits{A.}},
\bauthor{\bsnm{Wilhelm}, \binits{A.}},
\bauthor{\bsnm{Karge}, \binits{L.}},
\bauthor{\bsnm{Ostermann}, \binits{A.}},
\bauthor{\bsnm{Defendi}, \binits{I.}},
\bauthor{\bsnm{Schreyer}, \binits{A.}},
\bauthor{\bsnm{Petry}, \binits{W.}},
\bauthor{\bsnm{Gilles}, \binits{R.}}:
\batitle{{The new small-angle neutron scattering instrument SANS-1 at
  MLZ---characterization and first results}}.
\bjtitle{Nuclear Instruments and Methods in Physics Research Section A:
  Accelerators, Spectrometers, Detectors and Associated Equipment}
\bvolume{832},
\bfpage{297}--\blpage{305}
(\byear{2016}).
doi:\doiurl{10.1016/j.nima.2016.06.105}
\end{barticle}
\endbibitem

\bibitem{MG_gamma}
\begin{barticle}
\bauthor{\bsnm{Khaplanov}, \binits{A.}},
\bauthor{\bsnm{Piscitelli}, \binits{F.}},
\bauthor{\bsnm{Buffet}, \binits{J.-C.}},
\bauthor{\bsnm{Clergeau}, \binits{J.-F.}},
\bauthor{\bsnm{Correa}, \binits{J.}},
\bauthor{\bparticle{van} \bsnm{Esch}, \binits{P.}},
\bauthor{\bsnm{Ferraton}, \binits{M.}},
\bauthor{\bsnm{Guerard}, \binits{B.}},
\bauthor{\bsnm{Hall-Wilton}, \binits{R.}}:
\batitle{Investigation of gamma-ray sensitivity of neutron detectors based on
  thin converter films}.
\bjtitle{Journal of Instrumentation}
\bvolume{8}(\bissue{10}),
\bfpage{10025}
(\byear{2013})
\end{barticle}
\endbibitem

\end{thebibliography}

\newcommand{\BMCxmlcomment}[1]{}

\BMCxmlcomment{

<refgrp>

<bibl id="B1">
  <aug>
    <au><snm>Peggs</snm><fnm>S.</fnm></au>
  </aug>
  <source>ESS Technical Design Report - (ESS-2013-0001)</source>
  <url>http://eval.esss.lu.se/cgi-bin/public/DocDB/ShowDocument?docid=274.</url>
</bibl>

<bibl id="B2">
  <title><p>{SNS}</p></title>
  <source>http://neutrons.ornl.gov/facilities/SNS/</source>
  <url>http://neutrons.ornl.gov/facilities/SNS/</url>
</bibl>

<bibl id="B3">
  <title><p>{J-PARC}</p></title>
  <source>http://j-parc.jp/MatLife/en/index.html</source>
  <url>http://j-parc.jp/MatLife/en/index.html</url>
</bibl>

<bibl id="B4">
  <title><p>{European Spallation Source ESS ERIC}</p></title>
  <source>http://europeanspallationsource.se</source>
  <url>http://europeanspallationsource.se</url>
</bibl>

<bibl id="B5">
  <title><p>{The European Spallation Source Design}</p></title>
  <aug>
    <au><snm>Garoby</snm><fnm>R</fnm></au>
  </aug>
  <source>Physica Scripta</source>
  <pubdate>2018</pubdate>
  <volume>93</volume>
  <issue>1</issue>
  <fpage>014001</fpage>
  <url>http://stacks.iop.org/1402-4896/93/i=1/a=014001</url>
</bibl>

<bibl id="B6">
  <title><p>New perspectives from new generations of neutron
  sources</p></title>
  <aug>
    <au><snm>Mezei</snm><fnm>F</fnm></au>
  </aug>
  <source>Comptes Rendus Physique</source>
  <pubdate>2007</pubdate>
  <volume>8</volume>
  <issue>7</issue>
  <fpage>909</fpage>
  <lpage>920</lpage>
  <url>http://www.sciencedirect.com/science/article/pii/S1631070507002496</url>
  <note>Neutron scattering: a comprehensive tool for condensed matter
  research</note>
</bibl>

<bibl id="B7">
  <title><p>Progress for the European spallation source in
  Scandinavia</p></title>
  <aug>
    <au><snm>Vettier</snm><fnm>C</fnm></au>
    <au><snm>Carlile</snm><fnm>CJ</fnm></au>
    <au><snm>Carlsson</snm><fnm>P</fnm></au>
  </aug>
  <source>Nuclear Instruments and Methods in Physics Research Section A:
  Accelerators, Spectrometers, Detectors and Associated Equipment</source>
  <pubdate>2009</pubdate>
  <volume>600</volume>
  <issue>1</issue>
  <fpage>8</fpage>
  <lpage>9</lpage>
  <url>http://www.sciencedirect.com/science/article/pii/S0168900208016434</url>
</bibl>

<bibl id="B8">
  <title><p>{The European Spallation Source}</p></title>
  <aug>
    <au><snm>Lindroos</snm><fnm>M</fnm></au>
    <au><snm>Bousson</snm><fnm>S.</fnm></au>
    <au><snm>Calaga</snm><fnm>R.</fnm></au>
    <au><snm>Danared</snm><fnm>H</fnm></au>
    <au><snm>Devanz</snm><fnm>G.</fnm></au>
    <au><snm>Duperrier</snm><fnm>R.</fnm></au>
    <au><snm>Eguia</snm><fnm>J.</fnm></au>
    <au><snm>Eshraqi</snm><fnm>M</fnm></au>
    <au><snm>Gammino</snm><fnm>S.</fnm></au>
    <au><snm>Hahn</snm><fnm>H</fnm></au>
    <au><snm>Jansson</snm><fnm>A</fnm></au>
    <au><snm>Oyon</snm><fnm>C.</fnm></au>
    <au><snm>Pape Moller</snm><fnm>S.</fnm></au>
    <au><snm>Peggs</snm><fnm>S</fnm></au>
    <au><snm>Ponton</snm><fnm>A.</fnm></au>
    <au><snm>Rathsman</snm><fnm>K</fnm></au>
    <au><snm>Ruber</snm><fnm>R.</fnm></au>
    <au><snm>Satogata</snm><fnm>T.</fnm></au>
    <au><snm>Trahern</snm><fnm>G</fnm></au>
  </aug>
  <source>Nuclear Instruments and Methods in Physics Research, Section B: Beam
  Interactions with Materials and Atoms</source>
  <publisher>Elsevier</publisher>
  <pubdate>2011</pubdate>
  <volume>269</volume>
  <issue>24</issue>
  <fpage>3258</fpage>
  <lpage>-3260</lpage>
</bibl>

<bibl id="B9">
  <title><p>{The 3He Supply Problem}</p></title>
  <aug>
    <au><snm>Kouzes</snm><fnm>R T</fnm></au>
  </aug>
  <pubdate>2009</pubdate>
</bibl>

<bibl id="B10">
  <title><p>International collaboration on the development of neutron
  detectors</p></title>
  <source>http://www.icnd.org</source>
  <url>http://www.icnd.org</url>
</bibl>

<bibl id="B11">
  <title><p>Search for alternative techniques to helium-3 based detectors for
  neutron scattering applications</p></title>
  <aug>
    <au><snm>Zeitelhack</snm><fnm>K</fnm></au>
  </aug>
  <source>Neutron News</source>
  <pubdate>2012</pubdate>
  <volume>23</volume>
  <issue>4</issue>
  <fpage>10</fpage>
  <lpage>13</lpage>
  <url>http://dx.doi.org/10.1080/10448632.2012.725325</url>
</bibl>

<bibl id="B12">
  <title><p>Why new neutron detector materials must replace
  helium-3</p></title>
  <aug>
    <au><snm>Hurd</snm><fnm>AJ</fnm></au>
    <au><snm>Kouzes</snm><fnm>RT</fnm></au>
  </aug>
  <source>The European Physical Journal Plus</source>
  <pubdate>2014</pubdate>
  <volume>129</volume>
  <issue>10</issue>
  <fpage>1</fpage>
  <lpage>-3</lpage>
  <url>http://dx.doi.org/10.1140/epjp/i2014-14236-6</url>
</bibl>

<bibl id="B13">
  <title><p>{Multi-Grid detector for neutron spectroscopy: results obtained on
  time-of-flight spectrometer CNCS}</p></title>
  <aug>
    <au><snm>Anastasopoulos</snm><fnm>M.</fnm></au>
    <au><snm>Bebb</snm><fnm>R.</fnm></au>
    <au><snm>Berry</snm><fnm>K.</fnm></au>
    <au><snm>Birch</snm><fnm>J.</fnm></au>
    <au><snm>Bry{\'s}</snm><fnm>T.</fnm></au>
    <au><snm>Buffet</snm><fnm>J. C.</fnm></au>
    <au><snm>Clergeau</snm><fnm>J. F.</fnm></au>
    <au><snm>Deen</snm><fnm>P.P.</fnm></au>
    <au><snm>Ehlers</snm><fnm>G.</fnm></au>
    <au><snm>Esch</snm><fnm>P.</fnm></au>
    <au><snm>Everett</snm><fnm>S.M.</fnm></au>
    <au><snm>Guerard</snm><fnm>B.</fnm></au>
    <au><snm>Hall Wilton</snm><fnm>R.</fnm></au>
    <au><snm>Herwig</snm><fnm>K.</fnm></au>
    <au><snm>Hultman</snm><fnm>L.</fnm></au>
    <au><snm>H{\"o}glund</snm><fnm>C.</fnm></au>
    <au><snm>Iruretagoiena</snm><fnm>I.</fnm></au>
    <au><snm>Issa</snm><fnm>F.</fnm></au>
    <au><snm>Jensen</snm><fnm>J.</fnm></au>
    <au><snm>Khaplanov</snm><fnm>A.</fnm></au>
    <au><snm>Kirstein</snm><fnm>O.</fnm></au>
    <au><snm>Higuera</snm><fnm>IL</fnm></au>
    <au><snm>Piscitelli</snm><fnm>F.</fnm></au>
    <au><snm>Robinson</snm><fnm>L.</fnm></au>
    <au><snm>Schmidt</snm><fnm>S.</fnm></au>
    <au><snm>Stefanescu</snm><fnm>I.</fnm></au>
  </aug>
  <source>Journal of Instrumentation</source>
  <pubdate>2017</pubdate>
  <volume>12</volume>
  <issue>04</issue>
  <fpage>P04030</fpage>
  <url>http://stacks.iop.org/1748-0221/12/i=04/a=P04030</url>
</bibl>

<bibl id="B14">
  <title><p>{In-beam test of the Boron-10 Multi-Grid neutron detector at the
  IN6 time-of-flight spectrometer at the ILL}</p></title>
  <aug>
    <au><snm>Birch</snm><fnm>J</fnm></au>
    <au><snm>Buffet</snm><fnm>J C</fnm></au>
    <au><snm>Clergeau</snm><fnm>J F</fnm></au>
    <au><snm>Correa</snm><fnm>J</fnm></au>
    <au><snm>Esch</snm><fnm>P</fnm></au>
    <au><snm>Ferraton</snm><fnm>M</fnm></au>
    <au><snm>Guerard</snm><fnm>B</fnm></au>
    <au><snm>Halbwachs</snm><fnm>J</fnm></au>
    <au><snm>Hall Wilton</snm><fnm>R</fnm></au>
    <au><snm>Hultman</snm><fnm>L</fnm></au>
    <au><snm>H{\"o}glund</snm><fnm>C</fnm></au>
    <au><snm>Khaplanov</snm><fnm>A</fnm></au>
    <au><snm>Koza</snm><fnm>M</fnm></au>
    <au><snm>Piscitelli</snm><fnm>F</fnm></au>
    <au><snm>Zbiri</snm><fnm>M</fnm></au>
  </aug>
  <source>Journal of Physics: Conference Series</source>
  <pubdate>2014</pubdate>
  <volume>528</volume>
  <issue>1</issue>
  <fpage>012040</fpage>
  <url>http://stacks.iop.org/1742-6596/528/i=1/a=012040</url>
</bibl>

<bibl id="B15">
  <title><p>{Ionizing radiation detector - Patent no. 20110215251.}</p></title>
  <aug>
    <au><snm>Guerard</snm><fnm>B.</fnm></au>
    <au><snm>Buffet</snm><fnm>J.C.</fnm></au>
  </aug>
  <publisher>Google Patents</publisher>
  <pubdate>2011</pubdate>
  <url>https://www.google.com/patents/US20110215251</url>
  <note>US Patent App. 13/038,915</note>
</bibl>

<bibl id="B16">
  <title><p>{10B4C Multi-Grid as an Alternative to 3He for Large Area Neutron
  Detectors}</p></title>
  <aug>
    <au><snm>Birch</snm><fnm>J.</fnm></au>
    <au><snm>Buffet</snm><fnm>J. C.</fnm></au>
    <au><snm>Correa</snm><fnm>J.</fnm></au>
    <au><snm>Esch</snm><fnm>P.</fnm></au>
    <au><snm>Gu{\'e}rard</snm><fnm>B.</fnm></au>
    <au><snm>Hall Wilton</snm><fnm>R.</fnm></au>
    <au><snm>H{\"o}glund</snm><fnm>C.</fnm></au>
    <au><snm>Hultman</snm><fnm>L.</fnm></au>
    <au><snm>Khaplanov</snm><fnm>A.</fnm></au>
    <au><snm>Piscitelli</snm><fnm>F.</fnm></au>
  </aug>
  <source>IEEE Transactions on Nuclear Science</source>
  <pubdate>2013</pubdate>
  <volume>60</volume>
  <issue>2</issue>
  <fpage>871</fpage>
  <lpage>878</lpage>
</bibl>

<bibl id="B17">
  <title><p>{Study of a high spatial resolution 10 B -based thermal neutron
  detector for application in neutron reflectometry: the Multi-Blade
  prototype}</p></title>
  <aug>
    <au><snm>Piscitelli</snm><fnm>F</fnm></au>
    <au><snm>Buffet</snm><fnm>J C</fnm></au>
    <au><snm>Clergeau</snm><fnm>J F</fnm></au>
    <au><snm>Cuccaro</snm><fnm>S</fnm></au>
    <au><snm>Gu{\'e}rard</snm><fnm>B</fnm></au>
    <au><snm>Khaplanov</snm><fnm>A</fnm></au>
    <au><snm>Manna</snm><fnm>QL</fnm></au>
    <au><snm>Rigal</snm><fnm>J M</fnm></au>
    <au><snm>Esch</snm><fnm>PV</fnm></au>
  </aug>
  <source>Journal of Instrumentation</source>
  <pubdate>2014</pubdate>
  <volume>9</volume>
  <issue>03</issue>
  <fpage>P03007</fpage>
  <url>http://stacks.iop.org/1748-0221/9/i=03/a=P03007</url>
</bibl>

<bibl id="B18">
  <title><p>{Characterization of the Multi-Blade 10B-based detector at the
  CRISP reflectometer at ISIS for neutron reflectometry at ESS}</p></title>
  <aug>
    <au><snm>Piscitelli</snm><fnm>F.</fnm></au>
    <au><snm>Mauri</snm><fnm>G.</fnm></au>
    <au><snm>Messi</snm><fnm>F.</fnm></au>
    <au><snm>Anastasopoulos</snm><fnm>M.</fnm></au>
    <au><snm>Arnold</snm><fnm>T.</fnm></au>
    <au><snm>Glavic</snm><fnm>A.</fnm></au>
    <au><snm>H{\"o}glund</snm><fnm>C.</fnm></au>
    <au><snm>Ilves</snm><fnm>T.</fnm></au>
    <au><snm>Higuera</snm><fnm>IL</fnm></au>
    <au><snm>Pazmandi</snm><fnm>P.</fnm></au>
    <au><snm>Raspino</snm><fnm>D.</fnm></au>
    <au><snm>Robinson</snm><fnm>L.</fnm></au>
    <au><snm>Schmidt</snm><fnm>S.</fnm></au>
    <au><snm>Svensson</snm><fnm>P.</fnm></au>
    <au><snm>Varga</snm><fnm>D.</fnm></au>
    <au><snm>Hall Wilton</snm><fnm>R.</fnm></au>
  </aug>
  <source>Journal of Instrumentation</source>
  <pubdate>2018</pubdate>
  <volume>13</volume>
  <issue>05</issue>
  <fpage>P05009</fpage>
  <url>http://stacks.iop.org/1748-0221/13/i=05/a=P05009</url>
</bibl>

<bibl id="B19">
  <title><p>{Neutron reflectometry with the Multi-Blade 10B-based
  detector}</p></title>
  <aug>
    <au><snm>Mauri</snm><fnm>G.</fnm></au>
    <au><snm>Messi</snm><fnm>F.</fnm></au>
    <au><snm>Anastasopoulos</snm><fnm>M.</fnm></au>
    <au><snm>Arnold</snm><fnm>T.</fnm></au>
    <au><snm>Glavic</snm><fnm>A.</fnm></au>
    <au><snm>H{\"o}glund</snm><fnm>C.</fnm></au>
    <au><snm>Ilves</snm><fnm>T.</fnm></au>
    <au><snm>Lopez Higuera</snm><fnm>I.</fnm></au>
    <au><snm>Pazmandi</snm><fnm>P.</fnm></au>
    <au><snm>Raspino</snm><fnm>D.</fnm></au>
    <au><snm>Robinson</snm><fnm>L.</fnm></au>
    <au><snm>Schmidt</snm><fnm>S.</fnm></au>
    <au><snm>Svensson</snm><fnm>P.</fnm></au>
    <au><snm>Varga</snm><fnm>D.</fnm></au>
    <au><snm>Hall Wilton</snm><fnm>R.</fnm></au>
    <au><snm>Piscitelli</snm><fnm>F.</fnm></au>
  </aug>
  <source>Proceedings of the Royal Society of London A: Mathematical, Physical
  and Engineering Sciences</source>
  <publisher>The Royal Society</publisher>
  <pubdate>2018</pubdate>
  <volume>474</volume>
  <issue>2216</issue>
  <url>http://rspa.royalsocietypublishing.org/content/474/2216/20180266</url>
</bibl>

<bibl id="B20">
  <title><p>{The 10B based Jalousie neutron detector -- An alternative for 3He
  filled position sensitive counter tubes}</p></title>
  <aug>
    <au><snm>Henske</snm><fnm>M.</fnm></au>
    <au><snm>Klein</snm><fnm>M.</fnm></au>
    <au><snm>K{\"o}hli</snm><fnm>M.</fnm></au>
    <au><snm>Lennert</snm><fnm>P.</fnm></au>
    <au><snm>Modzel</snm><fnm>G.</fnm></au>
    <au><snm>Schmidt</snm><fnm>C.J.</fnm></au>
    <au><snm>Schmidt</snm><fnm>U.</fnm></au>
  </aug>
  <source>Nuclear Instruments and Methods in Physics Research Section A:
  Accelerators, Spectrometers, Detectors and Associated Equipment</source>
  <pubdate>2012</pubdate>
  <volume>686</volume>
  <fpage>151</fpage>
  <lpage>155</lpage>
  <url>http://www.sciencedirect.com/science/article/pii/S016890021200589X</url>
</bibl>

<bibl id="B21">
  <title><p>Diffraction measurements with a boron-based {GEM} neutron
  detector</p></title>
  <aug>
    <au><snm>Croci</snm><fnm>G.</fnm></au>
    <au><cnm>al.</cnm></au>
  </aug>
  <source>EPL</source>
  <pubdate>2014</pubdate>
  <volume>107</volume>
  <issue>1</issue>
  <fpage>12001</fpage>
  <url>http://dx.doi.org/10.1209/0295-5075/107/12001</url>
</bibl>

<bibl id="B22">
  <title><p>{Evolution in boron-based GEM detectors for diffraction
  measurements: from planar to 3D converters}</p></title>
  <aug>
    <au><snm>Albani</snm><fnm>G</fnm></au>
    <au><snm>Cippo</snm><fnm>EP</fnm></au>
    <au><snm>Croci</snm><fnm>G</fnm></au>
    <au><snm>Muraro</snm><fnm>A</fnm></au>
    <au><snm>Schooneveld</snm><fnm>E</fnm></au>
    <au><snm>Scherillo</snm><fnm>A</fnm></au>
    <au><snm>Hall Wilton</snm><fnm>R</fnm></au>
    <au><snm>Kanaki</snm><fnm>K</fnm></au>
    <au><snm>H{\"o}glund</snm><fnm>C</fnm></au>
    <au><snm>Hultman</snm><fnm>L</fnm></au>
    <au><snm>Birch</snm><fnm>J</fnm></au>
    <au><snm>Claps</snm><fnm>G</fnm></au>
    <au><snm>Murtas</snm><fnm>F</fnm></au>
    <au><snm>Rebai</snm><fnm>M</fnm></au>
    <au><snm>Tardocchi</snm><fnm>M</fnm></au>
    <au><snm>Gorini</snm><fnm>G</fnm></au>
  </aug>
  <source>Measurement Science and Technology</source>
  <pubdate>2016</pubdate>
  <volume>27</volume>
  <issue>11</issue>
  <fpage>115902</fpage>
  <url>http://stacks.iop.org/0957-0233/27/i=11/a=115902</url>
</bibl>

<bibl id="B23">
  <title><p>{Boron-coated straws as a replacement for 3He-based neutron
  detectors}</p></title>
  <aug>
    <au><snm>Lacy</snm><fnm>JL</fnm></au>
    <au><snm>Athanasiades</snm><fnm>A</fnm></au>
    <au><snm>Sun</snm><fnm>L</fnm></au>
    <au><snm>Martin</snm><fnm>CS</fnm></au>
    <au><snm>Lyons</snm><fnm>TD</fnm></au>
    <au><snm>Foss</snm><fnm>MA</fnm></au>
    <au><snm>Haygood</snm><fnm>HB</fnm></au>
  </aug>
  <source>Nuclear Instruments and Methods in Physics Research Section A:
  Accelerators, Spectrometers, Detectors and Associated Equipment</source>
  <pubdate>2011</pubdate>
  <volume>652</volume>
  <issue>1</issue>
  <fpage>359</fpage>
  <lpage>363</lpage>
  <url>http://www.sciencedirect.com/science/article/pii/S016890021001973X</url>
  <note>Symposium on Radiation Measurements and Applications (SORMA) \{XII\}
  2010</note>
</bibl>

<bibl id="B24">
  <title><p>{Efficiency and spatial resolution of the CASCADE thermal neutron
  detector}</p></title>
  <aug>
    <au><snm>K{\"o}hli</snm><fnm>M.</fnm></au>
    <au><snm>Allmendinger</snm><fnm>F.</fnm></au>
    <au><snm>H{\"a}u{\ss}ler</snm><fnm>W.</fnm></au>
    <au><snm>Schr{\"o}der</snm><fnm>T.</fnm></au>
    <au><snm>Klein</snm><fnm>M.</fnm></au>
    <au><snm>Meven</snm><fnm>M.</fnm></au>
    <au><snm>Schmidt</snm><fnm>U.</fnm></au>
  </aug>
  <source>Nuclear Instruments and Methods in Physics Research Section A:
  Accelerators, Spectrometers, Detectors and Associated Equipment</source>
  <pubdate>2016</pubdate>
  <volume>828</volume>
  <fpage>242</fpage>
  <lpage>-249</lpage>
  <url>http://www.sciencedirect.com/science/article/pii/S0168900216303722</url>
</bibl>

<bibl id="B25">
  <title><p>First measurements with new high-resolution gadolinium-{GEM}
  neutron detectors</p></title>
  <aug>
    <au><snm>Pfeiffer</snm><fnm>D.</fnm></au>
    <au><snm>Resnati</snm><fnm>F.</fnm></au>
    <au><snm>Birch</snm><fnm>J.</fnm></au>
    <au><snm>Etxegarai</snm><fnm>M.</fnm></au>
    <au><snm>Hall Wilton</snm><fnm>R.</fnm></au>
    <au><snm>H{\"o}glund</snm><fnm>C.</fnm></au>
    <au><snm>Hultman</snm><fnm>L.</fnm></au>
    <au><snm>Llamas Jansa</snm><fnm>I.</fnm></au>
    <au><snm>Oliveri</snm><fnm>E.</fnm></au>
    <au><snm>Oksanen</snm><fnm>E.</fnm></au>
    <au><snm>Robinson</snm><fnm>L.</fnm></au>
    <au><snm>Ropelewski</snm><fnm>L.</fnm></au>
    <au><snm>Schmidt</snm><fnm>S.</fnm></au>
    <au><snm>Streli</snm><fnm>C.</fnm></au>
    <au><snm>Thuiner</snm><fnm>P.</fnm></au>
  </aug>
  <source>Journal of Instrumentation</source>
  <pubdate>2016</pubdate>
  <volume>11</volume>
  <issue>05</issue>
  <fpage>P05011</fpage>
  <url>http://stacks.iop.org/1748-0221/11/i=05/a=P05011</url>
</bibl>

<bibl id="B26">
  <title><p>{The {$\mu$}TPC method: improving the position resolution of
  neutron detectors based on MPGDs}</p></title>
  <aug>
    <au><snm>Pfeiffer</snm><fnm>D.</fnm></au>
    <au><snm>Resnati</snm><fnm>F.</fnm></au>
    <au><snm>Birch</snm><fnm>J.</fnm></au>
    <au><snm>Hall Wilton</snm><fnm>R.</fnm></au>
    <au><snm>H{\"o}glund</snm><fnm>C.</fnm></au>
    <au><snm>Hultman</snm><fnm>L.</fnm></au>
    <au><snm>Iakovidis</snm><fnm>G.</fnm></au>
    <au><snm>Oliveri</snm><fnm>E.</fnm></au>
    <au><snm>Oksanen</snm><fnm>E.</fnm></au>
    <au><snm>Ropelewski</snm><fnm>L.</fnm></au>
    <au><snm>Thuiner</snm><fnm>P.</fnm></au>
  </aug>
  <source>Journal of Instrumentation</source>
  <pubdate>2015</pubdate>
  <volume>10</volume>
  <issue>04</issue>
  <fpage>P04004</fpage>
  <url>http://stacks.iop.org/1748-0221/10/i=04/a=P04004</url>
</bibl>

<bibl id="B27">
  <title><p>Amorphous silicon position sensitive neutron detector</p></title>
  <aug>
    <au><snm>Mireshghi</snm><fnm>A.</fnm></au>
    <au><snm>Cho</snm><fnm>G.</fnm></au>
    <au><snm>Drewery</snm><fnm>J.</fnm></au>
    <au><snm>Jing</snm><fnm>T.</fnm></au>
    <au><snm>Kaplan</snm><fnm>S. N.</fnm></au>
    <au><snm>Perez Mendez</snm><fnm>V.</fnm></au>
    <au><snm>Wildermuth</snm><fnm>D.</fnm></au>
  </aug>
  <source>IEEE Transactions on Nuclear Science</source>
  <pubdate>1992</pubdate>
  <volume>39</volume>
  <issue>4</issue>
  <fpage>635</fpage>
  <lpage>640</lpage>
</bibl>

<bibl id="B28">
  <title><p>The use of large area silicon sensors for thermal neutron
  detection</p></title>
  <aug>
    <au><snm>Schulte</snm><fnm>R.L.</fnm></au>
    <au><snm>Swanson</snm><fnm>F.</fnm></au>
    <au><snm>Kesselman</snm><fnm>M.</fnm></au>
  </aug>
  <source>Nuclear Instruments and Methods in Physics Research Section A:
  Accelerators, Spectrometers, Detectors and Associated Equipment</source>
  <pubdate>1994</pubdate>
  <volume>353</volume>
  <issue>1</issue>
  <fpage>123</fpage>
  <lpage>127</lpage>
  <url>http://www.sciencedirect.com/science/article/pii/0168900294916179</url>
</bibl>

<bibl id="B29">
  <title><p>Solid state neutron detectors</p></title>
  <aug>
    <au><snm>Petrillo</snm><fnm>C</fnm></au>
    <au><snm>Sacchetti</snm><fnm>F</fnm></au>
    <au><snm>Toker</snm><fnm>O</fnm></au>
    <au><snm>Rhodes</snm><fnm>N.J</fnm></au>
  </aug>
  <source>Nuclear Instruments and Methods in Physics Research Section A:
  Accelerators, Spectrometers, Detectors and Associated Equipment</source>
  <pubdate>1996</pubdate>
  <volume>378</volume>
  <issue>3</issue>
  <fpage>541</fpage>
  <lpage>551</lpage>
  <url>http://www.sciencedirect.com/science/article/pii/0168900296004871</url>
</bibl>

<bibl id="B30">
  <title><p>{Development of pulse shape analysis for noise reduction in
  Si-based neutron detectors}</p></title>
  <aug>
    <au><snm>Mauri</snm><fnm>G.</fnm></au>
    <au><snm>Mariotti</snm><fnm>M.</fnm></au>
    <au><snm>Casinini</snm><fnm>F.</fnm></au>
    <au><snm>Sacchetti</snm><fnm>F.</fnm></au>
    <au><snm>Petrillo</snm><fnm>C.</fnm></au>
  </aug>
  <source>Nuclear Instruments and Methods in Physics Research Section A:
  Accelerators, Spectrometers, Detectors and Associated Equipment</source>
  <pubdate>2018</pubdate>
  <volume>910</volume>
  <fpage>184</fpage>
  <lpage>193</lpage>
  <url>http://www.sciencedirect.com/science/article/pii/S0168900218312154</url>
</bibl>

<bibl id="B31">
  <title><p>Fast neutron sensitivity of neutron detectors based on Boron-10
  converter layers</p></title>
  <aug>
    <au><snm>Mauri</snm><fnm>G.</fnm></au>
    <au><snm>Messi</snm><fnm>F.</fnm></au>
    <au><snm>Kanaki</snm><fnm>K.</fnm></au>
    <au><snm>Hall Wilton</snm><fnm>R.</fnm></au>
    <au><snm>Karnickis</snm><fnm>E.</fnm></au>
    <au><snm>Khaplanov</snm><fnm>A.</fnm></au>
    <au><snm>Piscitelli</snm><fnm>F.</fnm></au>
  </aug>
  <source>Journal of Instrumentation</source>
  <pubdate>2018</pubdate>
  <volume>13</volume>
  <issue>03</issue>
  <fpage>P03004(arxiv:1712.05614)</fpage>
  <url>http://stacks.iop.org/1748-0221/13/i=03/a=P03004</url>
</bibl>

<bibl id="B32">
  <title><p>{The Multi-Blade Boron-10-based neutron detector for high intensity
  neutron reflectometry at ESS}</p></title>
  <aug>
    <au><snm>Piscitelli</snm><fnm>F.</fnm></au>
    <au><snm>Messi</snm><fnm>F.</fnm></au>
    <au><snm>Anastasopoulos</snm><fnm>M.</fnm></au>
    <au><snm>Bry{\'s}</snm><fnm>T.</fnm></au>
    <au><snm>Chicken</snm><fnm>F.</fnm></au>
    <au><snm>Dian</snm><fnm>E.</fnm></au>
    <au><snm>Fuzi</snm><fnm>J.</fnm></au>
    <au><snm>H{\"o}glund</snm><fnm>C.</fnm></au>
    <au><snm>Kiss</snm><fnm>G.</fnm></au>
    <au><snm>Orban</snm><fnm>J.</fnm></au>
    <au><snm>Pazmandi</snm><fnm>P.</fnm></au>
    <au><snm>Robinson</snm><fnm>L.</fnm></au>
    <au><snm>Rosta</snm><fnm>L.</fnm></au>
    <au><snm>Schmidt</snm><fnm>S.</fnm></au>
    <au><snm>Varga</snm><fnm>D.</fnm></au>
    <au><snm>Zsiros</snm><fnm>T.</fnm></au>
    <au><snm>Hall Wilton</snm><fnm>R.</fnm></au>
  </aug>
  <source>Journal of Instrumentation</source>
  <pubdate>2017</pubdate>
  <volume>12</volume>
  <issue>03</issue>
  <fpage>P03013</fpage>
  <url>http://stacks.iop.org/1748-0221/12/i=03/a=P03013</url>
</bibl>

<bibl id="B33">
  <source>{CRISP instrument manual 2010 -
  https://www.isis.stfc.ac.uk/Pages/crisp-instrument-manual-nov-2010.pdf}</source>
  <pubdate>2010</pubdate>
  <url>https://www.isis.stfc.ac.uk/Pages/crisp-instrument-manual-nov-2010.pdf</url>
</bibl>

<bibl id="B34">
  <source>{ISIS Neutron and Muon Source - https://www.isis.stfc.ac.uk}</source>
  <url>https://www.isis.stfc.ac.uk</url>
</bibl>

<bibl id="B35">
  <title><p>{The neutron tagging facility at Lund University}</p></title>
  <aug>
    <au><snm>Messi</snm><fnm>F.</fnm></au>
    <au><snm>Perrey</snm><fnm>H.</fnm></au>
    <au><snm>Fissum</snm><fnm>K.</fnm></au>
    <au><snm>Akkawi</snm><fnm>M.</fnm></au>
    <au><snm>Jebali</snm><fnm>RA</fnm></au>
    <au><snm>Annand</snm><fnm>J.R.M.</fnm></au>
    <au><snm>Bentley</snm><fnm>P.</fnm></au>
    <au><snm>Boyd</snm><fnm>L.</fnm></au>
    <au><snm>Cooper Jensen</snm><fnm>C.P.</fnm></au>
    <au><snm>DiJulio</snm><fnm>D.D.</fnm></au>
    <au><snm>Freita Ramos</snm><fnm>J.</fnm></au>
    <au><snm>Hall Wilton</snm><fnm>R.</fnm></au>
    <au><snm>Huusko</snm><fnm>A.</fnm></au>
    <au><snm>Ilves</snm><fnm>T.</fnm></au>
    <au><snm>Issa</snm><fnm>F.</fnm></au>
    <au><snm>Jalgen</snm><fnm>A.</fnm></au>
    <au><snm>Kanaki</snm><fnm>K.</fnm></au>
    <au><snm>Karnickis</snm><fnm>E.</fnm></au>
    <au><snm>Khaplanov</snm><fnm>A.</fnm></au>
    <au><snm>Koufigar</snm><fnm>S.</fnm></au>
    <au><snm>Maulerova</snm><fnm>V.</fnm></au>
    <au><snm>Mauri</snm><fnm>G.</fnm></au>
    <au><snm>Mauritzson</snm><fnm>N.</fnm></au>
    <au><snm>Pei</snm><fnm>W.</fnm></au>
    <au><snm>Piscitelli</snm><fnm>F.</fnm></au>
    <au><snm>Rofors</snm><fnm>E.</fnm></au>
    <au><snm>Scherzinger</snm><fnm>J.</fnm></au>
    <au><snm>Soderhielm</snm><fnm>H.</fnm></au>
    <au><snm>Soderstrom</snm><fnm>D.</fnm></au>
    <au><snm>Stefanescu</snm><fnm>I.</fnm></au>
  </aug>
  <source>arXiv:1711.10286 (submitted to IAEA Technical Report on Modern
  Neutron Detection (2017))</source>
  <pubdate>2017</pubdate>
</bibl>

<bibl id="B36">
  <title><p>General Electric (GE) Reuter Stokes</p></title>
  <source>https://www.industrial.ai</source>
  <url>https://www.industrial.ai</url>
</bibl>

<bibl id="B37">
  <title><p>{Tagging fast neutrons from an 241Am/9Be source}</p></title>
  <aug>
    <au><snm>Scherzinger</snm><fnm>J.</fnm></au>
    <au><snm>Annand</snm><fnm>J.R.M.</fnm></au>
    <au><snm>Davatz</snm><fnm>G.</fnm></au>
    <au><snm>Fissum</snm><fnm>K.G.</fnm></au>
    <au><snm>Gendotti</snm><fnm>U.</fnm></au>
    <au><snm>Hall Wilton</snm><fnm>R.</fnm></au>
    <au><snm>H{\aa}kansson</snm><fnm>E.</fnm></au>
    <au><snm>Jebali</snm><fnm>R.</fnm></au>
    <au><snm>Kanaki</snm><fnm>K.</fnm></au>
    <au><snm>Lundin</snm><fnm>M.</fnm></au>
    <au><snm>Nilsson</snm><fnm>B.</fnm></au>
    <au><snm>Rosborge</snm><fnm>A.</fnm></au>
    <au><snm>Svensson</snm><fnm>H.</fnm></au>
  </aug>
  <source>Applied Radiation and Isotopes</source>
  <pubdate>2015</pubdate>
  <volume>98</volume>
  <fpage>74</fpage>
  <lpage>79</lpage>
  <url>//www.sciencedirect.com/science/article/pii/S0969804315000044</url>
</bibl>

<bibl id="B38">
  <title><p>{A comparison of untagged gamma-ray and tagged-neutron yields from
  241AmBe and 238PuBe sources}</p></title>
  <aug>
    <au><snm>Scherzinger</snm><fnm>J.</fnm></au>
    <au><snm>Jebali</snm><fnm>RA</fnm></au>
    <au><snm>Annand</snm><fnm>J.R.M.</fnm></au>
    <au><snm>Fissum</snm><fnm>K.G.</fnm></au>
    <au><snm>Hall Wilton</snm><fnm>R.</fnm></au>
    <au><snm>Koufigar</snm><fnm>S.</fnm></au>
    <au><snm>Mauritzson</snm><fnm>N.</fnm></au>
    <au><snm>Messi</snm><fnm>F.</fnm></au>
    <au><snm>Perrey</snm><fnm>H.</fnm></au>
    <au><snm>Rofors</snm><fnm>E.</fnm></au>
  </aug>
  <source>Applied Radiation and Isotopes</source>
  <pubdate>2017</pubdate>
  <volume>127</volume>
  <fpage>98</fpage>
  <lpage>102</lpage>
  <url>http://www.sciencedirect.com/science/article/pii/S0969804316309861</url>
</bibl>

<bibl id="B39">
  <title><p>{Characterisation of the Spatial Resolution and the Gamma-ray
  Discrimination of Helium-3 Proportional Counters}</p></title>
  <aug>
    <au><snm>Rossi</snm><fnm>E</fnm></au>
  </aug>
  <source>Master's thesis</source>
  <pubdate>2017</pubdate>
</bibl>

<bibl id="B40">
  <title><p>{Measurement of Fast Neutron Spectrum with Helium-3 Proportional
  Counter}</p></title>
  <aug>
    <au><cnm>{Tsutomu IIJIMA and Takehiko MUKAIYAMA and Keisho
  SHIRAKATA}</cnm></au>
  </aug>
  <source>Journal of Nuclear Science and Technology</source>
  <publisher>Taylor & Francis</publisher>
  <pubdate>1971</pubdate>
  <volume>8</volume>
  <issue>4</issue>
  <fpage>192</fpage>
  <lpage>200</lpage>
</bibl>

<bibl id="B41">
  <title><p>NIST National Nuclear Data Center</p></title>
  <source>https://www.nndc.bnl.gov</source>
  <url>https://www.nndc.bnl.gov</url>
</bibl>

<bibl id="B42">
  <title><p>{Geant4 - A Simulation Toolkit}</p></title>
  <aug>
    <au><cnm>{S.~Agostinelli, J.~Allison \textit{et al}.}</cnm></au>
  </aug>
  <source>Nucl. Instrum. Meth. A</source>
  <pubdate>2003</pubdate>
  <volume>506</volume>
  <fpage>250</fpage>
  <lpage>303</lpage>
</bibl>

<bibl id="B43">
  <title><p>{Geant4 developments and applications}</p></title>
  <aug>
    <au><cnm>{J.~Allison \textit{et al}.}</cnm></au>
  </aug>
  <source>IEEE Trans. Nucl. Sci.</source>
  <pubdate>2006</pubdate>
  <volume>53</volume>
  <fpage>07</fpage>
</bibl>

<bibl id="B44">
  <title><p>{Recent Developments in Geant4}</p></title>
  <aug>
    <au><cnm>{J.~Allison, K.~Amako \textit{et al}.}</cnm></au>
  </aug>
  <source>Nucl. Instrum. Meth. A</source>
  <pubdate>2016</pubdate>
  <volume>835</volume>
  <fpage>186</fpage>
  <lpage>225</lpage>
</bibl>

<bibl id="B45">
  <title><p>{Geant4 Based Simulations for Novel Neutron Detector
  Development}</p></title>
  <aug>
    <au><cnm>{T.~Kittelmann \textit{et al}.}</cnm></au>
  </aug>
  <source>J. Phys. Conf. Ser.</source>
  <pubdate>2014</pubdate>
  <volume>513</volume>
  <fpage>022017</fpage>
</bibl>

<bibl id="B46">
  <title><p>Polycrystalline neutron scattering for {Geant4}:
  {NXSG4}</p></title>
  <aug>
    <au><snm>Kittelmann</snm><fnm>T.</fnm></au>
    <au><snm>Boin</snm><fnm>M.</fnm></au>
  </aug>
  <source>Computer Physics Communications</source>
  <pubdate>2015</pubdate>
  <volume>189</volume>
  <issue>0</issue>
  <fpage>114</fpage>
  <lpage>118</lpage>
</bibl>

<bibl id="B47">
  <title><p>Using backscattering to enhance efficiency in neutron
  detectors</p></title>
  <aug>
    <au><snm>Kittelmann</snm><fnm>T.</fnm></au>
    <au><snm>Kanaki</snm><fnm>K.</fnm></au>
    <au><snm>Klinkby</snm><fnm>E.</fnm></au>
    <au><snm>Cai</snm><fnm>X. X.</fnm></au>
    <au><snm>Cooper Jensen</snm><fnm>C. P.</fnm></au>
    <au><snm>Hall Wilton</snm><fnm>R.</fnm></au>
  </aug>
  <source>IEEE TNS</source>
  <pubdate>2017</pubdate>
  <volume>64</volume>
  <issue>6</issue>
  <fpage>1562</fpage>
  <lpage>1573</lpage>
</bibl>

<bibl id="B48">
  <title><p>{GEANT 4 simulation of neutron transport and scattering in
  media}</p></title>
  <aug>
    <au><snm>Deiev</snm><fnm>O. S.</fnm></au>
  </aug>
  <source>Voprosy Atomnoj Nauki i Tekhniki</source>
  <publisher>Ukraine</publisher>
  <pubdate>2013</pubdate>
  <issue>3-85/60</issue>
  <fpage>236</fpage>
  <lpage>241</lpage>
  <url>http://inis.iaea.org/search/search.aspx?orig_q=RN:46029253</url>
</bibl>

<bibl id="B49">
  <title><p>{The new small-angle neutron scattering instrument SANS-1 at
  MLZ---characterization and first results}</p></title>
  <aug>
    <au><snm>M{\"u}hlbauer</snm><fnm>S.</fnm></au>
    <au><snm>Heinemann</snm><fnm>A.</fnm></au>
    <au><snm>Wilhelm</snm><fnm>A.</fnm></au>
    <au><snm>Karge</snm><fnm>L.</fnm></au>
    <au><snm>Ostermann</snm><fnm>A.</fnm></au>
    <au><snm>Defendi</snm><fnm>I.</fnm></au>
    <au><snm>Schreyer</snm><fnm>A.</fnm></au>
    <au><snm>Petry</snm><fnm>W.</fnm></au>
    <au><snm>Gilles</snm><fnm>R.</fnm></au>
  </aug>
  <source>Nuclear Instruments and Methods in Physics Research Section A:
  Accelerators, Spectrometers, Detectors and Associated Equipment</source>
  <pubdate>2016</pubdate>
  <volume>832</volume>
  <fpage>297</fpage>
  <lpage>305</lpage>
  <url>http://www.sciencedirect.com/science/article/pii/S0168900216306714</url>
</bibl>

<bibl id="B50">
  <title><p>Investigation of gamma-ray sensitivity of neutron detectors based
  on thin converter films</p></title>
  <aug>
    <au><snm>Khaplanov</snm><fnm>A</fnm></au>
    <au><snm>Piscitelli</snm><fnm>F</fnm></au>
    <au><snm>Buffet</snm><fnm>J C</fnm></au>
    <au><snm>Clergeau</snm><fnm>J F</fnm></au>
    <au><snm>Correa</snm><fnm>J</fnm></au>
    <au><snm>Esch</snm><fnm>P</fnm></au>
    <au><snm>Ferraton</snm><fnm>M</fnm></au>
    <au><snm>Guerard</snm><fnm>B</fnm></au>
    <au><snm>Hall Wilton</snm><fnm>R</fnm></au>
  </aug>
  <source>Journal of Instrumentation</source>
  <pubdate>2013</pubdate>
  <volume>8</volume>
  <issue>10</issue>
  <fpage>P10025</fpage>
  <url>http://stacks.iop.org/1748-0221/8/i=10/a=P10025</url>
</bibl>

</refgrp>
} 


\end{backmatter}
\end{document}